\documentclass[11pt]{article}
\usepackage{benstyle}
\pdfoutput=1

\usepackage{enumerate}
\usepackage[dvipsnames]{xcolor}
\usepackage{graphicx}
\usepackage[hidelinks]{hyperref}

\title{Do log factors matter? On optimal wavelet approximation and the foundations of compressed sensing}
\author{Ben Adcock$^*$, Simone Brugiapaglia$^\dag$ and Matthew King--Roskamp$^*$ \\ 
\\
$^*$ Department of Mathematics\\ Simon Fraser University\\ Canada\\
\\
$^\dag$ Department of Mathematics and Statistics\\ Concordia University\\ Canada}

\begin{document}

\pagenumbering{arabic}
\setcounter{page}{1}

\maketitle

\begin{abstract}
A signature result in compressed sensing is that Gaussian random sampling achieves stable and robust recovery of sparse vectors under optimal conditions on the number of measurements.  However, in the context of image reconstruction, it has been extensively documented that sampling strategies based on Fourier measurements outperform this purportedly optimal approach.  Motivated by this seeming paradox, we investigate the problem of optimal sampling for compressed sensing.  Rigorously combining the theories of wavelet approximation and infinite-dimensional compressed sensing, our analysis leads to new error bounds in terms of the total number of measurements $m$ for the approximation of piecewise $\alpha$-H\"{o}lder functions. Our theoretical findings suggest that Fourier sampling outperforms random Gaussian sampling when the H\"older exponent $\alpha$ is large enough. Moreover, we establish a provably optimal sampling strategy.  This work is an important first  step towards the resolution of the claimed paradox, and provides a clear theoretical justification for the practical success of compressed sensing techniques in imaging problems.
\end{abstract}



\section{Introduction}

Compressed sensing asserts that a vector $x \in \bbC^N$ with at most $s$ nonzero components can be recovered from $m$ suitably-chosen linear measurements $y = A x$, where $A \in \bbC^{m \times N}$ and $y \in \bbC^m$, with $m$ satisfying
\be{
\label{Gaussmeascondintro}
m \geq c \cdot s \cdot \log(N/s).
}
This can be achieved, for instance, by using a random Gaussian matrix for $A$ and by solving the \textit{basis pursuit} problem
\be{
\label{BP}
\min_{z \in \bbC^N} \nm{z}_{\ell^1}\ \mbox{subject to $A z  =y$}.
} 
In practice, \R{Gaussmeascondintro} can represent a significant saving in the number of measurements over classical approaches, and for this reason compressed sensing has found use in many different applications in science and engineering.  In fact, compressed sensing is \textit{optimal} for the recovery of sparse vectors.  No stable method (that is, one which is robust to perturbations in $x$) can recover sparse vectors from asymptotically fewer than $s \log(N/s)$ measurements.  

Imaging lies at the foundation of compressed sensing, and has been one of its key beneficiaries \cite{adcock2021compressive}.  Natural images have approximately sparse wavelet coefficients, and compressed sensing allows for the reconstruction of an image up to its best $s$-term approximation error from a few as $m \approx c \cdot s \cdot \log(N/s)$ measurements. Magnetic Resonance Imaging (MRI), for instance, was one of the original motivations for compressed sensing -- indeed, it was considered in the seminal paper of Cand\`es, Romberg \& Tao \cite{CandesRombergTao} -- and has been one of its most fruitful areas of application \cite{Lustig3,Lustig}. Such methods were approved for commercial use in MRI by the US FDA in 2017 \cite{FesslerOptMRI}. 
In the realm of optical imaging, the single-pixel camera \cite{SinglePixelCamera} was one of the first empirical demonstrations of compressed sensing principles, and its various progenitors such as lensless imaging \cite{BoominathanEtAlLensless,Lensless} continue to be active areas of investigation \cite{GehmBradyCSOpt}. Other modalities, including X-ray CT \cite{GraffSidkyCSMedImag}, infrared imaging \cite{WillettCompOpt}, spectral imaging \cite{CSCodedAp}, light-field imaging \cite{MarwahEtAlLightField}, ghost imaging \cite{CompGhostImaging}, STORM \cite{STORMImaging}, holography \cite{CompHolo}, fluorescence microscopy \cite{Candes_PNAS}, NMR \cite{HollandEtAlNMR1,KazimierczukEtAl}, radio interferometry \cite{CSRadioInterferometry}, to name but a few, have all benefitted from compressed sensing approaches.

\subsection{A paradox}

Many imaging modalities such as MRI acquire Fourier samples of an image, and not measurements according to a random Gaussian matrix.  The best known measurement condition for $s$-term recovery from Fourier measurements is $m \approx c \cdot s \cdot \log^3(s) \cdot\log^2(N)$ (see \S\ref{ss:previous}), which has a worse scaling in $s$ and $N$ than the optimal condition \R{Gaussmeascondintro} for Gaussian measurements.

\begin{figure}[t]
\centering
\begin{tabular}{@{\hspace{0pt}}c@{\hspace{3pt}}c@{\hspace{3pt}}c@{\hspace{0pt}}}
\includegraphics[width = 4.9cm]{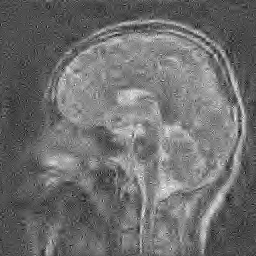} &
\includegraphics[width = 4.9cm]{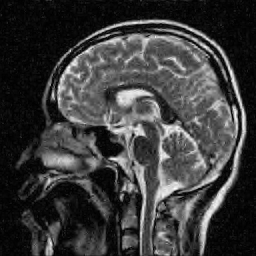} &
\includegraphics[width = 4.9cm]{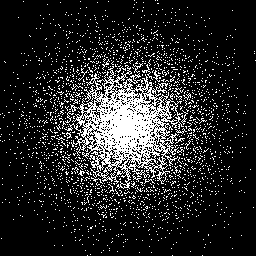}
\\
\includegraphics[width = 4.9cm]{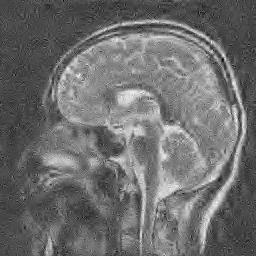} &
\includegraphics[width = 4.9cm]{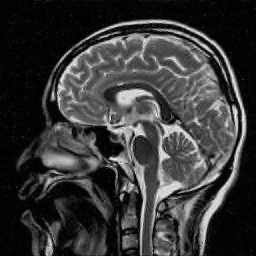} &
\includegraphics[width = 4.9cm]{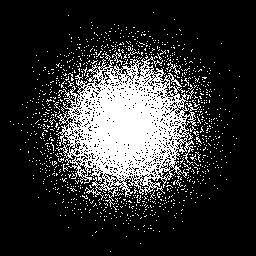}
\end{tabular}
\caption{Reconstruction of a brain image (original image can be found here \cite{BrainImg}) of resolution $N = 256 \times 256$ from $m/N = 15\%$ (top row) and $m/N = 20\%$ (bottom row) samples.  Left: reconstruction from Gaussian measurements.  Middle: reconstruction from Fourier measurements.  Right: the Fourier sampling strategies used.  Each white dot represents a frequency in $k$-space sampled.}
\label{f:FourierGauss}
\end{figure}

However, in practice, Fourier measurements outperform Gaussian measurements for recovering images.  A typical example is shown in Fig.~\ref{f:FourierGauss}.  With the same total number of measurements, reconstructing from a suitably-chosen set of Fourier samples gives a significantly better reconstruction.  As is standard in imaging, in this figure a db4 wavelet basis is used as a sparsifying transform.

Motivated by this paradox, the focus of this paper is the wavelet approximation of piecewise smooth functions via compressed sensing.  Specifically, we investigate the following three questions:

\begin{enumerate}
\item[(Q1)] Is a random Gaussian sampling an optimal sampling strategy for wavelet approximation of piecewise smooth functions?  Specifically, does it achieve optimal approximation rates in terms of $m$?
\item[(Q2)] If not, what is an optimal sampling strategy?
\item[(Q3)] How close to optimal is Fourier sampling?  In particular, why is it that Fourier sampling often outperforms random Gaussian sampling, even though the latter is optimal for recovering sparse vectors?
\end{enumerate}
The observation made in Fig.~\ref{f:FourierGauss} that Fourier sampling outperforms random Gaussian sampling for imaging has been well-documented \cite{OptimalSamplingQuest,BastounisEtAlSIAMNews,AsymptoticCS}, but not yet rigorously explained.  However, it arguably lies at the heart as to why compressed sensing has proved so effective for these applications.  In particular, it explains why modalities such as MRI have benefitted significantly from compressed sensing principles, despite their measurements being seemingly suboptimal.\footnote{This also provides some explanation as to why attempts to modify devices such as MR scanners to produce Gaussian-like measurements (see, for example, \cite{HaldarRandomEncoding,VanderEtAlSpreadSpectrum,QuEtAlSpreadSpectrum}) have not been widely adopted.}

\subsection{Our contributions}\label{ss:whatwedo}

We address these questions using the language of nonlinear approximation theory of piecewise smooth functions. Consider a function $f : [0,1] \rightarrow \bbR$ that is piecewise $\alpha$-H\"older continuous ($\alpha \geq 1$). As such, $f$ can have a finite number of discontinuities and the parameter $\alpha$ measures its smoothness between two consecutive points of discontinuity (see Definition~\ref{def:alpha-Holder}). In this paper, we consider the class of piecewise $\alpha$-H\"older continuous functions as an idealized one-dimensional model for images, where discontinuities correspond to edges and intervals of smoothness correspond to areas with a gradual grayscale variation (in analogy with cartoon-like images \cite{candes2004new}). See \S \ref{s:conclusions} for some discussion on the two-dimensional case.

In this setting, we first show that random Gaussian sampling combined with a  decoder based on orthonormal wavelets and basis pursuit \R{BP} gives an approximation $\tilde{f}_m$ satisfying
\be{
\label{GaussErr}
\nmu{f - \tilde{f}_m }_{L^2} \leq C (\log(m))^{\alpha} / m^{\alpha},
}
for some $C >0$. Here, by Gaussian sampling we mean a measurement strategy of the form $y = Ax$, where $x$ collects the wavelet coefficients of $f$ and $A$ has i.i.d.\ random Gaussian entries. Moreover, the wavelet coefficients of $\tilde{f}_m$ are computed by solving problem \eqref{BP}. The recovery error bound \eqref{GaussErr} is optimal `up to a logarithmic factor', and for small $\alpha$ one may not be inclined to worry.  Yet this is hardly satisfactory for moderate to large $\alpha$. Although we are not able to rigorously prove that the suboptimal factor $(\log(m))^{\alpha}$ cannot be removed from \eqref{GaussErr} (and that (Q1) admits a negative answer), we have strong reasons to conjecture that this is the case (see Remark~\ref{rmk:lower_bound_Gauss} for a more detailed discussion).  

On the other hand, in answer to (Q2) we show that there exists a sampling strategy which, when combined with the same decoder (orthonormal wavelets and basis pursuit), achieves the optimal error bound
\be{
\label{OptErr}
\nmu{f - \tilde{f}_m }_{L^2} \leq C / m^{\alpha}.
}
(In this section, the symbol $C$ denotes possibly different constants.) Unfortunately, this strategy does not employ Fourier measurements, rendering it inapplicable in many practical problems.  Focusing on this case, we prove that it is possible to construct a Fourier sampling strategy that achieves an error bound
\be{
\label{FourErr}
\nmu{f - \tilde{f}_m }_{L^2} \leq C (\log(m))^{13/4+\delta}/ m^{\alpha},
}
for any $0 < \delta < 1$ and where the coefficients of $\tilde{f}_m$ are computed by solving a weighted $\ell^1$-minimization problem.  This result pertains to question (Q3).  Its main implication is that for sufficiently regular functions (specifically, $\alpha > 13/4$) Fourier sampling outperforms random Gaussian sampling \R{GaussErr}. (For the sake of clarity, we note that the recovery error estimate \eqref{OptErr} is \emph{nonuniform}, i.e.\ it holds with high probability for any fixed $f$, whereas \eqref{FourErr} and \eqref{GaussErr} are  \emph{uniform}, i.e.\ they hold with high probability for all picewise $\alpha$-H\"older continuous functions $f$; a uniform result analogous to \eqref{OptErr} holds with an extra factor $\sqrt{\log(m)}$.) Like with \R{GaussErr} we do not claim that the $13/4$ factor in \R{FourErr} is sharp -- we expect it can be improved (see \S \ref{s:conclusions} for some further discussion).

Let us return to the question posed in the title. Do log factors matter? Our results lead to answer this question with a sound ``Yes''. In fact, the different log factors in \eqref{GaussErr}, \eqref{OptErr}, and \eqref{FourErr} are the key to address (Q1), (Q2), and (Q3) from a rigorous perspective. We now also highlight several key features of our analysis:
\begin{enumerate}
\item We, for arguably the first time, rigorously connect compressed sensing theory to the classical theory of nonlinear approximation using wavelets.
\item We work in the infinite-dimensional setting.  Compressed sensing is customarily presented in the finite-dimensional setting of vectors and matrices, whereas the concern of wavelet approximation theory is, of course, functions in function spaces.  Here we work directly with functions $f : [0,1] \rightarrow \bbR$ using the framework of \textit{infinite-dimensional} compressed sensing \cite{BAACHGSCS}.  This setup avoids errors due to discretization (e.g.\ related the wavelet crime).  Our analysis also takes care to incorporate all sources of approximation error, for instance, those due to truncation.\footnote{In particular, by `Fourier measurements' we mean samples of the continuous Fourier transform of $f$, not its discrete Fourier transform.  Not only is this more convenient for the analysis, it is also more relevant in practice, since modalities such as MRI are based on the continuous Fourier transform \cite{BAGSAIEP}.}
\item Our approach is essentially a \textit{black box}. 
Given a smoothness parameter $\alpha$ and a budget of measurements $m$, our recipe determines the correct samples to acquire, and from them finds an approximation $\tilde{f}_m$ satisfying the above bounds.  No other inputs are required.  In particular, unfeasible conditions such as bounds on the expansion error of $f$ (which are quite common in the compressed sensing literature) are not required.
\item In order to construct $\tilde{f}_m$, our recipe uses Daubechies' wavelets and solves a standard, finite-dimensional (weighted) $\ell^1$ minimization problem such as basis pursuit \R{BP}.  This is very similar to standard implementations of compressed sensing in most practical applications.  In particular, we do not make use of any exotic `structure-promoting' decoders, these typically being difficult to implement in large-scale problems (see \S\ref{ss:modelbaseddiscuss}).
\item Our results \R{GaussErr} and \R{FourErr} are uniform in $f$.  That is, given $m$, we construct a sampling strategy and decoder that guarantees these error bounds (with high probability) for all piecewise $\alpha$-H\"older continuous functions.  As noted above, \R{GaussErr} is nonuniform (that is, specific to the function $f$), but we prove a uniform version with an additional $\sqrt{\log(m)}$ factor.
\item Our result for Fourier sampling relies on recent advances in compressed sensing theory based on local, as opposed to global, structure.  Specifically, we use the principles of \textit{local sparsity in levels} and \textit{multilevel random subsampling} \cite{AHPRBreaking} to finely tune the sampling strategy to give the approximation result \R{FourErr}.  To the best of our knowledge, this result cannot be achieved using standard, sparsity-based, compressed sensing theory.
\end{enumerate}

\subsection{Structure is key}\label{ss:discussion}

How are the results \R{OptErr} and \R{FourErr} possible?  The answer lies with the structure of wavelet coefficients.  Random Gaussian measurements exploit the approximate sparsity of wavelet coefficients.  In particular, a piecewise $\alpha$-H\"older continuous function has a best $s$-term wavelet approximation error decaying like $s^{-\alpha}$ (see Theorem \ref{t:waveletsterm}).  The bound \R{GaussErr} follows almost directly from this and the measurement condition \R{Gaussmeascondintro}. 

By contrast, the Fourier and optimal sampling strategies exploit both the sparsity and the distribution of the wavelet coefficients.  In particular, they exploit the following properties:
\bull{
\item Asymptotically, all but $\ord{\log(s)}$ of the significant wavelet coefficients are located at coarse scales.  That is, the sparsity at coarse scales satisfies $s_{\mathrm{coarse}} = s - \ord{\log(s)}$.  
\item The coarse scales are \textit{saturated}.  Coefficients at these scales are \textit{nonsparse}, i.e.\ all coefficients contribute to the best $s$-term approximation.  
\item At fine scales, wavelet coefficients are \textit{sparse}, with the number of significant coefficients being roughly $s_{\mathrm{fine}} = \ord{\log(s)}$. 
}
Since the coarse scales are saturated, they can be recovered efficiently using `classical' noncompressive measurements.  Specifically, only $m_{\mathrm{coarse}} = s_{\mathrm{coarse}}$ suitably-chosen measurements are required to recover the corresponding coefficients.  Conversely, the fine scales are recovered using compressive measurements, requiring $m_{\mathrm{fine}} \approx c \cdot s_{\mathrm{fine}} \cdot \log(N/s_{\mathrm{fine}})$ measurements, where $N \approx s^{2 \alpha + 1}$ is the range of indices in which the largest $s$ wavelet coefficients live.  Hence, as $s \rightarrow \infty$, we have
\bes{
m = m_{\mathrm{coarse}} + m_{\mathrm{fine}} = s + o(s).
}
This, in combination with the $s^{-\alpha}$ approximation rate, leads to the optimal bound \R{GaussErr}.

This works for the optimal sampling strategy since we have the luxury to choose the measurements. The situation is significantly more complicated for Fourier sampling.  In this case, we exploit the following fundamental property of wavelets:

\pbk
(P) Wavelets are concentrated in frequency.  The Fourier transform of a wavelet at scale $k$ is essentially supported in a dyadic band $B_k$ in frequency.

\pbk
Let $s_k$ be the sparsity of the wavelet coefficients at scale $k$.  At coarse scales, $s_k = 2^k$, where $2^k$ is the size of the scale, and at fine scales $s_k = \ord{ \log(s)}$.  The Fourier sampling strategy therefore proceeds as follows.  At coarse scales, it fully samples the band $B_k$, using $m_k = |B_k| = 2^k = s_k$ measurements.  At fine scales, it randomly subsamples from the band $B_k$, using $m_k = c \cdot s_k \cdot \mathrm{polylog}(s_k,N)$ measurements.  Hence, as $s \rightarrow \infty$, we have
\bes{
m = m_1+\ldots+m_r = s + o(s),
}
where $r \approx \log_2(N)$ is the maximum scale and where $N \approx m^{\max\{2\alpha+1,\frac{\alpha}{\alpha-1/2}\}}$.
As with the optimal strategy, the key here is that Fourier measurements can efficiently recover the coarse scale wavelet coefficients.

\subsection{Is Fourier sampling optimal?}\label{ss:FourOptdiscuss}

Having read the above argument, the reader may wonder why \R{FourErr} involves a logarithmic factor at all? The challenge stems from the word `essentially' appearing in property (P).  While wavelets at scale $k$ are concentrated in the band $B_k$, they are not fully supported there.  Hence, the pleasant scenario of simply choosing $m_k$ measurements per band according to the corresponding $s_k$ is not realized in practice.  Dealing with these inter-scale \textit{interferences} is a major technical hurdle, and culminates in the log term in \R{FourErr}.  We discuss this term in more detail in \S\ref{ss:logfactor_Fourier} and, in particular, the prospect of reducing the exponent $13/4$, in \S\ref{s:conclusions}.

\begin{figure}[t]
\centering
\includegraphics[width = 0.47\textwidth]{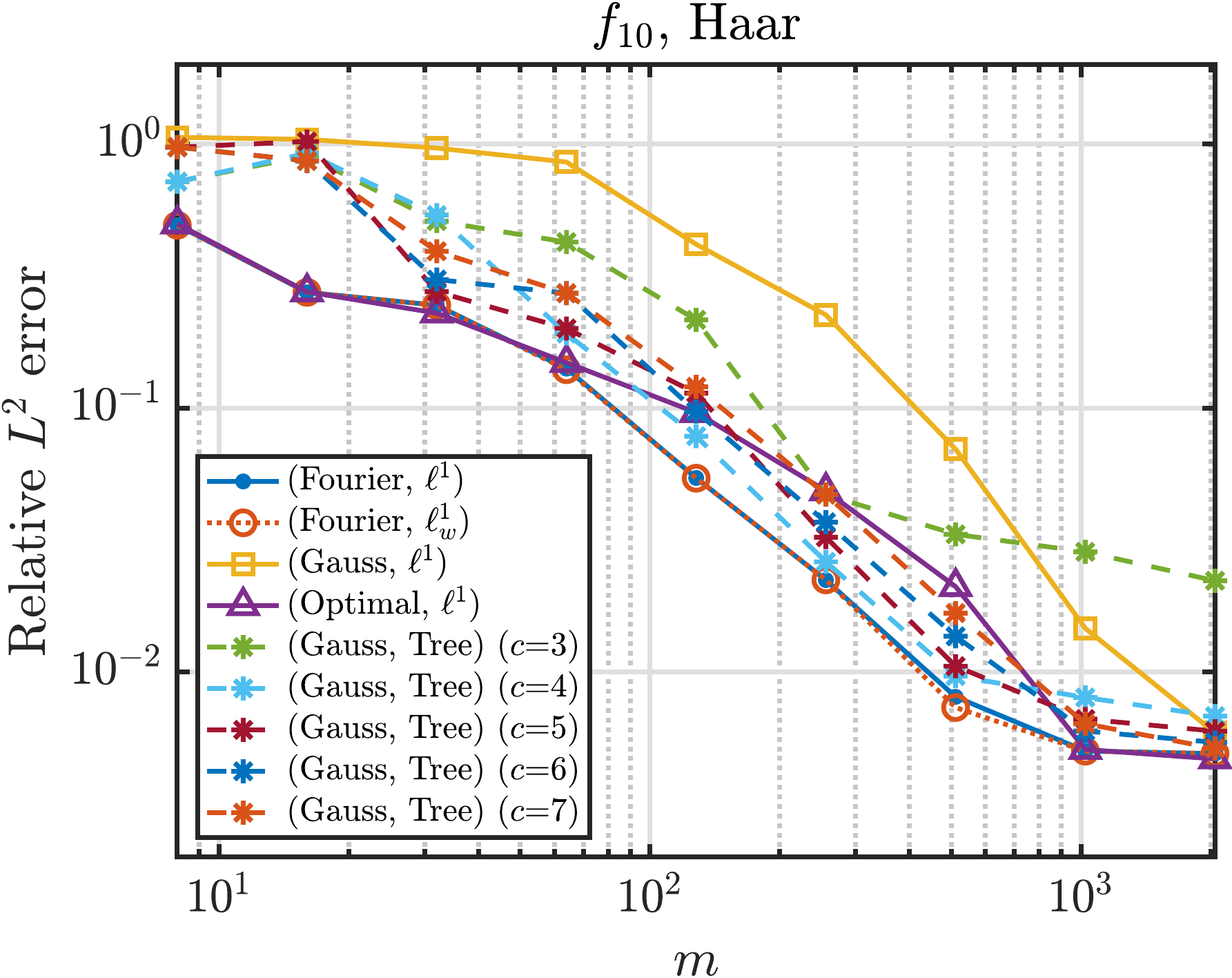}
\includegraphics[width = 0.47\textwidth]{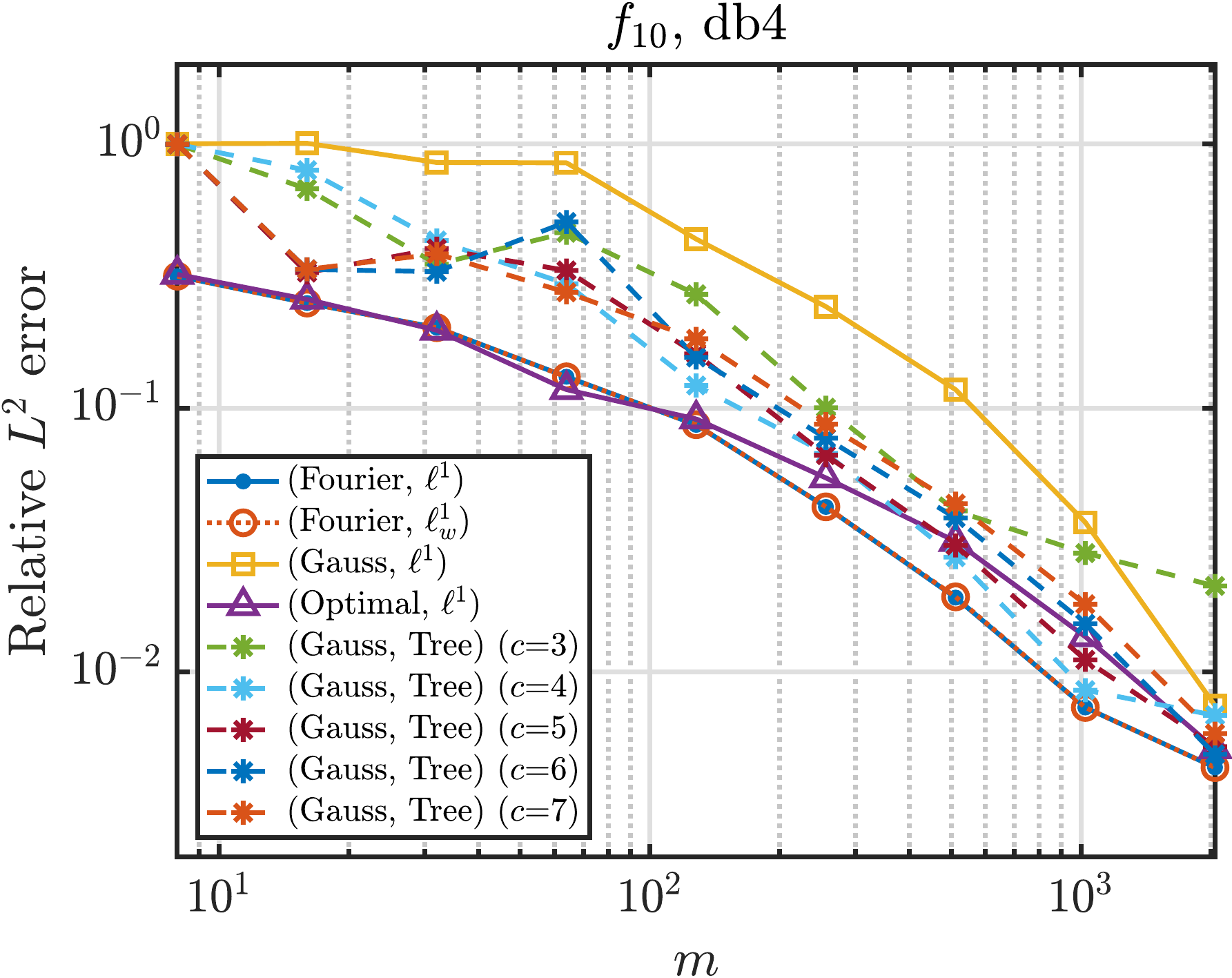}
\caption{Comparison of different sampling and recovery strategies for the approximation of a piecewise $\alpha$-H\"{o}lder function with 10 discontinuities, defined as in \eqref{eq:def_fk}, using Haar (left) and db4 wavelets (right). The strategies (Gauss,~$\ell^1$) and (Optimal,~$\ell^1$) correspond to \eqref{GaussErr} and \eqref{OptErr}, respectively. The strategies  (Fourier,~$\ell^1$) and (Fourier,~$\ell^1_w$) correspond to \eqref{FourErr}. Recovery is performed via \eqref{BP} for (Gauss,~$\ell^1$), (Optimal,~$\ell^1$), and (Fourier,~$\ell^1$). A weighted version of \eqref{BP} is considered for (Fourier,~$\ell^1_w$). (Gauss,~Tree) corresponds to random Gaussian sampling combined with tree-structured CoSaMP recovery, as proposed in \cite{BaranuikModelCS}, where $c$ is a tuning parameter of the method. Further details are provided in Appendix~\ref{s:NumExp}.}
\label{fig:intro_comp}
\end{figure}

Nevertheless, our main result \R{FourErr} suggests that Fourier sampling will, asymptotically, outperform Gaussian sampling for sufficiently large $\alpha$.  In practice, this always appears to be the case.  Unexpectedly in fact, Fourier sampling also outperforms the theoretically optimal strategy in practice.  Several examples of this are shown in Fig.~\ref{fig:intro_comp}.
Note that \R{BP} is used in the case of Gaussian sampling. For Fourier sampling, although \R{FourErr} is based on recovery via weighted $\ell^1$ minimization, results using both weighted and unweighted $\ell^1$ minimization are shown in Fig.~\ref{fig:intro_comp}.
 
Our focus in this paper is on proving theoretical statements such as \R{FourErr}.  As part of this, we prescribe a specific Fourier sampling strategy (by this, we mean a specific set of frequencies to sample) suitable for the class of piecewise $\alpha$-H\"older functions.  Unsurprisingly, in practice, further improvements can be achieved by empirically tuning the sampling strategy to better capture the structure of the images being reconstructed.  We refer to \cite{AsymptoticCS} for a demonstration of the significant practical benefits of doing this in various different modalities.

\subsection{On structure-promoting decoders}\label{ss:modelbaseddiscuss}

As discussed in \S\ref{ss:discussion}, the methodology employed in this paper designs a sampling strategy based on the \textit{structured sparsity} of the wavelet coefficients.  There is a line of work in compressed sensing that seeks to exploit structured sparsity by changing the decoder, e.g.\ by replacing the $\ell^1$-norm in \R{BP} by a different convex penalty, or by using iterative or greedy algorithms \cite{EldarDuarteCSReview,DuarteEldarStructuredCS}.  For example, \cite{BaranuikModelCS} considers a modification of the CoSaMP algorithm that promotes the \textit{connected tree} structure of wavelet coefficients.  Theoretical analysis shows that for random Gaussian sampling this decoder achieves recovery using asymptotically fewer than the $m \approx C s \log(N/s)$ measurements required by \R{BP}.  However, this analysis only applies to Gaussian measurements, and certainly not Fourier sampling, as is common in practice in imaging.  Moreover, even with Gaussian measurements, this approach is typically outperformed by Fourier sampling with vanilla $\ell^1$-minimization.  Fig.~\ref{fig:intro_comp} gives a typical example (see Appendix \ref{s:NumExp} for further examples).
Computationally, the latter is also significantly faster.  

From this, it was concluded in \cite{OptimalSamplingQuest} that the types of sparsity structures inherent to imaging are more effectively exploited in the sampling strategy than in the decoder.  The results \R{OptErr} and \R{FourErr} support this conclusion.  Further examples demonstrating this phenomenon in imaging scenarios are shown in \cite{OptimalSamplingQuest,AsymptoticCS}.  Whether one can do both, i.e.\ combine structured sampling strategies with decoders that promote structured sparsity, and witness even further benefits, is an interesting question for future work.

\subsection{Extensions}

The objects considered in this paper are piecewise $\alpha$-H\"older functions of one variable.  Although wavelet approximation theory is more commonly studied in Besov spaces \cite{DeVoreNLACTA}, we have chosen this class for the ease of presentation.  Possible extensions, including to more than one dimension and other -let families such as shearlets, are discussed in \S\ref{s:conclusions}.

\subsection{Relation to previous work}\label{ss:previous}

The idea of sampling the coarse and fine wavelet scales of an image differently was arguably first considered by Tsaig \& Donoho \cite{DonohoCSext}.  Romberg, in his seminal paper on compressed sensing for imaging, used DCT measurements to capture the coarse scales and noiselet measurements for the fine scales \cite{RombergCompImg}.  Later, Cand\`es \& Romberg \cite{Candes_Romberg} considered Fourier sampling in the dyadic bands $B_k$ according to structured wavelet sparsity.  However, their analysis assumes direct sensing of individual wavelet scales, which is infeasible in practice.  This idea was also used in \cite{WangAcre} to design empirical Fourier sampling strategies for MRI.  Similar ideas have also been pursued in \cite{BoyerBlockStructured}. 

In the context of Fourier sampling, it has long been known that fully sampling the low frequencies is crucial for high-quality compressed sensing reconstruction \cite{Lustig3}.
Partial theoretical justifications have been given in \cite{ChauffertEtAlVDS} and \cite{ModelsCS}.  Furthermore, in \cite{AHPWavelet} (one dimension) and \cite{AHKM2DWavelets} (two dimensions) it has been shown that low frequency Fourier samples optimally recover the coarse scale wavelet coefficients.

In terms of theoretical results, for two-dimensional discrete Fourier sampling with Haar wavelets, it was shown in \cite{KrahmerWardCSImaging} that
\be{
\label{KWFHRIP}
m \approx c \cdot s \cdot \log^3(s) \cdot \log^2(N),
}
measurements (chosen randomly according to an inverse square law) suffice for recovery of an approximately $s$-sparse vector of Haar wavelet coefficients.  Unfortunately, even if this were extended to continuous Fourier measurements with higher-order wavelets, it would lead to a highly suboptimal error bound $\nmu{f-\tilde{f}_m}_{L^2} \leq C (\log(m))^{5 \alpha} / m^{\alpha}$.

Instead, our approach relies on the generalization of compressed sensing \cite{AHPRBreaking} based on local structure.  This framework provides precise estimates relating local quantities (sparsities and numbers of measurements), which allows us to adjust the sampling strategy to the local sparsity structure.  We also make use of ideas and theoretical results from \cite{BastounisHansen}, \cite{LiAdcockRIP} and \cite{TraonmilinGribonvalRIP}.

\subsection{Outline}

We commence in \S\ref{s:prelims} with some requisite material, including nonlinear approximation using wavelets and some standard compressed sensing theory.  Our main results are stated in \S\ref{s:mainres} and discussed in detail in \S\ref{s:discussion}.  \S\ref{s:GaussErrProof} and \S\ref{s:OptErrProof} give the proofs corresponding to \R{GaussErr} and \R{OptErr} respectively.  The remainder of the paper is devoted to the proof of \R{FourErr}.  In \S\ref{s:FourWaveSetup} we reformulate wavelet approximation from Fourier samples as a finite compressed sensing problem.  Next in \S\ref{s:CSlocal} we recap the framework of \cite{AHPRBreaking}.  Finally, \S\ref{s:FourErrProofs} gives the proof of \R{FourErr}.  We conclude in \S\ref{s:conclusions} by listing some open problems.

\section{Preliminaries}\label{s:prelims}

\subsection{Notation}

We denote $\mathbb{N} = \{1,2,3,\ldots\}$ and $\mathbb{N}_0 = \{0\} \cup \mathbb{N}$. We work primarily in the Hilbert space $L^2([0,1])$ of square-integrable functions on $[0,1]$. Write $\ip{\cdot}{\cdot}_{L^2([0,1])}$ and $\nm{\cdot}_{L^2([0,1])}$ (or, in short, $\ip{\cdot}{\cdot}_{L^2}$ and $\nm{\cdot}_{L^2}$) for the corresponding inner product or norm.  We use $c$, $C$ to denote arbitrary numerical constants.  We write $c_{\alpha}$, $C_{\alpha}$, and so forth for constants that depend on a parameter $\alpha$. By convention, capital letters are used in error estimates and lower case letters are used for assumptions on, for instance, the number of measurements $m$.  Throughout, the value of such constants may change from line to line.  We make no attempt to track constants. The shorthand notation $A \lesssim B$ stands for $A \leq C B$ for some constant $C>0$ independent of $A$ and $B$ and $A \gtrsim B$ is defined analogously. Moreover, $A \asymp B$ means that $A \lesssim B$ and $A \gtrsim B$ hold simultaneously.

If $\Omega \subseteq \{1,\ldots,N\}$, we use $P_{\Omega}$ to denote either the $N \times N$ matrix of the projection onto the space of vectors supported on $\Omega$, or the $m \times N$ matrix ($m = |\Omega|$) which selects the entries of a vector in $\Omega$.  Its meaning will be clear from the context.  If $\Omega = \{1,\ldots,M\}$ we write $P_M$, and if $\Omega = \{ N_1+1,\ldots,N_2 \}$ we write $P^{N_1}_{N_2}$.

\subsection{Wavelets and nonlinear approximation}
\label{ss:wavelets_and_nonlinear}

We consider Daubechies' wavelets with $p \geq 1$ vanishing moments.  Write $\varphi$ and $\psi$ for the corresponding scaling function and mother wavelet, respectively.  Given such a wavelet, we define the \textit{smoothness parameter} $q \geq 0$ as the largest number such that
\be{
\label{smoothness}
| \hat{\varphi}(\omega) | \lesssim (1+|\omega|)^{-1-q},\qquad | \hat{\psi}(\omega) | \lesssim (1+|\omega|)^{-1-q},\qquad \forall \omega \in \bbR.
}
Here $\hat{\cdot}$ denotes the Fourier transform -- see Appendix \ref{s:Fourier}.  The exact values for $q$ can be found, for example, in \cite[p.\ 226]{MR1162107}. 
Note that asymptotically one has $q \rightarrow \infty$ as $p \rightarrow \infty$ \cite[p.\ 226]{MR1162107}.

Since our interest lies with approximation on $[0,1]$, we use the orthonormal wavelet basis of $L^2([0,1])$ constructed via periodization.  We refer to this as the \textit{periodized Daubechies' wavelet basis with $p$ vanishing moments} and denote it by $\{ \phi_n \}_{n \in \bbN}$.  See Appendix~\ref{s:wavelets} for the construction.  Given $f \in L^2([0,1])$, we write $d = (d_n)^{\infty}_{n=1} \in \ell^2(\bbN)$ for its infinite vector of wavelet coefficients, i.e.
\bes{
d_{n} = \ip{f}{\phi_n},\quad n = 1,2,\ldots,
}
so that $f = \sum^{\infty}_{n=1} d_n \phi_n$.

\rem{
There are several other strategies for constructing an orthonormal wavelet basis of $L^2([0,1])$ \cite[Sec.\ 7.5]{mallat09wavelet}.  Periodization is the simplest, and hence we use it throughout.  Periodization treats the endpoints as additional discontinuities, which is often undesirable in practice.  However,  this is no limitation in our setting, since the concern of this paper lies with the asymptotic rate of nonlinear approximation of piecewise smooth functions, which is unaffected by the addition of a finite number of discontinuities.  So-called \textit{boundary-adapted} wavelets avoid this issue, and could be used in what follows.  However, these require some rather intricate technical modifications to the proofs (for related work, see \cite{AAHWalshWavelet}).
}

Let $\{ \phi_n \}^{\infty}_{n=1}$ be an orthonormal basis of $L^2([0,1])$ (not necessarily of wavelet type).  For $s \geq 1$, the \textit{linear $s$-term approximation} of a function $f = \sum^{\infty}_{n=1} d_n \phi_n \in L^2([0,1])$ is
\bes{
f^{l}_s = \sum^{s}_{k=1} d_n \phi_n,
}
and the \textit{best $s$-term approximation} of $f$ is
\bes{
f^{nl}_s = \sum^{s}_{n=1} d_{\pi(n)} \phi_{\pi(n)},
}
where $\pi : \bbN \rightarrow \bbN$ is a bijection that rearranges the coefficients $d_n$ in nonincreasing order of absolute value, i.e.\ $|d_{\pi(1)}| \geq |d_{\pi(2)}| \geq \ldots$.  The \textit{linear} and \textit{best $s$-term approximation errors} are
\be{
\label{linnonlin}
e_{s}(f)_{L^2} = \nmu{f - f^{l}_s}_{L^2} = \sqrt{\sum_{k > s} |d_k |^2},\qquad
\sigma_{s}(f)_{L^2} = \nmu{f - f^{nl}_s}_{L^2} = \sqrt{\sum_{k > s} | d_{\pi(k)} |^2 },
}
respectively.

The concern of this paper is nonlinear wavelet approximation in suitable spaces of piecewise regular functions.  We now define these spaces:  

\defn{
Let $0 < \alpha \leq 1$ and $- \infty < a < b < \infty$.  The H\"older semi-norm of index $\alpha$ of a function $f \in C([a,b])$ is
\bes{
|f|_{C^{\alpha}([a,b])} = \sup_{\substack{x,y \in [a,b] \\ x \neq y}} \left \{ \frac{|f(x) - f(y)|}{|x-y|^{\alpha}} \right \}.
}
A function $f$ is \textit{$\alpha$-H\"older continuous} if $|f|_{C^{\alpha}([a,b])}  < \infty$.
}

\defn{
Let $- \infty < a < b < \infty$, $d \in \bbN_0$, $0 < \beta \leq 1$ and $\alpha = d + \beta $.  The \textit{H\"older space} $C^{\alpha}([a,b])$ consists of those functions $f$ that are $d$-times continuously differentiable on $[a,b]$ and for which the $d^{\rth}$ derivative is $\beta$-H\"older continuous.  This is a Banach space with norm
\bes{
\nm{f}_{C^{\alpha}([a,b])} = \sum^{d}_{j=0} \nmu{f^{(j)}}_{C([a,b])} + |f^{(d)}|_{C^{\beta}([a,b])},
}
where $\nm{g}_{C([a,b])} = \sup_{x \in [a,b]} |g(x)|.$
}

\defn{
\label{def:alpha-Holder}
For $\alpha = d+ \beta$, where $d\in \bbN_0$ and $0 < \beta \leq 1$, the space $PC^{\alpha}([a,b])$ consists of all functions which are discontinuous at (at most) a finite number of points in $[a,b]$ and are $\alpha$-H\"older continuous in between any two consecutive discontinuities.  
}

We define the norm of such a function as the maximum of its $C^{\alpha}$-norm over all intervals of smoothness, and write $\nm{f}_{PC^{\alpha}([a,b])}$, or $\nm{f}_{PC^{\alpha}}$ when the domain of $f$ is clear.  For convenience, we also define
$$
\cN(f) = |\{\text{discontinuities of $f$ in $(a,b)$}\}| + 1, \qquad f \in PC^{\alpha}([a,b]).
$$ 
The following well-known result summarizes the effectiveness of wavelets for approximating piecewise smooth functions (see \S\ref{s:proof_waveletsterm} for a proof):

\thm{
\label{t:waveletsterm}
Suppose that $f \in PC^{\alpha}([0,1])$, where $\alpha = d+ \beta \geq 1/2$ for $d\in \bbN_0$ and $0 < \beta \leq 1$.  Consider the periodized Daubechies wavelet basis with $p > d$ vanishing moments with coarsest scale $j_0$ given by \R{j0def}.  Then there exists a constant $C_{p,\alpha} > 0$ such that, for all $s \geq 1$,
\bes{
e_{s}(f)_{L^2} \leq C_{p,\alpha} \sqrt{\cN(f)} \nm{f}_{PC^{\alpha}}s^{-1/2},
}
and, if $s/\log_2(s) \geq 64 \alpha p^2 \cN(f)$, 
$$
\sigma_{s}(f)_{L^2} \leq C_{p,\alpha}  \nm{f}_{PC^{\alpha}} s^{-\alpha}.
$$
}

\subsection{Standard compressed sensing theory}

Recall that a vector $x \in \bbC^N$ is \textit{$s$-sparse} if it has at most $1 \leq s \leq N$ nonzero entries.  Standard compressed sensing theory concerns the recovery of such a vector from measurements $y = A x \in \bbC^{m}$.  Recovery can be carried out via a number of different procedures.  However, in this paper we focus on convex optimization approaches, such as the basis pursuit \R{BP}.  The overarching goal of compressed sensing is to derive conditions on the matrix $A \in \bbC^{m \times N}$ under which any $s$-sparse $x$ can be recovered via \R{BP} from $m \approx s$ measurements, up to log factors.  

A particularly useful tool in this endeavour is the Restricted Isometry Property:

\defn{
\label{d:RIP}
Let $1 \leq s \leq N$.  The $s^{\rth}$ \textit{Restricted Isometry Constant (RIC)} $\delta_s$ of a matrix $A \in \bbC^{m \times N}$ is the smallest $\delta \geq 0$ such that
\be{
\label{RIP}
(1-\delta) \| x \|^2_{\ell^2} \leq \| A x \|^2_{\ell^2} \leq (1+\delta) \| x \|^2_{\ell^2},\quad \mbox{for all $s$-sparse $x$}.
}
If $0 < \delta_s < 1$ then $A$ is said to have the \textit{Restricted Isometry Property (RIP)} of order $s$.
}

The following result is well known (see, for example, \cite{CandesRIP}):

\thm{
\label{t:CSRIPstandard}
Suppose that $A \in \bbC^{m \times N}$ satisfies the RIP of order $2s$ with constant
\be{
\label{RIP_delta_bound}
\delta_{2s} < \sqrt{2}-1.
}
Let $x \in \bbC^{N}$ and $y = A x$.  Then any minimizer $\hat{x} \in \bbC^N$ of \R{BP} satisfies
\ea{
\label{stablerecovery}
\| \hat{x} - x \|_{\ell^2} &\leq C \frac{\sigma_{s}(x)_{\ell^1} }{\sqrt{s}} ,
}
where $\sigma_{s}(x)_{\ell^1} = \min \{ \nm{x- z}_{\ell^1} : \mbox{$z$ is $s$-sparse} \}$ and the constant $C>0$ depends on $\delta_{2s}$ only.
}

We remark that the constant $\sqrt{2}-1$ in \R{RIP_delta_bound} be improved (see \cite{cai2014sparse}); however, this will be of little consequence for this paper.  We also note that one can prove a recovery guarantee similar to this theorem for measurements that are also corrupted by noise.  However, our focus in this paper is on the noise-free setting. Furthermore, it is well known random Gaussian matrices satisfy the RIP with high probability.

\thm{
\label{t:RIPGauss}
Let $0 < \delta, \varepsilon < 1$, $1 \leq s \leq N$ and 
\be{
\label{Gaussmeascond}
m \geq c \cdot \left( s \cdot \log(\E N/s) + \log(2 \varepsilon^{-1}) \right ).
}
Let $A \in \bbR^{m \times N}$ have i.i.d.\ entries drawn from the normal distribution with mean zero and variance $1/m$.  Then, with probability at least $1-\varepsilon$, $A$ has the RIP of order $s$ with $\delta_s \leq \delta$.
}

The scaling $m \geq c \cdot s \cdot \log(N/s)$ is essentially optimal.  Any method which satisfies the error bound \R{stablerecovery} must also have $m \geq c \cdot s \cdot \log(N/s)$ \cite[Chpt.\ 10]{FoucartRauhutCSbook}.

\section{Main results}\label{s:mainres}

We now present our main results.  To state these, we use the language of \textit{encoders} and \textit{decoders}, terminology employed in the compressed sensing context, e.g., in \cite{CDDCSkterm}.

\subsection{Encoding and decoding}

An $m$-term \textit{encoder} is a linear mapping $\cE_{m} : L^2([0,1]) \rightarrow \bbC^m$.  An $m$-term \textit{decoder} is a mapping $\cD_{m} : \mathrm{Ran}(\cE_m) \rightarrow L^2([0,1])$.  Note that $\cE_m$ is by assumption linear, whereas $\cD_m$ will typically be nonlinear.  A \textit{Fourier encoder} is an encoder of the form
\bes{
\cE_m(f) = \left ( \hat{f}(\omega_i) \right )^{m}_{i=1},
}
where $\hat{f}$ is the Fourier transform of $f$ and $\omega_i \in \bbR$.

Random Gaussian measurements are popular in compressed sensing.  Since in this paper we consider the approximation of functions (rather than finite vectors), we define a \textit{Gaussian encoder} as any encoder of the form
\bes{
\cE_m(f) = A \left ( \ip{f}{\phi_j} \right )^{N}_{j=1},
}
where $\{ \phi_j \}^{N}_{j=1}$ is an orthonormal system in $L^2([0,1])$ and $A \in \bbR^{m \times N}$ is a random Gaussian matrix. Note that $\{ \phi_j \}$ will typically coincide with the orthonormal basis that is used for decoding, i.e.\ a wavelet basis.

Given an encoder-decoder pair $(\cE_m,\cD_m)$ and $f \in L^2([0,1])$, we write $\tilde{f}_m = \cD_m \left ( \cE_m (f) \right )$.

\subsection{Statements}

We commence with Gaussian encoders:

\thm{
\label{t:GaussErr}
Let $0 < \varepsilon < 1$, $\alpha > 1/2$ and $\cN_* \in \mathbb{N}$. Then there exist $c,c_\alpha, C_\alpha >0$ such that the following holds. For any $m \in \mathbb{N}$ satisfying
\bes{
m \geq c \log(2 / \varepsilon) ,\quad m / (\log(m))^2 \geq c_{\alpha} \cN_*,
}
the Gaussian encoder-decoder pair $(\cE_m,\cD_m)$ based on periodized Daubechies' wavelets with $p = \lceil \alpha \rceil$ vanishing moments satisfies
\be{
\label{GaussErrBd}
\nmu{f - \tilde{f}_m}_{L^2} 
\leq C_{\alpha} \nm{f}_{PC^{\alpha}} 
\left(\frac{\log(m)}{m}\right)^{\alpha},
}
for all $f \in PC^{\alpha}([0,1])$ with $\cN(f) \leq \cN_*$, with probability at least $1-\varepsilon$. The decoder requires the solution of a basis pursuit problem of size at most $m \times \lfloor m^{ 2 \alpha +1 } / (\log(m))^{2 \alpha + 2}\rfloor$. 
}

Although we do not rigorously show that the log factor in the error bound \eqref{GaussErrBd} cannot be improved (i.e., that under the same conditions of Theorem~\ref{t:GaussErr} a lower bound of the form $\nmu{f - \tilde{f}_m}_{L^2} 
\geq C_{\alpha} \nm{f}_{PC^{\alpha}} 
(\log(m))^\alpha / {m}^{\alpha}$ holds), we conjecture that this is the case. The rationale behind this conjecture is discussed in Remark~\ref{rmk:lower_bound_Gauss}. In view of this observation, the next result suggests that the Gaussian encoder is far from optimal:

\thm{
\label{t:OptErr}
Let $0 < \varepsilon < 1$, $\alpha >1/2$ and $\cN_* \in \mathbb{N}$. Then there exist constants $c,c_\alpha, C_\alpha >0$ such that the following holds. For any $m \in \mathbb{N}$ satisfying 
\bes{
m \geq c \log(2 / \varepsilon) ,\quad m / (\log(m))^2 \geq c_{\alpha} \cN_*,
}
there exists an encoder-decoder pair $(\cE_m,\cD_m)$ such that, with probability at least $1-\varepsilon$,
\be{
\label{OptErrBd}
\nmu{f - \tilde{f}_m}_{L^2} 
\leq C_{\alpha} \nm{f}_{PC^{\alpha}} \frac{\sqrt{\log(m)}}{m^{\alpha}},
}
for all $f \in PC^{\alpha}([0,1])$ with $\cN(f) \leq \cN_*$.  The decoder uses periodized Daubechies' wavelets with $p = \lceil \alpha \rceil$ vanishing moments and requires the solution of a basis pursuit problem of size at most $m \times \lfloor m^{ 2 \alpha +1 } / (\log(m))^3 \rfloor$.

}

In view of Theorem \ref{t:waveletsterm}, this pair is optimal up to $\sqrt{\log(m)}$; a significant improvement on the Gaussian case.  This factor can be removed altogether at the price of a nonuniform recovery guarantee.  This is discussed in \S\ref{ss:nonunifopt}.

\rem{
The above result ceases to hold for $\alpha = 1/2$.  If $\alpha = 1/2$, then one can trivially find an encoder-decoder pair satisfying
\bes{
\nmu{f - \tilde{f}_m}_{L^2} \leq C \sqrt{\cN(f)} \nm{f}_{PC^{\alpha}} m^{-1/2}.
}
Indeed, one merely directly senses the first $m$ wavelet coefficients.  This gives  $\nmu{f - \tilde{f}_m}_{L^2}  = e_{m}(f)_{L^2}$, and the bound follows from Theorem \ref{t:waveletsterm}.  
}

Next we present our main result for Fourier encoders, which relies on the framework of compressed sensing with local structure (see \S\ref{s:CSlocal}).

\thm{
\label{t:FourErr}
Let $0 < \delta < 1$, $\alpha >1/2$, $p \geq \lceil \alpha \rceil$, and $\cN_* \geq 1$. Then, there exist constants $c_{p,\alpha,\delta},C_{p,\alpha} > 0$ such that the following holds. For any $m  \in \mathbb{N}$ such that
\bes{
m \geq c_{p,\alpha,\delta} (\cN_*)^2,\qquad (\log(m))^{5} \geq  \log(\varepsilon^{-1}),
}
there exists a Fourier encoder-decoder pair $(\cE_m,\cD_m)$ such that, with probability at least $1-\varepsilon$,
\be{
\label{FourErrBd1}
\nmu{f - \tilde{f}_m}_{L^2} 
\leq C_{p,\alpha} \nm{f}_{PC^{\alpha}} 
  \frac{(\log(m))^{\frac{1}{4} + \frac{(6+\delta)(q+\alpha+1/2)}{2(q+1)}}}{ m^{\alpha}}.
}
for all $f \in PC^{\alpha}([0,1])$ with $\cN(f) \leq \cN_*$.  The decoder uses periodized Daubechies' wavelets with $p$ vanishing moments, smoothness parameter $q$ and requires solution of a weighted square-root LASSO problem of size no more than $m \times \lfloor m^{\sigma}\rfloor$, where $\sigma= \max\{\frac{\alpha}{\alpha-1/2},2\alpha +1\}$. 
In particular, if $p$ is the smallest integer so that the smoothness parameter
\be{
\label{qminFourErr}
q \geq 6 \frac{\alpha-1/2}{\delta} + \alpha - \frac{3}{2},
}
then the encoder-decoder pair satisfies
\be{
\label{FourErrBd2}
\nmu{f - \tilde{f}_m}_{L^2} 
\leq C_{p,\alpha} \nm{f}_{PC^{\alpha}} 
  \frac{(\log(m))^{\frac{13}{4} + \delta}}{ m^{\alpha}}.
  }
for all $f \in PC^{\alpha}([0,1])$ with $\cN(f) \leq \cN_*$.

}

The reader will notice that the decoder in the case solves a so-called \textit{weighted square-root LASSO} problem.  This problem takes the form
\be{
\label{wSRLASSOblurb}
\min_{z \in \bbC^N} \nm{z}_{\ell^1_w} + \lambda \nm{A z - y}_{\ell^2},
} 
where $\nm{z}_{\ell^1_w} = \sum^{N}_{i=1} w_i |z_i|$ is a weighted $\ell^1$-norm and $\lambda > 0$ is a parameter.  The reason for solving this problem instead of \R{BP} is discussed in \S\ref{ss:weightedSRLASSO}.

We stress that the encoder-decoder pairs in Theorems~\ref{t:GaussErr} and \ref{t:OptErr} are determined completely by $m$ and the smoothness $\alpha$. In Theorem~\ref{t:FourErr} the Fourier encoder-decoder pair also depends on the number vanishing moments $p$ and the auxiliary parameter $\lambda$.  In particular, neither $\cN(f)$ nor $\cN_*$ need to be known for any of these strategies.

\rem{
\label{r:unifrecov}
These results are uniform in the sense that for each fixed (and sufficiently large) $m$, a single random draw of the matrix used in the encoding stage guarantees the corresponding approximation error bound for all $f \in PC^{\alpha}([0,1])$ with $\cN(f) \leq \cN_*$.  The parameter $\cN_*$ is included primarily for convenience.  It, in combination with the condition on $m$, allows for a uniform error bound that is independent the number of discontinuities $\cN(f)$.
}

\section{Discussion}
\label{s:discussion}

We now discuss these results in some more detail.

\subsection{The idea of Theorems \ref{t:GaussErr} and \ref{t:OptErr}}
\label{ss:logfactors}

Theorem \ref{t:GaussErr} is proved as follows.  To find a quasi-best $s$-term wavelet approximation one needs to search within the first $N$ wavelets, where $N \asymp s^{2 \alpha + 1}$.  The measurement condition \R{Gaussmeascond} now gives $m \asymp s \cdot \log(s)$, or equivalently, $s \asymp m / \log(m)$.  After modifying the proof of Theorem \ref{t:waveletsterm}, we show that $\sigma_{s}(P_N d)_{\ell^1} \lesssim s^{-(\alpha-1/2)}$, where $d \in \ell^2(\bbN)$ is the vector of wavelet coefficients of $f$.  The error bound now follows immediately from this and Theorem \ref{t:CSRIPstandard}.

\begin{remark}
\label{rmk:lower_bound_Gauss}
We conjecture that under then same setting of Theorem~\ref{t:GaussErr}, the error bound \eqref{GaussErrBd} cannot be improved. This is due to the fact that if an encoder-decoder pair is such that \eqref{stablerecovery} holds uniformly for every vector $x \in \mathbb{C}^N$, then we necessarily have $m \geq c s \log(eN/s)$  (see, e.g., \cite[Theorem 11.6]{FoucartRauhutCSbook}). This leads to conjecture that the condition $s \lesssim m/\log(m)$ is also necessary. Therefore, the optimal rate satisfies $1/s^{\alpha} \gtrsim (\log(m)/m)^\alpha$, which suggests that the log factor $(\log(m))^{\alpha}$ cannot be reduced. A possible way to avoid the necessary condition $m \geq c s \log(eN/s)$ would be to assume that \eqref{stablerecovery} holds uniformly only for a restricted class of coefficient vectors having a certain structure, as opposed to all vectors of $\mathbb{C}^N$. However, this requirement seems to be incompatible with the setting of Theorem~\ref{t:GaussErr}, due to ``structure agnostic'' nature of random Gaussian sampling and of the basis pursuit decoder.
\end{remark}

In contrast, Theorem \ref{t:OptErr} constructs a sampling strategy that directly measures the coarse wavelet scales and then randomly samples the fine scales.  Specifically, the first $N_1 \approx m/2$ wavelet coefficients are directly sampled, and then for the wavelet coefficients in the range $\{N_1+1,\ldots,N_2\}$, where $N_2 \approx m^{2 \alpha+1}$, we use random Gaussian sampling.  By Theorem \ref{t:CSRIPstandard}, the error is effectively determined by the term
\be{
\label{OptErressentialbd}
\sigma_{s} \left ( P^{N_1}_{N_2} d \right )_{\ell^1} \Big / \sqrt{s},
}
where $s$ is any number satisfying $s \lesssim m / \log(m)$.  As discussed in \S\ref{ss:discussion}, there are very few significant coefficients in the range $\{N_1+1,\ldots,N_2\}$, roughly $\ord{\cN(f) \log(m)}$ in total.  Due to the choice of $N_1$, this means that $\sigma_{s} \left ( P^{N_1}_{N_2} d \right )_{\ell^1} \lesssim m^{-(\alpha-1/2)}$ whenever $ \cN(f) \log(m) \lesssim s \lesssim m / \log(m)$.  However, in order for \R{OptErressentialbd} to attain the optimal algebraic rate $m^{-\alpha}$, we need to make $s$ as large as possible, i.e.\ $s \approx m/\log(m)$.  This leads directly to the $\sqrt{\log(m)}$ factor.

This approach improves significantly over the log term of the Gaussian case, but does not remove it completely.  The reason can be traced to the \textit{$(\ell^2,\ell^1)$-instance optimality} of Gaussian measurements \cite[Chpt.\ 10]{FoucartRauhutCSbook}; that is, the fact that the bound \R{stablerecovery} bounds the $\ell^2$-norm reconstruction error in terms of the $\ell^1$-norm best $s$-term error.  As we show in \S\ref{ss:nonunifopt}, this log term can be removed by exploiting \textit{nonuniform $(\ell^2,\ell^2)$-instance optimality} of Gaussian measurements.  In this setting, \R{OptErressentialbd} is replaced by $\sigma_{s} \left ( P^{N_1}_{N_2} d \right )_{\ell^2}$ and the optimal rate $m^{-\alpha}$ follows by the same arguments.

\subsection{The log factor in Theorem \ref{t:FourErr}}
\label{ss:logfactor_Fourier}
We defer a more detailed explanation of the strategy behind this theorem to \S\ref{ss:FourErrIdea}.  However, let us briefly comment on how the  log term arises.  First, since we use Fourier measurements, the measurement condition that ensures a RIP-type property involves a significantly larger log factor than \R{Gaussmeascond}, roughly of size $(\log(m))^6$.  As above, because of the lack of $(\ell^2,\ell^2)$-instance optimality, we need to take a much larger sparsity to ensure the $m^{-\alpha}$ rate than is strictly necessary to capture the large wavelet coefficients corresponding to the discontinuities.  This effectively leads to a log term of the order $\sqrt{(\log(m))^6} = (\log(m))^3$ in the error bound.

Where do the remaining log terms come from?  The source is the interference between wavelet scales and frequency bands discussed in \S\ref{ss:FourOptdiscuss}.  Coarse scale wavelets have small but nonzero components in the frequency bands corresponding to the fine scales. To control the effect of this interference, we need to fully sample a few more frequency bands than the number of saturated wavelet scales.  The precise number of fully sampled bands corresponding to the saturated scales depends on the wavelet smoothness $q$, and gives rise to the log factor seen in \R{FourErrBd1}.

Nevertheless, we expect this log factor can be improved.  In \S\ref{s:conclusions} we discuss how this might be achieved.

\subsection{The decoder in Theorem \ref{t:FourErr}}\label{ss:weightedSRLASSO}

As noted, in Theorem \ref{t:FourErr} we solve \R{wSRLASSOblurb} instead of basis pursuit.  There are two aspects to this choice: the weights and the unconstrained formulation.  

\pbk
\textit{1) Weights.}\ These are incorporated to control the effect of the interferences discussed above.  The weights are constant on each wavelet scale, and in the $k^{\rth}$ scale, are taken as roughly $\sqrt{s/s_k}$, where $s_k$ is the sparsity in that scale.

\pbk
\textit{2) Unconstrained formulation.}\ Unlike in Theorems \ref{t:GaussErr} and \ref{t:OptErr}, in Theorem \ref{t:FourErr} we have to deal with the effect of truncating the wavelet expansion $\sum^{\infty}_{n=1} d_n \phi_n$ to its first $N$ terms.  This truncation introduces an error proportional to $\nm{P^{\perp}_N d}_{\ell^2}$, which can be considered as noise in the measurement vector $y$.  Such an error is, of course, unknown \textit{a priori}.

Unfortunately, standard decoders in compressed sensing for noisy measurements such as \textit{quadratically-constrained basis pursuit (QCBP)} 
\be{
\label{QCBP}
\min_{z \in \bbC^M} \nm{z}_{\ell^1}\ \mbox{subject to $\nm{A z - y}_{\ell^2} \leq \eta$},
}
typically require explicit upper bounds on the noise \cite{BASBCSmodel,FoucartRauhutCSbook}.\footnote{There are some theoretical results for QCBP in the presence of unknown noise \cite{BASBCSmodel,DeVore2009,Foucart2014,Wojtaszczyk2010}.  However, except in specific cases, these involve additional factors (so-called \textit{quotients}) which are difficult to estimate.}  To avoid this, one may consider unconstrained optimization problems.  One standard choice is the unconstrained LASSO
\bes{
\min_{z \in \bbC^N} \nm{z}_{\ell^1_w} + \lambda \nm{A z - y}^2_{\ell^2}.
} 
Unfortunately, it is well known that the optimal tuning parameter $\lambda$ in LASSO depends on the $\ell^2$-norm of the noise, rendering it unsuitable for this problem.

The square-root LASSO -- a little known variant of its more famous cousin -- was conceived specifically to overcome this issue \cite{Belloni2011}.  It was introduced to the compressed sensing setting in \cite{ABBCorrecting}.  Therein it was shown that the optimal tuning parameter is independent of the noise, thus rendering it suitable for our purposes.
We note in passing that \R{wSRLASSOblurb} can be solved efficiently via standard algorithms.  For instance, Chambolle-Pock's primal-dual algorithm \cite{Chambolle2011}.

\pbk
It is common in Fourier imaging to construct a measurement matrix $A = P_{\Omega} F \Phi^*$, where $F \in \bbC^{N \times N}$ and $\Phi \in \bbC^{N \times N}$ are the discrete Fourier and wavelet transforms respectively, and $P_{\Omega} \in \bbC^{m \times N}$ restricts to the frequencies sampled.  This formulation is readily amenable to fast computations.  However, this approach commits a discretization error, since (as noted in  \S\ref{ss:whatwedo}) the measurements are of the continuous Fourier transform of $f$.  To avoid this error, we employ infinite-dimensional compressed sensing \cite{BAACHGSCS}, and formulate $A$ as $P_{\Omega} U P_N$, where $U$ is the cross-Grammian between the Fourier and wavelet bases of $L^2([0,1])$.  We note in passing that this matrix also admits fast computations in FFT time \cite{GataricPoonFast}.

\subsection{An optimal nonuniform recovery guarantee}\label{ss:nonunifopt}

Our main results are uniform for each fixed $m$ (see Remark \ref{r:unifrecov}).  The log factor $\sqrt{\log(m)}$ present in the error estimates of Theorem~\ref{t:OptErr} can be removed at the price of having a nonuniform recovery guarantee, i.e.\ a guarantee that holds with high probability for a fixed function $f \in PC^\alpha([0,1])$ and not uniformly for every $f \in PC^\alpha([0,1])$.

The key element to prove this is a compressed sensing result concerning the so-called \textit{nonuniform $(\ell^2,\ell^2)$-instance optimality} of random Gaussian measurements. The following is a direct consequence of \cite[Thm.\ 11.23]{FoucartRauhutCSbook}. 

\thm{
\label{t:nonuniform_instance_opt}
There exist constants $0 < c_1 < 1$, $c_2,C > 0$ such that the following holds.  Let $0 < \varepsilon < 1$, $x \in \mathbb{C}^N$ be a fixed vector, $A \in \mathbb{R}^{m\times N}$ have i.i.d.\ entries drawn from the normal distribution with mean zero and variance $1/m$, and assume that $c_1 m \leq N$ and 
$$
m \geq c_2 ( s \log(\E N/m) + \log(5/\varepsilon)).
$$
Then any minimizer $\hat{x} \in \mathbb{C}^N$ of \eqref{BP} satisfies
$$
\|x-\hat{x}\|_{\ell^2} \leq C \sigma_s(x)_{\ell^2},
$$
with probability at least $1-\varepsilon$, where $\sigma_{s}(x)_{\ell^2} = \min \{ \nm{x-z}_{\ell^2} : \mbox{$z$ is $s$-sparse} \}$.
}

This result implies the following theorem, analogous to Theorem~\ref{t:OptErr}.  The only differences with respect to the assumptions of Theorems~\ref{t:OptErr} are the lack of uniformity with respect to $f\in PC^\alpha([0,1])$ and the slightly larger dimension of the basis pursuit problem. 
\thm{
\label{t:OptErrNU}
Let $0 < \varepsilon < 1$ and $\alpha >1/2$. Then there exist constants $c,c_\alpha, C_\alpha >0$ such that the following holds.  For each $f \in PC^\alpha([0,1])$ and $m \in \mathbb{N}$ satisfying 
\bes{
m \geq c \log(5 / \varepsilon) ,\quad m / \log^2(m) \geq c_{\alpha} \cN(f),
}
there exists an encoder-decoder pair $(\cE_m,\cD_m)$ such that, with probability at least $1-\varepsilon$,
\be{
\label{OptErrNUBd}
\nmu{f - \tilde{f}_m}_{L^2} 
\leq C_{\alpha} \nm{f}_{PC^{\alpha}} {m^{-\alpha}},
}
The decoder uses periodized Daubechies' wavelets with $p = \lceil \alpha \rceil$ vanishing moments and requires the solution of a basis pursuit problem of size at most $m \times \lfloor m^{ 2 \alpha +1 } / (\log(m))^2 \rfloor$. 
}

\subsection{Computational cost and the size of the optimization problems}
\label{ss:discussNsize}

It is important to consider the size of the optimization problem that needs to be solved in each case.  This size is $m \times N$, where $N \leq \lfloor m^{2 \alpha + 1} / (\log(m))^{2 \alpha+2} \rfloor$ in Theorem \ref{t:GaussErr}, $N \leq \lfloor m^{2 \alpha + 1} / (\log(m))^{3} \rfloor$ in Theorem \ref{t:OptErr} and $N \leq \lfloor \max \{ m^{\frac{\alpha}{\alpha-1/2}} , m^{2 \alpha+1} \} \rfloor$ in Theorem \ref{t:FourErr}.  These choices have been made to balance the various error terms, and in particular, to remove any dependence on the number of discontinuities $\cN(f)$.

Since the size of $N$ affects the memory and computational time required for decoding, it is worth dwelling on precisely how $N$ affects the approximation error.  We have the following:

\thm{
\label{t:GaussErr2}
Consider the setup of Theorem \ref{t:GaussErr}, except where the basis pursuit problem is of size at most $m \times N$ for some $m \leq N \leq m^{2 \alpha +1}$.  Then the error bound \R{GaussErrBd} is replaced by
\bes{
\nmu{f - \tilde{f}_m}_{L^2} 
\leq C_{\alpha} \nm{f}_{PC^{\alpha}} \left ( 
\left(\frac{\log(m)}{m}\right)^{\alpha} + \sqrt{\frac{\cN(f)}{N}} \right ).
}
}

For the optimal encoder-decoder, we have:

\thm{
\label{t:OptErr2}
Consider the setup of Theorem \ref{t:OptErr}, except where the basis pursuit problem is of size at most $m \times N$ for some $m \leq N \leq m^{2 \alpha +1}$.  Then the error bound \R{OptErrBd} is replaced by
\bes{
\nmu{f - \tilde{f}_m}_{L^2} 
\leq C_{\alpha} \nm{f}_{PC^{\alpha}} \left ( \frac{\sqrt{\log(m)}}{m^{\alpha}} + \sqrt{\frac{\cN(f)}{N}} \right ).
}
}

Similarly, we have a nonuniform result analogous to Theorem~\ref{t:OptErr2}.

\thm{
\label{t:OptErrNU2}
Consider the setup of Theorem~\ref{t:OptErrNU}, except where the basis pursuit problem is of size at most $m \times N$ for some $m \leq N \leq m^{2 \alpha +1}$.  Then the error bound \R{OptErrNUBd} is replaced by
\bes{
\nmu{f - \tilde{f}_m}_{L^2} 
\leq C_{\alpha} \nm{f}_{PC^{\alpha}} \left ( m^{-\alpha} + \sqrt{\frac{\cN(f)}{N}} \right ).
}
}

Finally, for the Fourier encoder-decoder, we have:

\thm{
\label{t:FourErr2}
Consider the setup of Theorem \ref{t:FourErr}, except where the square-root LASSO problem is of size at most $m \times N$ for some $m \leq N \leq m^{\sigma}$, where $\sigma= \max\{\frac{\alpha}{\alpha-1/2},2\alpha +1\}$.  Then the error bounds \R{FourErrBd1} and \R{FourErrBd2} are replaced by
\be{
\label{FourErrBd1alt}
\nmu{f - \tilde{f}_m}_{L^2} \leq C_{p,\alpha} \nm{f}_{PC^{\alpha}} \left ( m^{-\alpha} + N^{-(\alpha-1/2)} + \frac{\cN(f)}{\sqrt{N}} \right ) (\log(m))^{\frac{1}{4} + \frac{(6+\delta)(q+\alpha+1/2)}{2(q+1)}} ,
}
and
\be{
\label{FourErrBd2alt}
\nmu{f - \tilde{f}_m}_{L^2} \leq C_{p,\alpha} \nm{f}_{PC^{\alpha}} \left ( m^{-\alpha} + N^{-(\alpha-1/2)} + \frac{\cN(f)}{\sqrt{N}} \right ) (\log(m))^{\frac{13}{4} + \delta} ,
}
respectively.
}

Notice that Theorems \ref{t:GaussErr}, \ref{t:OptErr}, \ref{t:FourErr}, and \ref{t:OptErrNU} are just corollaries of Theorems \ref{t:GaussErr2}, \ref{t:OptErr2}, \ref{t:FourErr2}, and \ref{t:OptErrNU2} respectively, obtained by setting $N = \lfloor m^{2 \alpha + 1} / (\log(m))^{2 \alpha+2} \rfloor$, $N = \lfloor m^{2 \alpha + 1} / (\log(m))^{3} \rfloor$, $N = \lfloor \max \{ m^{\frac{\alpha}{\alpha-1/2}} , m^{2 \alpha+1} \} \rfloor$ or $N = \lfloor m^{2 \alpha + 1} / (\log(m))^{2} \rfloor$.  For instance, in the case of Theorem \ref{t:OptErr} this choice of $N$ gives
\bes{
 \sqrt{\frac{\cN(f)}{N}} \leq \frac{\sqrt{\log(m)}}{m^{\alpha}} \sqrt{ \frac{\cN(f)(\log(m))^2}{m} } \leq C_{\alpha} \frac{\sqrt{\log(m)}}{m^{\alpha}},
}
which yields the corresponding error bound in Theorem \ref{t:OptErr} (note here we use the assumed condition $m / \log^2(m) \geq c_{\alpha} \cN_{*}$).

However, the above results also show that the error bounds can be obtained using asymptotically smaller $N$ -- respectively $N = \lfloor m^{2 \alpha} / (\log(m))^{2 \alpha} \rfloor$, $N = \lfloor m^{2 \alpha} / \log(m) \rfloor $ or $N = \lfloor \max \{ m^{\frac{\alpha}{\alpha-1/2}} , m^{2 \alpha} \} \rfloor$ -- at the expense of having the additional factor depending on $\cN(f)$ in the error bound.  For instance, making this choice in Theorem \ref{t:OptErr2} results in the error bound
\bes{
\nmu{f - \tilde{f}_m}_{L^2} 
\leq C_{\alpha} \nm{f}_{PC^{\alpha}} \left ( 1 + \sqrt{\cN(f)} \right) \frac{\sqrt{\log(m)}}{m^{\alpha}} .
}

\section{Proof of Theorem \ref{t:waveletsterm}}
\label{s:proof_waveletsterm}

While numerous variants on Theorem \ref{t:waveletsterm} exist in the literature on wavelet approximation, we include a short proof in order to make the dependence on $\cN(f)$ explicit, and also for exposition, since the key ideas in the proof will be used in subsequent arguments.  

In what follows, we will make use of the periodic extension $f^\mathrm{ext}$ of a function $f$, defined as
$$
f^\mathrm{ext}(x+k) = f(x), \quad \forall x \in [0,1), \; \forall k \in \mathbb{Z}. 
$$
Notice that for every $f \in L^\infty([0,1])$ and $g \in L^1(\mathbb{R})$, this satisfies
$$
\langle f^\mathrm{ext} , g \rangle_{L^2(\mathbb{R})} 
= \langle f , g^\mathrm{per} \rangle_{L^2([0,1])}.
$$
where $g^{\mathrm{per}}$ is the periodization of $g$, defined by
\be{
\label{def_periodization}
g^{\mathrm{per}}(x) = \sum_{k \in \bbZ} g(x+k).
}
We first require the following lemma:
\lem{
\label{l:PLipalpha_coeff}
Suppose that $f \in PC^{\alpha}([0,1])$, where $\alpha = d + \beta$ for some $d \in \bbN_0$ and $0 < \beta \leq 1$.  Consider the periodized Daubechies' wavelet basis with $p > d$ vanishing moments and coarsest scale $j_0$ given by \R{j0def}.  Then, there exists a constant $C_{p,\alpha} >0$ such that
\bes{
| \ip{f}{\psi^{\mathrm{per}}_{j,n}} | \leq C_{p,\alpha} \nm{f}_{PC^\alpha} 2^{-(\alpha+1/2)j},
}
whenever the interval $[(n-p+1)/2^j , (n+p)/2^j ] = \overline{\mathrm{supp}(\psi_{j,n})}$ contains no discontinuities of $f^{\mathrm{ext}}$ and
\bes{
| \ip{f}{\psi^{\mathrm{per}}_{j,n}} | \leq C_{p,\alpha} \nm{f}_{PC^\alpha} 2^{-j/2},
}
otherwise. The constant can be chosen as 
$$
C_{p,\alpha} 
= \max\left\{\frac{1}{d!} \int_{-p+1}^p |\psi(x)||x|^\alpha dx,
\sqrt{2p-1}\right\}.
$$
}
\prf{
 First, we observe that for every interval $I\subseteq \mathbb{R}$ the periodic extension of $f$ satisfies $f^{\mathrm{ext}} \in PC^\alpha(I)$ and $\|f^{\mathrm{ext}}\|_{PC^\alpha(I)}\leq \|f\|_{PC^\alpha([0,1])}$.

Let us consider the case where $I:=\overline{\mathrm{supp}(\psi_{j,n})}$ contains no discontinuities of $f^{\mathrm{ext}}$. In that case, $f^\mathrm{ext} \in C^\alpha(I)$ and, using Taylor's theorem it is not difficult to show that there exists a polynomial $T$ of degree $d$ such that 
$$
|f^\mathrm{ext}(x) - T(x)| 
\leq \frac{1}{d!} \|f^\mathrm{ext}\|_{C^\alpha(I)} |x- n/2^{j}|^\alpha, \quad \forall x \in I.
$$
As a consequence, by using the fact that $\psi$ has $p>d$ vanishing moments, we have 
$$
|\langle f,\psi_{j,n}^\mathrm{per}\rangle|
= |\langle f^\mathrm{ext},\psi_{j,n}\rangle|
= |\langle f^\mathrm{ext}-T,\psi_{j,n}\rangle|
\leq \frac{1}{d!} \|f^\mathrm{ext}\|_{C^\alpha(I)} 2^{-(\alpha+1/2)j}\int_{-p+1}^p |\psi(x)||x|^\alpha dx,
$$
which implies the first estimate.

When $I:=\overline{\mathrm{supp}(\psi_{j,n})}$ contains at least one discontinuity of $f^\mathrm{ext}$, we have
$$
|\langle f,\psi_{j,n}^\mathrm{per}\rangle|
= |\langle f^\mathrm{ext},\psi_{j,n}\rangle|
\leq \|f^\mathrm{ext}\|_{C(I)} \|\psi_{j,n}\|_{L^1}
\leq \|f\|_{C([0,1])} 2^{-j/2} \sqrt{2p-1}.
$$
This yields the second estimate and concludes the proof.
}

We are now in a position to prove Theorem~\ref{t:waveletsterm}.

\prf{[Proof of Theorem~\ref{t:waveletsterm}]
Consider $e_s(f)_{L^2}$.  Without loss of generality, we may assume that $s \geq 2^{j_0+2}$ (up to choosing $C_{p,\alpha}$ large enough).  Hence there exists a $j_1 \geq 2$ such that $2^{j_0+j_1} \leq s < 2^{j_0+j_1+1}$.  Due to the ordering \R{DBbasis}, the linear approximation of $f$ of order $s$ includes at least all the scaling function coefficients $\ip{f}{\varphi^{\mathrm{per}}_{j_0,n}}_{L^2}$ and the wavelet coefficients $\ip{f}{\psi^{\mathrm{per}}_{j,n}}_{L^2}$ with $j=j_0,\ldots,j_0+j_1-1$.  Therefore
\bes{
\left ( e_s(f)_{L^2} \right )^2 
\leq \sum_{j \geq j_0+j_1} \sum^{2^j-1}_{n=0} | \ip{f}{\psi^{\mathrm{per}}_{j,n}} |^2 .
}
Since $f$ has $\cN(f)-1$ discontinuities, the extension $f^{\mathrm{ext}}$ has at most $2p \cN(f)$ discontinuities on the interval $[-p+1,p]$, and therefore there are at most $(2p)^2 \cN(f)$ wavelets $\psi_{j,n}$ at any fixed scale $j$ whose support contains a discontinuity of $f^{\mathrm{ext}}$. 
Since there are no more than $2^j$ wavelets at each scale $j$ supported in smooth regions of $f$, we deduce that
\bes{
\left ( e_s(f)_{L^2} \right )^2 
\leq C_{p,\alpha} \nm{f}^2_{PC^\alpha} \sum_{j \geq j_0+j_1} \left ( \cN(f) 2^{-j} + 2^{-2\alpha j} \right ) \leq C_{p,\alpha} \nm{f}^2_{PC^\alpha} \cN(f) 2^{-j_0-j_1},
}
where we have also used the fact that $\alpha \geq 1/2$.  Since $s \leq 2^{j_0+j_1+1}$ the result now follows.

Now consider $\sigma_s(f)_{L^2}$.  The idea is to judiciously choose $s$ wavelet coefficients so as to obtain the desired error decay rate.  To this end, note that, by assumption, $s \geq 2^{j_0+1}$ and suppose that $\{I_j\}_{j \geq j_0}$ is a collection of index sets with $I_j \subseteq \{0,\ldots,2^{j}-1\}$ that satisfies
\be{
\label{Indsetsizecond}
\left | \bigcup_{j \geq j_0} I_j \right | \leq s - 2^{j_0},
}
(the subtraction of $2^{j_0}$ takes into account the inclusion of the scaling coefficients).  Then, the resulting approximation is $s$-sparse and, using the orthonormality of the wavelet basis in $L^2([0,1])$, we obtain
\begin{align}
\nonumber
\left(\sigma_s(f)_{L^2}\right)^2 & \leq
\Bigg\|f - \Bigg(\sum_{0 \leq k < 2^{j_0}} \langle f, \varphi_{j_0,k}^{\mathrm{per}}\rangle \varphi_{j_0,k}^{\mathrm{per}}
+\sum_{j = j_0}^\infty \sum_{k \in I_j} \ip{f}{\psi^{\mathrm{per}}_{j,k}} \psi_{j,k}^{\mathrm{per}} \Bigg)\Bigg\|_{L^2}^2\\
& = 
 \sum^{\infty}_{j=j_0} \sum_{\substack{0 \leq k < 2^j \\ k \notin I_j}} | \ip{f}{\psi^{\mathrm{per}}_{j,k}} |^2.
 \label{sigmasfIjbd}
\end{align}
We now define the index sets $\{I_j\}_{j \geq j_0}$ .  First, for $j_1,j_2 \in \bbN$ such that $1 \leq j_1 \leq j_2$, which will be chosen in a moment, let
\be{
\label{Ijchoice}
\begin{split}
I_j &= \{ 0,\ldots, 2^{j}-1 \},\quad  j_0 \leq j < j_0+j_1,
\\
I_j &= \emptyset,\quad j \geq j_0+j_2,
\end{split}
}
and for $j_0+j_1 \leq j < j_0+j_2$ let $I_j$ be the index set consisting of those values of $n$ where the support of $\psi_{j,n}$ contains a discontinuity of $f^{\mathrm{ext}}$. 
Recall that, in this last case, $|I_j| \leq (2p)^2 \cN(f)$ from earlier in the proof.  Notice that
\be{
\label{Iconstructionsize}
\left | \bigcup_{j \geq j_0} I_j \right | 
= \sum^{j_0+j_1-1}_{j=j_0} 2^j + \sum^{j_0+j_2-1}_{j=j_0+j_1} | I_j| \leq 2^{j_0+j_1} + (2p)^2 \cN(f) (j_2-j_1).
}
We now examine the error.  From \R{sigmasfIjbd} and the definition of the $I_j$, we have
\bes{
\left(\sigma_s(f)_{L^2}\right)^2 
\leq \sum^{j_0+j_2-1}_{j=j_0+j_1} \sum_{\substack{0 \leq k < 2^j \\ k \notin I_j}} |\ip{f}{\psi^{\mathrm{per}}_{j,k}}|^2 + \left ( e_{\tilde{s}}(f)_{L^2} \right )^2,
}
where $\tilde{s} = 2^{j_0+j_2}$.  Hence, using Lemma \ref{l:PLipalpha_coeff} and the first part of the theorem, we deduce that
\begin{align*}
\left(\sigma_s(f)_{L^2} \right)^2 
& \leq C_{p,\alpha} \nm{f}^2_{PC^{\alpha}}  \left ( \sum^{j_0+j_2-1}_{j=j_0+j_1} 2^{-2 \alpha j} + \mathcal{N}(f) 2^{-(j_0+j_2)} \right ) \\
& \leq C_{p,\alpha} \nm{f}^2_{PC^{\alpha}}  \left ( 2^{-2 \alpha (j_0+j_1)} + \cN(f) 2^{-(j_0+j_2)} \right ).
\end{align*}
Now recall that, by assumption, $s \geq 2^{j_0+2}$.  Set $j_1 = \lfloor \log_2(s/2) \rfloor - j_0$ so that $s/4 \leq 2^{j_0+j_1} \leq s/2$, and let $j_2 = \lfloor (2 \alpha + 1)\log_2( s) \rfloor - j_0$ so that $s^{2\alpha + 1}/2 \leq 2^{j_0+j_2} \leq s^{2\alpha + 1}$.  This gives 
\bes{
\left(\sigma_s(f)_{L^2}\right)^2 
\leq C_{p,\alpha} \nm{f}^2_{C^{\alpha}}  s^{-2 \alpha} (1 + \cN(f) s^{-1}).
}
Notice that $j_0+j_2 > j_0+j_1$ since $\alpha \geq 1/2$ and $s \geq 2^{j_0+1}\geq 2$, hence this choice is valid.  

It remains to verify that \R{Indsetsizecond} holds for this choice of index set. Substituting the values of $j_1$ and $j_2$ into \R{Iconstructionsize} and observing that $j_2-j_1 \leq 2\alpha \log_2(s) +2 \leq 4 \alpha \log_2(s)$ (since $s \geq 2^{j_0+2} \geq 4$ and $\alpha \geq 1/2$), we see that
\bes{
\left | \bigcup_{j \geq j_0} I_j \right |  
\leq \frac{s}{2} + 16\alpha p^2 \cN(f) \log_2(s).
}
Therefore, recalling again that $s \geq 2^{j_0+2}$, condition \R{Indsetsizecond} holds since
$$
\frac{s}{\log_2(s)} \geq 64 \alpha p^2 \cN(f)
\quad \Longrightarrow \quad
\frac{s}{2} + 16\alpha p^2 \cN(f) \log_2(s) \leq \frac{s}{2}+\frac{s}{4} \leq   s - 2^{j_0},
$$
where the last inequality holds because $s/4 \geq 2^{j_0+2} \geq 2^{j_0}$. Moreover, the assumption on $s$ implies that $\cN(f) s^{-1} \leq 1$. This completes the proof.
}

\section{Proof of Theorems \ref{t:GaussErr} and \ref{t:GaussErr2}}\label{s:GaussErrProof}

We now move on to the proofs of our main theorems.  For each, we first give a recipe that describes the inputs (the number of measurements $m$ and the smoothness parameter $\alpha$), the various parameters for the encoder and decoder, and then the encoder and decoder themselves.

\subsection{Recipe}

The following recipe applies to Theorem \ref{t:GaussErr}:

\pbk \textbf{Inputs:} Number of measurements $m$, smoothness parameter $\alpha$.

\pbk \textbf{Parameters:} Let
\bull{
\item $p = \lceil \alpha \rceil$, $j_0$ be as in \R{j0def};
\item $\{ \varphi^{\mathrm{per}}_{j_0,k} \} \cup \{\psi^{\mathrm{per}}_{j,k} \}$ be the periodized Daubechies' wavelet basis with $p$ vanishing moments;
\item $r = \lfloor \log_2( m^{2 \alpha+1} / (\log(m))^{2 \alpha+2} ) \rfloor - j_0$;
\item $N = 2^{j_0+ r}$;
\item $A \in \bbR^{m \times N}$ be a random Gaussian matrix. 
}

\noindent \textbf{Encoder:} Define $\cE_m(f) = A \left ( \ip{f}{\phi_j} \right )^{N}_{j=1}$, where $\{ \phi_j \}^{N}_{j=1}$ are the periodized Daubechies' wavelets up to scale $j_0+r-1$ (see \R{DBbasis}).

\pbk \textbf{Decoder:} Given measurements $y = \cE_m(f)$, define $\cD_m(y)$ as
\bes{
\cD_m(y) = \sum^{N}_{j=1} \tilde{d}_j \phi_j,
}
where $\tilde{d} = (\tilde{d}_j)^{N}_{j=1} \in \bbC^N$ is any solution of the basis pursuit problem
\bes{
\min_{z \in \bbC^N} \nm{z}_{\ell^1}\ \mbox{subject to $A z = y$.}
}

\pbk \textbf{The case of Theorem \ref{t:GaussErr2}:} We now assume $N$ is an input rather than a parameter, and change the definition of $r$ to $r = \lfloor \log_2(N) \rfloor - j_0$.  Note that
\bes{
N/2 \leq 2^{j_0+r} \leq N.
}
Hence, if necessary, we replace $N$ by $2^{j_0+r}$ so that the encoder-decoder pair includes all the wavelets up to scale $j_0+r-1$.  This pair is then defined in exactly the same way as above.

\subsection{Proof of Theorem \ref{t:GaussErr2}}

Notice that $y = \cE_m(f) = A P_N d$ for this encoder, where $d$ is the infinite vector of wavelet coefficients of $f$.  Hence, $\tilde{d}$ is a minimizer of the problem
\bes{
\min_{z \in \bbC^N} \nm{z}_{\ell^1}\ \mbox{subject to $A z = A P_N d$.}
}
Define $s$ by
$$
s = \left\lfloor \frac{m}{2 c_1(2\alpha+1)\log(m)}\right\rfloor.
$$
Recall that $N \leq m^{2 \alpha + 1}$ by assumption and also that $s \geq  \E$ for all $m \geq c_{\alpha}$. Hence
\eas{
c_1(s \log(\E N/s) + \log(2/\varepsilon)) & \leq c_1 \frac{m}{2 c_1 (2\alpha+1) \log(m)} \log(m^{2 \alpha + 1}) + c_1 \log(2 / \varepsilon)
 \leq \frac{m}{2} + \frac{m}{2} = m,
}
where in the final step we recall that $m \geq c \log(2/\varepsilon)$ by assumption. Then, by Theorems \ref{t:CSRIPstandard} and \ref{t:RIPGauss} we have
$$
\|P_N d - \tilde{d}\|_{\ell^2} \leq C \frac{\sigma_s(P_N d)_{\ell^1}}{\sqrt{s}},
$$
with probability at least $1-\varepsilon$.   

We next estimate $\sigma_s(P_N d)_{\ell^1}$ using arguments analogous to those employed in the proof of Theorem~\ref{t:waveletsterm}.  First, we select all coefficients in scales $j_0, \ldots, j_0+\bar{r} -1$ where $j_0+\bar{r} = \lfloor\log_2(s/2)\rfloor$ (note that $j_0+\bar{r} \geq j_0 + 1$ as soon as $s \geq 2^{j_0+3}$, which is guaranteed by choosing $m \geq c_\alpha$). 
Next, in the remaining levels, we select all the coefficients corresponding to wavelets intersecting the discontinuities of $f^\mathrm{ext}$. 
Then there are at most $(2p)^2 \cN(f)$ of such coefficients at any given scale. Hence, the total number of such coefficients is bounded by
$$
(2p)^2\cN(f)(r - \bar{r}) 
\leq (2p)^2 \cN(f) \log_2(N) 
\leq c'_\alpha \cN(f) \log(m),
$$ 
which is at most $s/2$ for $m / \log^2(m) \geq c_{\alpha} \cN(f)$.  
Hence Lemma \ref{l:PLipalpha_coeff} gives
\eas{
\sigma_s(P_N d)_{\ell^1} 
&\leq C_{\alpha}\|f\|_{PC^{\alpha}}
\left(
\sum_{j = j_0+\bar{r}}^{j_0+r} 2^{-(\alpha+1/2)j} \cdot 2^j 
\right)
\\
&\leq C_{\alpha}\|f\|_{PC^{\alpha}} 2^{-(\alpha-1/2)(j_0+\bar{r})}
\\
&\leq C_{\alpha}\|f\|_{PC^{\alpha}} s^{-(\alpha-1/2)}.
}
In the last inequality we have used the fact that $\alpha > 1/2$. Therefore
$$
\|P_N d - \tilde{d}\|_{\ell^2} 
\leq C_{\alpha}\|f\|_{PC^{\alpha}} s^{-\alpha}
\leq C_{\alpha} \|f\|_{PC^{\alpha}} \left(\frac{\log(m)}{m}\right)^\alpha.
$$
By Theorem \ref{t:waveletsterm}, the linear approximation error satisfies
$$
e_N(f)_{L^2} 
\leq C_\alpha \|f\|_{PC^{\alpha}} \sqrt{\cN(f)/N} 
$$
Observing that
$$
\|f-\tilde{f}_m\|_{L^2} \leq \|P_N d - \tilde{d}\|_{\ell^2} + e_N(f)_{\ell^2}.
$$
concludes the proof.

\subsection{Proof of Theorem \ref{t:GaussErr}}
In this case $N = 2^{j_0+r}$, where $r = \lfloor \log_2( m^{2 \alpha+1} / (\log(m))^{2 \alpha+2} ) \rfloor - j_0$.  Observe that
\bes{
\frac{m^{2 \alpha+1}}{2 (\log(m))^{2 \alpha+2}} \leq N \leq \frac{m^{2 \alpha+1}}{ (\log(m))^{2 \alpha+2}}.
}
We now apply Theorem \ref{t:GaussErr2} and use the fact that $c_{\alpha} \cN(f) \leq  m / (\log(m))^2$ by assumption.

\section{Proof of Theorems \ref{t:OptErr}, \ref{t:OptErrNU}, \ref{t:OptErr2} and \ref{t:OptErrNU2}}\label{s:OptErrProof}

We now move on to the optimal sampling strategy.  Recall from \S\ref{ss:logfactors} that the key here is to directly sample up to a certain scale ($\bar{r}+j_0$ below) and then subsample the remaining scales with a random Gaussian matrix.

\subsection{Recipe}

We first consider Theorems \ref{t:OptErr} and \ref{t:OptErrNU} (the recipe is the same for both):

\pbk \textbf{Inputs:} number of measurements $m$, smoothness parameter $\alpha$.

\pbk \textbf{Parameters:} Let
\bull{
\item $p = \lceil \alpha \rceil$, $j_0$ be as in \R{j0def};
\item $\{ \varphi^{\mathrm{per}}_{j_0,k} \} \cup \{\psi^{\mathrm{per}}_{j,k} \}$ be the periodized Daubechies' wavelet basis with $p$ vanishing moments;
\item $r = \lfloor(2 \alpha +1) \log_2(m) \rfloor - j_0$;
\item $\bar{r} = \lfloor \log_2(m/2) \rfloor - j_0$;
\item $N_1 = 2^{j_0+\bar{r}}$, $N_2 = 2^{j_0+r}$;
\item $m_2 = m - m_1$, where $m_1 = N_1 = 2^{j_0+\bar{r}}$;
\item $A = (a_{il})^{m_2,N_2-N_1}_{i,l=1} \in \bbR^{m_2 \times (N_2-N_1)}$ be a random Gaussian matrix.
}
This choice of parameters requires that $\bar{r} > 0$ and that $\bar{r} < r$. However, we note that the former holds for every $m \geq 2^{j_0+2}$ and the latter holds for every $\alpha > 1/2$.

\pbk \textbf{Encoder:} Write $\cE_{m}(f) = \left ( \begin{array}{c} \cE^{(1)}_{m_1}(f) \\ \cE^{(2)}_{m_2}(f) \end{array} \right )$, where 
\bes{
\cE^{(l)}_{m_l}(f)
= \left ( e^{(l)}_{i}(f) \right )^{m_l}_{i=1} \in \bbC^{m_l},
\qquad l = 1,2.
}
For $l=1$, define
\eas{
e^{(1)}_{k+1}(f) 
&= \ip{f}{\varphi_{j_0,k}^\mathrm{per}}_{L^2},
\quad k=0,\ldots,2^{j_0}-1,
\\
e^{(1)}_{2^{j}+k+1}(f) 
&= \ip{f}{\psi_{j,k}^\mathrm{per}}_{L^2},
\quad k = 0,\ldots,2^{j}-1,\ j = j_0,\ldots,j_0+\bar{r}-1.
}
For $l = 2$, we let
\bes{
\chi_{i} = 
\sum^{j_0+r-1}_{j = j_0+\bar{r}} \sum^{2^j-1}_{k=0} a_{i,2^{j-\bar{r}}+k} \psi^{\mathrm{per}}_{j,k},\qquad i = 1,\ldots,m_2,
}
and define
\bes{
e^{(2)}_{i}(f) = \ip{f}{\chi_i}_{L^2},\quad i = 1,\ldots,m_2.
}

\pbk \textbf{Decoder:} Given measurements $\cE_m(f)$ as above, define
\bes{
\cD_m(\mathcal{E}_m(f)) 
= \sum^{2^{j_0}-1}_{k=0} e^{(1)}_{k+1}(f) \varphi^{\mathrm{per}}_{j_0,k} 
+ \sum^{j_0+\bar{r}-1}_{j = j_0} \sum^{2^j-1}_{k=0} e^{(1)}_{2^j+k+1}(f) \psi^{\mathrm{per}}_{j,k} 
+ \sum^{j_0+r-1}_{j=j_0+\bar{r}} \sum^{2^j-1}_{k=0} \tilde{d}_{2^{j-\bar{r} -j_0}+k} \psi^{\mathrm{per}}_{j,k},
}
where the vector $\tilde{d} = (\tilde{d}_i )^{N_2-N_1}_{i=1}$ is any minimizer of the basis pursuit problem
\be{
\label{2levelmin}
\min_{z \in \bbC^{N_2-N_1}} \nm{z}_{\ell^1}\ \mbox{subject to $A z = \cE^{(2)}_{m_2}(f)$}.
}

\pbk \textbf{The case of Theorems \ref{t:OptErr2} and \ref{t:OptErrNU2}:} As previously, we let $N$ be an input rather than a parameter, and change the definition of $r$ to $r = \lfloor \log_2(N) \rfloor - j_0$.  We now set $N_2 = 2^{j_0+r}$ and note that
\bes{
N/2 \leq N_2 \leq N,
}
so that the encoder-decoder pair includes all the wavelets up to the scale $j_0+r-1$.  This pair is then defined in exactly the same way.

\subsection{Proof of Theorem \ref{t:OptErr2}}

Let $d \in \ell^2(\bbN)$ be the wavelet coefficients of $f \in L^2([0,1])$.  By Parseval's identity, and the fact that $\tilde{f}_m$ recovers the coarse scale coefficients exactly, we have
\be{
\label{2levelmainerr}
\nmu{f - \tilde{f}_m}_{L^2} \leq \nmu{P^{N_1}_{N_2} d - \tilde{d}}_{\ell^2}+ \nmu{P^{\perp}_{N_2} d}_{\ell^2}.
}
Let $\bar{d} = (\bar{d}_i)^{N_2-N_1}_{i=1} = P^{N_1}_{N_2} d$.  Then,
\bes{
e^{(2)}_{i}(f) = \sum^{N_2-N_1}_{l=1} a_{i,l} \bar{d}_i,
}
and therefore $\cE^{(2)}_{m_2}(f) = A \bar{d}$.  Hence \R{2levelmin} is equivalent to
\bes{
\min_{z \in \bbC^{N_2-N_1}} \nm{z}_{\ell^1}\ \mbox{subject to $A z = A \bar{d}$}.
}
Since $\tilde{d}$ is a minimizer, we deduce from Theorems \ref{t:CSRIPstandard} and \ref{t:RIPGauss} that there exists $C >0$ such that
\be{
\label{2levelCSbit}
\nmu{P^{N_1}_{N_2} d - \tilde{d}}_{\ell^2} 
\leq C \frac{\sigma_{s}\left(P^{N_1}_{N_2} d\right)_{\ell^1}}{\sqrt{s}},
}
with probability at least $1-\varepsilon$, where $s \in \mathbb{N}$ is any number such that
\be{
\label{m2cond}
m_2 \geq c_1 \left( s \log(\E(N_2-N_1)/s) + \log(2 / \varepsilon) \right ).
}
We now choose $s$ as follows:
\be{
\label{eq:cond_s}
s = \left \lfloor \frac{m}{4 c_1 (2\alpha +1) \log(m)} \right \rfloor.
}
Observe that $m_1 \leq m/2$ by construction and, consequently, $m_2 \geq m/2$.  Also, since $N_2 \leq N \leq m^{2 \alpha + 1}$, we have
\eas{
c_1 \left( s \log(\E(N_2-N_1)/s) + \log(2 / \varepsilon) \right ) & \leq c_1 \frac{m}{4 c_1 (2\alpha +1) \log(m)} \log(\E m^{2 \alpha + 1} / s) + c_1 \log(2/\varepsilon)
\\
& \leq \frac{m}{4} + \frac{m}{4} = m/2 \leq m_2,
}
hence this choice of $s$ is valid.  Here, in the second step we used the facts that $s \geq \E$ for $m / \log(m) \geq c_{\alpha}$ and $c_1 \log(2/\varepsilon) \leq m/4$ for $m \geq c \log(2/\varepsilon)$.

We now consider \R{2levelCSbit}.  Recall that $P^{N_1}_{N_2} d$ is the vector of wavelet coefficients at scales $j_0+\bar{r},\ldots,j_0+r-1$.  As previously, we select all coefficients corresponding to wavelets intersecting the discontinuities of $f^\mathrm{ext}$.  The total number of such coefficients is at most
\begin{align*}
(2p)^2 \cN(f) (r-\bar{r}) 
& \leq (2p)^2\cN(f) \log_2(N)
\leq 12(\alpha+1)^3 \cN(f) \log(m),
\end{align*}
where we used that $p = \lceil \alpha \rceil \leq \alpha+1$ and the assumption $N \leq m^{2\alpha +1}$.  Recalling \R{eq:cond_s} and the assumption on $m$, we note that the number of such coefficients does not exceed $s$. Indeed, under \R{eq:cond_s}, we have
\bes{
\frac{m}{\log^2(m)} \geq 48 c_2(\alpha+1)^4 \cN_*
\quad
\Longrightarrow
\quad
12(\alpha+1)^3 \cN(f) \log(m) \leq  \frac{1}{2} \cdot \frac{m}{c_2(2\alpha+1)\log(m)} \leq  s.
}
Hence, we may exclude all these slowly decaying coefficients, and, using Lemma~\ref{l:PLipalpha_coeff}, bound the best $s$-term approximation error by 
\begin{align*}
\sigma_{s}\left(P^{N_1}_{N_2} d\right)_{\ell^1} 
& \leq C_{\alpha} \nm{f}_{PC^{\alpha}} \sum^{j_0+r-1}_{j=j_0+\bar{r}} \sum^{2^j-1}_{k=0} 2^{-(\alpha+1/2)j} \\
&\leq C_{\alpha} \nm{f}_{PC^{\alpha}} 2^{-(\alpha-1/2)(j_0+\bar{r})}
\leq C_{\alpha} \nm{f}_{PC^{\alpha}} m^{-(\alpha-1/2)}.
\end{align*}
In the penultimate step we have used the fact that $\alpha > 1/2$ and that $2^{j_0+\bar{r}} \geq m/4$.  This gives
\bes{
\nmu{P^{N_1}_{N_2} d - \tilde{d}}_{\ell^2} 
\leq C_{\alpha} \nm{f}_{PC^{\alpha}} \frac{m^{-(\alpha-1/2)}}{\sqrt{s}} \leq C_{\alpha} \nm{f}_{PC^{\alpha}}  \sqrt{\log(m)} m^{-\alpha}.
}
To complete the proof, we now use \R{2levelmainerr}, after noting that $\nmu{P^{\perp}_{N_2} d}_{\ell^2} = e_{N_2}(f)_{L^2}$, and therefore
\be{
\label{N2linerr}
\nmu{P^{\perp}_{N_2} d}_{\ell^2} 
\leq C_{\alpha}  \nm{f}_{PC^{\alpha}} \sqrt{\cN(f)/N_2} \leq C_{\alpha} \nm{f}_{PC^{\alpha}}  \sqrt{\cN(f)/N} ,
}
by Theorem \ref{t:waveletsterm}.

\subsection{Proof of Theorem~\ref{t:OptErrNU2}}
Th argument is similar to the proof of Theorem~\ref{t:OptErr2}. First, we have
$$
\|f-\tilde{f}_m\|_{L^2} \leq \|P_{N_1}^{N_2} d - \tilde{d}\|_{\ell^2} + \|P^\perp_{N_2}d\|_{\ell^2},
$$
where $d, \tilde{d} \in \mathbb{C}^{N_2-N_1}$ are as defined therein. Theorem~\ref{t:nonuniform_instance_opt} implies that
\be{
\label{eq:estimate_OptErrNU2}
\|P_{N_1}^{N_2} d - \tilde{d}\|_{\ell^2} \leq C \sigma_s(P_{N_1}^{N_2}d)_{\ell^2},
}
with probability at least $1-\varepsilon$, provided $N_2- N_1 \geq c_1 m_2$ and
$$
m_2 \geq c_2 (s \log (e(N_1-N_1)/m) + \log(5/\varepsilon)).
$$
Note that $N_2- N_1 \geq m^{2\alpha +1}/2 - m/2$. Hence, since $m_1 \geq m/4$ by construction and, consequently, $m_2 \leq 3 m/4$, condition $N_2-N_1 \geq c_1 m_2$ holds whenever $m \geq c$, which is implied by the assumptions on $m$. Now, we let 
\bes{
s = \left \lfloor \frac{m}{4 c_2 (2\alpha +1) \log(m)} \right \rfloor.
}
Arguing as in Theorem~\ref{t:OptErr2}, the assumptions $m/\log(m) \geq c_\alpha$ and $m \geq c \log(5/\varepsilon)$ give
$$
 c_2 (s \log (e(N_1-N_1)/m) + \log(5/\varepsilon))  \leq c_2 (s \log (e(N_1-N_1)/s) + \log(5/\varepsilon)) \leq m_2.
$$
Hence this choice is valid.
Now, arguing as before, we deduce that
$$
\sigma_s(P_{N_2}^{N_1} d)_{\ell^2} \leq C_\alpha \|f\|_{PC^\alpha} m^{-\alpha}.
$$
Combining this with \R{N2linerr} now yields the result.

\subsection{Proof of Theorem \ref{t:OptErr}}

We let $N = 2^{j_0+r}$, where $r = \lfloor \log_2( m^{2 \alpha+1} / (\log(m))^{3} ) \rfloor - j_0$.  Then
\bes{
\frac{m^{2 \alpha+1}}{2 (\log(m))^{3}} \leq N \leq \frac{m^{2 \alpha+1}}{ (\log(m))^{3}}.
}
We now apply Theorem \ref{t:GaussErr2} with this choice of $N$ and using the fact that $\cN(f) \leq c_{\alpha} m / (\log(m))^2$ by assumption.

\subsection{Proof of Theorem~\ref{t:OptErrNU}}

We let $N = 2^{j_0+r}$, where $r = \lfloor \log_2(m^{2\alpha+1}/(\log(m))^2)\rfloor-j_0$. With this choice, we have
$$
\frac{m^{2\alpha+1}}{2 \log^2(m)} \leq N \leq \frac{m^{2\alpha+1}}{\log^2(m)}.
$$
We now apply Theorem \ref{t:OptErrNU2} and use the fact that $m/\log^2(m) \geq c_\alpha \cN(f)$.

\section{Wavelet approximation from Fourier samples}\label{s:FourWaveSetup}

The remainder of this paper is devoted to the proof of Theorems \ref{t:FourErr} and \ref{t:FourErr2}.  This requires some significant additional work.  In this section, we first formulate the approximation of wavelet coefficients from Fourier samples as a compressed sensing problem.  Next, in \S\ref{s:CSlocal} we present the framework for compressed sensing with local structure.  Proofs of the two main theorems are presented in \S\ref{s:FourErrProofs}.

In order to align with notation used in previous works, we now make a minor change in notation. In particular, we will denote the dimension of the truncated wavelet space by $M$ and the dimension of the truncated Fourier space by $N$.

\subsection{Formulation as a compressed sensing problem}\label{ss:formulateCS}

We follow the approach of \cite{BAACHGSCS}.
Let $\{ \gamma_i \}_{i \in \bbN}$ be the Fourier basis \R{FourBasis} and $\{ \phi_j \}_{j \in \bbN}$ be the periodized Daubechies' wavelet basis with $p$ vanishing moments \R{DBbasis}.  Define the infinite cross-Gramian matrix
\be{
\label{Udef}
U = \left ( \ip{\phi_j}{\gamma_i} \right )_{i,j \in \bbN}.
}
Notice that $U$ is a bounded, unitary operator on $\ell^2(\bbN)$, since both sets of functions are orthonormal bases for $L^2([0,1])$.  Recalling the notation introduced in \S\ref{ss:wavelets_and_nonlinear}, if $f \in L^2([0,1])$ is the function to recover, write $d = (d_i)_{i \in \bbN}$ for its wavelet coefficients, so that $f = \sum_{i \in \bbN} d_i \phi_i$.  Observe that
\be{
\label{Ucb}
b = U d,
}
where $b = \left ( \ip{f}{\gamma_i} \right )_{i \in \bbN}$ is the infinite vector of Fourier coefficients of $f$.

The Fourier encoder must use only $m$ Fourier samples, or equivalently, select $m$ rows of the infinite linear system \R{Ucb}.  Let $P_{\Omega} \in \bbC^{m \times \infty}$ be the matrix that selects such rows, where $\Omega \subset \bbZ$, $|\Omega| = m$ is the set of frequencies.  Then we consider the $m \times \infty$ linear system
\be{
\label{Ainf}
P_{\Omega} U z = P_{\Omega} b,
}
where $z \in \ell^2(\bbN)$.  This system is not suitable for computations, however, since the matrix $P_{\Omega} U$ has infinitely-many columns.  To handle this, we introduce an additional parameter $M \geq 1$ and replace \R{Ainf} by the $m \times M$ linear system
\be{
\label{Adef}
P_{\Omega} U P_M z = P_{\Omega} b,
}
where $z \in \bbC^M$ and $y = P_{\Omega} b$.  Note that a solution $z \in \bbC^M$ to this linear system is an approximation to the first $M$ wavelet coefficients of $f$, i.e.\ the vector $P_M d$.  Indeed,
\bes{
P_{\Omega} U P_M z = P_{\Omega} U  P_M d + e,\qquad \mbox{where $e = P_{\Omega} U P^{\perp}_M d$},
}
hence this problem is now a typical compressed sensing problem: the recovery of a vector $P_M d$ from measurements taken according to a matrix $P_{\Omega} U P_M \in \bbC^{m \times M}$ corrupted by noise $e$.

\subsection{Sampling and frequency bands}\label{ss:sampfreqbands}

Having done this, we need to prescribe the set of samples $\Omega$.  For this, we follow the approach of \cite{Candes_Romberg} and divide frequency space $\bbZ$ into \textit{dyadic} bands $B_k$.  These are defined as follows:
\eas{
B_1 &= \{ - 2^{j_0}+1,\ldots,2^{j_0} \},
\\
B_{k+1} &= \{ - 2^{j_0+k} + 1,\ldots,-2^{j_0+k-1} \} \cup \{ 2^{j_0+k-1}+1,\ldots,2^{j_0+k} \},\quad k = 1,2,\ldots .
}
Here $j_0$ is as in \R{j0def}.  Observe that $\cup_{k \in \bbN} B_k = \bbZ$, and that
\bes{
|B_1| = 2^{j_0+1},\quad |B_{k+1}| = 2^{j_0+k},\quad k=1,2,\ldots .
}
Since we typically consider the Fourier basis \R{FourBasisreindex} indexed over $\bbN$ instead of $\bbZ$ we now note that the $B_k$ are equivalent to the partition of $\bbN$ into subsets (subsequently referred to as \textit{levels})
\bes{
\{N_{k-1}+1,\ldots,N_k\},\quad k = 1,2,\ldots,
}
where
\be{
\label{Nkvalues}
N_0 = 0, \qquad N_{k} = 2^{j_0+k},\quad k = 1,2,\ldots.
}
With this in hand, we may now define the sampling scheme $\Omega$.  The idea is to select $m_1$ samples from the first level, $m_2$ samples from the second level, and so forth, up to some maximal level $r \geq 1$, where the numbers $m_1,\ldots,m_r$ satisfy $\sum^{r}_{k=1} m_k = m$.  A judicious choice of these numbers is a key ingredient in the proof of Theorems \ref{t:FourErr} and \ref{t:FourErr2}.  For $k=1,\ldots,r$, let $\Omega_k \subseteq \{N_{k-1}+1,\ldots,N_k\}$ be the subset of frequencies chosen in the $k^{\rth}$ level.  Then we write
\bes{
\Omega = \Omega_1 \cup \ldots \cup \Omega_r.
}
In what follows, $\Omega_k$ will be chosen randomly (see the next section).

\section{Compressed sensing with local structure}\label{s:CSlocal}

As mentioned, a crucial part of the framework for wavelet approximation from Fourier samples is the notion of \textit{local structure} in compressed sensing.  The sampling scheme introduced previously is an instance of this principle, in the sense that different local numbers of samples can be chosen in different frequency bands.  However, in order to perform wavelet approximation from Fourier samples we also need a notion of \textit{local sparsities}.

In this section, we formalize these notions, and tie them together with a recovery guarantee which generalizes the classical compressed sensing result (Theorem \ref{t:CSRIPstandard}).  This follows the framework of \cite{AHPRBreaking}.  Since it is unnecessary for the moment, in this section we do not assume the sampling/recovery is performed using the Fourier and wavelet bases, as in \S\ref{s:FourWaveSetup}.  We simply assume the existence of a unitary operator $U : \ell^2(\bbN) \rightarrow \ell^2(\bbN)$, an infinite sequence $d \in \ell^2(\bbN)$ to recover, and a collection of possible measurements $b = U d$ (recall \R{Ucb}).

\subsection{Definitions}

We commence with a series of definitions:

\defn{
\label{def:sampling_pattern}
Let $r \geq \tilde{r} \geq 1$, $\mb{N} = (N_1,\ldots,N_r)$, where $1 \leq N_1 < N_2 < \ldots < N_r < \infty$ and $\mb{m} = (m_1,\ldots,m_r)$ where $m_k = N_{k}-N_{k-1}$ for $k = 1,\ldots,\tilde{r}$ and $m_k < N_k - N_{k-1}$ for $k=\tilde{r}+1,\ldots,r$, with $N_0 = 0$.  An \textit{$(\mb{N},\mb{m})$-multilevel random subsampling pattern with saturation $\tilde{r}$} is a subset $\Omega \subset \bbN$ is of the form $\Omega = \Omega_1 \cup \cdots \cup \Omega_r$, where
\bes{
\Omega_k = \{ N_{k-1}+1,\ldots,N_k \},\quad k = 1,\ldots,\tilde{r},
}
and, for each $k=\tilde{r}+1,\ldots,r$, $\Omega_k = \{ t_{k,1},\ldots,t_{k,m_k} \}$ where the $t_{k,i}$ are chosen independently and uniformly at random from the indices $\{ N_{k-1}+1,\ldots,N_k \}$.
}

This formalizes the sampling strategy introduced in \S\ref{ss:sampfreqbands}.  Note that the $N_k$ need not be given by \R{Nkvalues} in general, although they will be whenever we consider the Fourier-wavelet problem.  We refer to $N = N_r$ as the \textit{sampling bandwidth}, and the subsets $\{ N_{k-1}+1,\ldots,N_k \}$ as \textit{levels}.  We also write $m = m_1+\ldots+m_r$ for the total number of measurements.

Given such a sampling pattern $\Omega$, we now proceed similarly to \S\ref{ss:formulateCS}, and write
\bes{
P_{\Omega} U P_M z = P_{\Omega} U P_M d + P_{\Omega} U P^{\perp}_M d.
}
For technical reasons, we also renormalize the rows of this linear system.  Let
\be{
\label{Ddef}
D = \diag(d_i)^{\infty}_{i=1} \in \bbC^{N \times \infty},\qquad d_{i} = \left \{ \begin{array}{cl}\sqrt{\frac{N_k - N_{k-1}}{m_k}} & N_{k-1} < i \leq N_k,\ k =1,2,\ldots,r \\ 0 & i > N \end{array} \right ..
}
We then replace this system with
\be{
\label{finitesytemtosolve}
A z = A P_M d + e,
}
where
\bes{
A = P_{\Omega} D U P_M \in \bbC^{m \times M},\qquad e = P_{\Omega} D U P^{\perp}_M d \in \bbC^{m}.
}
The purpose of this normalization is to ensure that
\be{
\label{AstarAexp}
\bbE(A^* A) = P_M U^* P_N U P_M.
}
For suitable $M$ and $N$, the matrix on the right-hand side is well-conditioned, which is important for the proof.  In order to ensure this, we define the following:

\defn{
Let $U : \ell^2(\bbN) \rightarrow \ell^2(\bbN)$ be unitary, $0 < \theta < 1$ and $N \geq M \geq 1$.  Then $U$ has the \textit{balancing property} with constant $\theta$ if
\be{
\label{balprop}
\nm{P_M U^* P_N U P_M - P_M }_{\ell^2} \leq 1 -\theta.
}
}

We next define an appropriate local version of sparsity:

\defn{
Let $r \geq 1$, $\mb{M} = (M_1,\ldots,M_r)$, where $1 \leq M_1 < M_2 < \ldots < M_r < \infty$ and $\mb{s} = (s_1,\ldots,s_r)$, where $s_k \leq M_k - M_{k-1}$ for $k=1,\ldots,r$, with $M_0 = 0$.  A vector $x = (x_i)^{M}_{i=1} \in \bbC^M$ is \textit{$(\mb{s},\mb{M})$-sparse} if
\bes{
\left | \supp(x) \cap \{ M_{k-1}+1,\ldots,M_k \} \right | \leq s_k,\quad k = 1,\ldots,r.
}
We write $\Sigma_{\mb{s},\mb{M}} \subseteq \bbC^M$ for the set of $(\mb{s},\mb{M})$-sparse vectors.
}

Note that we take $M_r = M$, where $M$ is the truncation parameter introduced above.  We refer to this as the \textit{sparsity bandwidth}.  We also write $s = s_1+\ldots+s_r$ for the total sparsity.

Although the sparsity levels can in general be arbitrary, when we consider the recovery of wavelet coefficients we will make the following specific choices:
\be{
\label{Mkvalues}
M_0 =0,\qquad M_{k} = 2^{j_0+k},\quad k = 1,2,\ldots.
}
Since we assume the wavelets are ordered in the standard way \R{DBbasis}, this means that the $k^{\rth}$ sparsity level (i.e.\ the indices $\{ M_{k-1}+1,\ldots,M_k \}$) corresponds exactly to the wavelet coefficients at scale $j = j_0+k-1$.

We require one additional concept.  It is well known \cite{Candes_Plan,Candes_Romberg} that recovery guarantees in compressed sensing from measurements taken according to a unitary matrix are determined by its so-called coherence, defined as follows:

\defn{
The \textit{coherence} of a matrix $B = (b_{ij})^{N,M}_{i,j=1} \in \bbC^{N \times M}$ is
\bes{
\mu(B) = N \max_{\substack{i=1,\ldots,N \\ j = 1,\ldots,M}} |b_{ij} |^2. 
}
}

In our setting, we define so-called \textit{local} coherences.  Specifically, let $U$ be the infinite matrix defined in \R{Udef}.  Then the leading $N \times M$ section $P_N U P_M$ can be expressed in the block form
\be{
\label{Ublock}
P_N U P_M= \left ( \begin{array}{cccc} U^{(1,1)} & U^{(1,2)} & \cdots & U^{(1,r)} \\ U^{(2,1)} & U^{(2,2)} & \cdots & U^{(2,r)} \\ \vdots & \vdots & \ddots & \vdots \\ U^{(r,1)} & U^{(r,2)} & \cdots & U^{(r,r)} \end{array} \right ),
}
where, for $k,l = 1,\ldots,r$, the matrix $U^{(k,l)}$ is the block of $U$ defined by the $k^{\rth}$ sampling level and $l^{\rth}$ sparsity level:
\bes{
U^{(k,l)} = P^{N_{k-1}}_{N_k} U P^{M_{l-1}}_{M_l} \in \bbC^{(N_k-N_{k-1}) \times (M_l - M_{l-1})}.
}
We define $(k,l)^{\rth}$ \textit{local coherence} of $U$ as the coherence of the corresponding block, i.e.\ $\mu \left( U^{(k,l)} \right )$.

\subsection{The weighted-square root LASSO decoder}

As discussed in \S\ref{ss:weightedSRLASSO} we will not employ QCBP \R{QCBP} to solve \R{finitesytemtosolve}.  The reason for this is now clear.   Standard recovery guarantees for \R{QCBP} require the bound $\nm{e}_{\ell^2} \leq \eta$, where $e$ is the noise term.  However, in \R{finitesytemtosolve} this term depends on the expansion tail $P^{\perp}_M d$, which is generally unknown.
Instead, we consider the \textit{weighted square-root LASSO} decoder
\be{
\label{infminsrlasso}
\min_{z \in \bbC^M} \lambda \nm{z}_{\ell^1_w} +\nm{A z - y}_{\ell^2}.
}
This was introduced in \cite{ABBCorrecting} for compressed sensing in the context of high-dimensional function approximation.
Here $y = A P_M d + e \in \bbC^m$, $w = (w_i)^{M}_{i=1}$ is a vector of positive weights,  $\nm{z}_{\ell^1_w} = \sum^{M}_{i=1} w_i |z_i|$ is the weighted $\ell^1$-norm, and $\lambda > 0$ is a parameter.  As we show below, the optimal choice of $\lambda$ is independent of the noise level $\nm{e}_{\ell^2}$, rendering this decoder suitable for our purposes.  

These weights in \R{infminsrlasso} are used to control the off-diagonal blocks in $P_N U P_M$ -- or, equivalently, the interferences between wavelet scales (see \S\ref{ss:FourOptdiscuss}) -- so as to give the best possible measurement condition.
As it transpires, it is sufficient for the weights $w_i$ to be constant on the sparsity levels.  Hence, from now on we assume that
\be{
\label{weightsinf1}
w_i = w^{(k)},\quad M_{k-1} < i \leq M_k,\qquad k = 1,\ldots,r.
}
We write $\mb{w} = (w^{(1)},\ldots,w^{(r)})$ for the vector of these weights.

\subsection{A levels-based compressed sensing guarantee}

We now present a recovery guarantee for \R{infminsrlasso}.  For this, we also define the $\ell^1_w$-norm best $(\mb{s},\mb{M})$-term approximation error:
\bes{
\sigma_{\mb{s},\mb{M}}(x)_{\ell^1_w} = \min \left \{ \nm{x - z}_{\ell^1_w} : z \in \Sigma_{\mb{s},\mb{M}} \right \},\qquad x \in \bbC^M,
}
Here $\Sigma_{\bm{s},\mb{M}}$ is the set of $(\mb{s},\mb{M})$-sparse vectors.

\thm{
\label{t:infrecovmain1}
Let $0 < \varepsilon < 1$, $1 \leq \tilde{r} \leq r$, $U : \ell^2(\bbN) \rightarrow \ell^2(\bbN)$ be unitary, $N\geq M \geq 1$, $\mb{M} = (M_1,\ldots,M_r)$ be sparsity levels, where $M_r = M$, and $\mb{s} = (s_1,\ldots,s_r)$ be local sparsities with $s_l \geq 1$, $\forall l$, $\mb{N} = (N_1,\ldots,N_r)$ be sampling levels, where $N_r = N$, $\mb{m} = (m_1,\ldots,m_r)$ be local numbers of measurements, and $A = P_{\Omega} D U P_M \in \bbC^{m \times M}$ arise from the $(\mb{N},\mb{m})$-multilevel sampling scheme with saturation $\tilde{r}$  of Definition~\ref{def:sampling_pattern}, where $D$ is as in \R{Ddef}.  Let $x \in \bbC^M$, $y = A x + e$ and suppose that
\begin{enumerate}[(A)]
\item  $N$ and $M$ are such that the balancing property \R{balprop} holds with constant $\theta$,
\item the weights $w$ are as in \R{weightsinf1}, with $\mb{w}$ such that
\be{
\label{weightsinf2}
c_1 \sqrt{s/s_k} \leq w^{(k)} \leq c_2 \sqrt{s/s_k},
}
for constants $c_1,c_2 > 0$,
\item the vector $\mb{m}$ of local numbers of measurements satisfies
\eas{
m_k &= N_k - N_{k-1},\quad k=1,\ldots,\tilde{r},
\\
m_{k} &\gtrsim \theta^{-2} \cdot \left(\frac{c_2}{c_1}\right)^2  \cdot \left ( \sum^{r}_{l=1} s_l \mu \left ( U^{(k,l)} \right ) \right ) \cdot L,\quad k = \tilde{r}+1,\ldots,r,
}
where $L = r^2 \cdot \log(m) \cdot \log^2(c_2^2 r s/(c_1^2 \theta)) \cdot \log(M) + r \cdot \log(\varepsilon^{-1})$,
\item the parameter $\lambda$ satisfies
\bes{
0 < \lambda \leq \frac{3 \sqrt{\theta}}{5 \sqrt{2}} \frac{1}{c_2\sqrt{r s}}.
}
\end{enumerate}
Then, with probability at least $1-\varepsilon$, any minimizer $\hat{x}$ of \R{infminsrlasso} satisfies
\be{
\label{inferr2}
\nm{\hat{x} - x}_{\ell^2} \lesssim  \frac{(c_2/c_1)^{1/2}}{r^{1/4} c_1 \sqrt{s}}\sigma_{\mb{s},\mb{M}}(x)_{\ell^1_w}  +  \frac{(c_2/c_1)^{1/2}}{r^{1/4} c_1 \sqrt{s} \lambda} \nm{e}_{\ell^2} .
}
}

The proof of this theorem is given in Appendix \ref{a:recovproof}.

\section{Proof of Theorems \ref{t:FourErr} and \ref{t:FourErr2}}\label{s:FourErrProofs}

\subsection{The idea}\label{ss:FourErrIdea}

Let $m$ and $\alpha$ be the inputs.  The proof is based on making a judicious choice of the \textit{encoder parameters} (number of levels $r$, the number of saturated levels $\tilde{r}$ and the local numbers of measurements $\mb{m} = (m_1,\ldots,m_r)$) and the \textit{decoder parameters} (weights $\mb{w} = (w^{(1)},\ldots,w^{(r)})$ and $\lambda$) in terms of $m$ and $\alpha$ so that there exists a vector of local sparsities $\mb{s} = (s_1,\ldots,s_r)$ such that conditions (A)--(D) of Theorem \ref{t:infrecovmain1} hold, and so that the right-hand side of \R{inferr2} behaves like $m^{-\alpha} (\log(m))^t$, where $t$ is the exponent given in Theorem \ref{t:FourErr}.

\begin{figure}[t]
\centering
\includegraphics[width = 7cm]{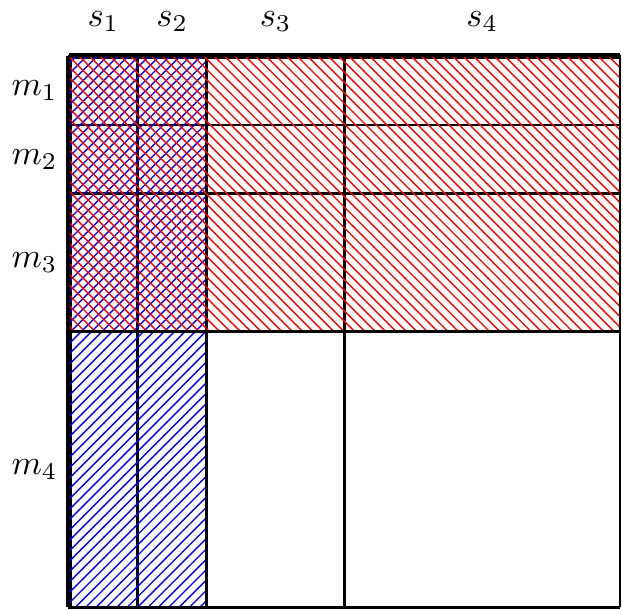}
\caption{An illustration of the idea of the proof. The diagram shows the matrix $U$ and its first $r = 4$ sampling and sparsity levels. The first $\tilde{r} = 3$ sampling levels are fully sampled, i.e.\ $m_1 = m_2 = 2^{j_0+1}$ and $m_3 = 2^{j_0+2}$, as indicated by the shaded rows (in red in the colour version). The first $\bar{r} = 2$ sparsity levels are saturated, i.e.\ $s_1 = s_2 =  2^{j_0+1}$, as indicated by the shaded columns (in blue in the colour version).  The remaining $r - \tilde{r} = 1$ sampling levels are randomly subsampled, and the local sparsities in the remaining $r - \bar{r} = 2$ sparsity levels are constant with $s_3 = s_4 = s_*$.\label{f:idea_ML_Fourier}}
\end{figure}

We first show that the choices made give that $\theta \asymp 1$, $c_1,c_2 \asymp 1$ and $r,\log(s),\log(M) \asymp \log(m)$.  The next key is the measurement condition (C) of Theorem~\ref{t:infrecovmain1}.  By estimating the local coherences and using the values of $N_k$ in \eqref{Nkvalues}, we show that the $\mb{m}$ must satisfy
\be{
\label{meascondidea}
\begin{split}
m_1 & = 2^{j_0+1}
\\
m_k &= 2^{j_0+k-1},\quad k=2,\ldots,\tilde{r},
\\
m_{k} &\asymp \left ( \sum^{k}_{l=1} s_l 2^{-(2q+1)(k-l)} + \sum^{r}_{l=k+1} s_l 2^{-(2p+1)(l-k)} \right ) \cdot (\log(m))^6,\quad k = \tilde{r}+1,\ldots,r,
\end{split}
}
where $p \geq 1$ is the number of vanishing moments and $q \geq 0$ is the smoothness.  Next, we make the following selection for the $\mb{s}$:
\be{
\label{sparsitiesdidea}
\begin{split}
s_{1} & = 2^{j_0+1}
\\
s_k &= 2^{j_0+k-1},\quad k=2,\ldots,\bar{r},
\\
s_k & = s_*,\quad k = \bar{r}+1,\ldots,r.
\end{split}
}
In other words, $s_k$ is equal to the size of the corresponding level for $k \leq \bar{r}$, and beyond this the $s_k$'s are constant. Note that we do not require $\bar{r} = \tilde{r}$.  A judicious choice of $\bar{r} < \tilde{r}$ is crucial in the proof.
As in previous proofs, $s_*$ is chosen large enough to capture all the discontinuities of $f$ at each scale.  This gives
\be{
\label{errorstepidea}
\frac{\sigma_{\mb{s},\mb{M}}(x)_{\ell^1_w}}{r^{1/4} \sqrt{s}} \lesssim \nm{f}_{PC^{\alpha}} \frac{2^{-(\alpha-1/2)\bar{r}}}{r^{1/4} \sqrt{s_*}},
}
(the other term in \R{inferr2} is bounded in a similar fashion to previous results).  We now relate this to the total number of measurements $m$.  Using the values \R{sparsitiesdidea} in \R{meascondidea}, we see that 
\bes{
m_{k} \asymp  \left ( 2^{2(q+1)\bar{r} - (2q+1) k}  + s_{*} \right ) \cdot (\log(m))^6,\qquad k = \tilde{r}+1,\ldots,r,
}
and therefore
\be{
\label{mchoiceidea}
m \asymp  2^{\tilde{r}} + \left ( 2^{2(q+1) \bar{r} - (2q+1) \tilde{r} } + s_{*} \log(m)  \right ) \cdot (\log(m))^6.
}
We now match terms, to get
\be{
\label{paramchoice}
2^{\tilde{r}} \asymp m,\qquad 2^{\bar{r}} \asymp m  (\log(m))^{-\frac{3}{q+1}},\qquad s_{*} \asymp m (\log(m))^{-7}.
}
Substituting this into \R{errorstepidea} now yields the desired $m^{-\alpha} (\log(m))^t$ rate. The idea of the proof is illustrated  in Figure~\ref{f:idea_ML_Fourier}.

Let us make several remarks.  First, the various bounds involving $\lesssim$ and $\gtrsim$ hide unknown constants depending on the wavelet.  To avoid having such constants in the decoder, we introduce the parameter $0 < \delta < 1$ and slightly increase the log factors in \R{paramchoice} by an amount depending on $\delta$.

Second, note that it is critical that the number of full sampling levels $\tilde{r}$ be allowed to exceed the number of saturated sparsity levels $\bar{r}$.  If $\tilde{r} = \bar{r}$ then the $\log(m)$ term in the error bound would not decrease with increasing $q$, which is the key in Theorem \ref{t:FourErr}.  As discussed in \S\ref{ss:discussion}, we are in effect heavily leveraging the ability of Fourier measurements to recover saturated wavelet scales efficiently, with a number of measurements equal to the size of the scale.

\subsection{Recipe}

\label{sec:recipe_Fourier}

\pbk We now provide the recipe.  Henceforth, we assume that the $N_k$ and $M_k$ are given by \R{Nkvalues} and \R{Mkvalues} respectively.  In particular, $N_k = M_k$, $\forall k$.  The following applies to Theorem \ref{t:FourErr}:

\pbk \textbf{Inputs:} number of measurements $m$, smoothness parameter $\alpha$, number of vanishing moments $p$, parameter $0 < \delta < 1$.

\pbk
Note that $p$ only needs to be provided as an input for \R{FourErrBd1}.  For \R{FourErrBd2} it is chosen in the recipe so that \R{qminFourErr} holds.

\pbk \textbf{Encoder parameters:} Let
\bull{
\item $q$ be the wavelet smoothness parameter, $j_0$ be as in \R{j0def};
\item $r = \lfloor \max\{ 2 \alpha + 1 , \frac{\alpha}{\alpha-1/2} \} \log_2(m) \rfloor -j_0$;
\item $\tilde{r} = \lfloor \log_2(m/2) \rfloor -j_0$;
\item $m_k = N_k - N_{k-1}$, $k=1,\ldots,\tilde{r}$, and
\bes{
m_k = \left\lfloor \frac14 \left ( m^{2q+2} 2^{-(2q+1)(k+j_0+2) } + \frac{m}{4(r-\tilde{r})} \right )  \right \rfloor,\qquad k = \tilde{r}+1,\ldots,r.
}
}

\pbk
Observe that $1 \leq \bar{r} \leq \tilde{r} \leq r$ for all $m \geq c_{\alpha,p}$, and that
\bes{
\sum^{r}_{k=1} m_k \leq 2^{\tilde{r}} + \frac14 m^{2q+2} 2^{-(2q+1)(\tilde{r}+j_0+2)} + \frac{m}{4} \leq m,
}
since $m/2^{j_0+2} \leq 2^{\tilde{r}} \leq m/2$. Hence these are valid parameters for the encoder.

\pbk \textbf{Decoder parameters:} Let
\bull{
\item $\{ \varphi^{\mathrm{per}}_{j_0,n} \} \cup \{\psi^{\mathrm{per}}_{j,n} \}$, the periodized Daubechies' wavelet basis with $p$ vanishing moments;
\item $\bar{L} = (\log(m))^{6+\delta}$;
\item $\bar{r} = \lfloor \log_2(m/\bar{L}^{\frac{1}{2(q+1)}}) \rfloor - j_0$;
\item $w = (w_i)^{M}_{i=1}$, where
\bes{
w_i = \sqrt{\frac{m}{2^k \bar{L}^{\frac{1}{2(q+1)}}}},\quad M_{k-1} < i \leq M_k,\quad k=1,\ldots,\bar{r},
}
\bes{
w_i = \sqrt{\bar{L}^{\frac{2q+1}{2q+2}} r},\quad M_{k-1} < i \leq M_k,\quad k=\bar{r}+1,\ldots,r,
}
and
\bes{
\lambda  = 1/\sqrt{r m};
}
\item $\Omega$ be the corresponding $(\mathbf{N},\mathbf{m})$-multilevel random subsampling pattern with saturation $\tilde{r}$ (see Definition~\ref{def:sampling_pattern});
\item $A \in \bbC^{m \times M}$ given by $A = P_{\Omega} D U P_M$, where $U$ is the Fourier-wavelets matrix \R{Udef} and $D$ is as in \R{Ddef}.
}

\pbk \textbf{Encoder:} Let 
$$
\cE_{m}(f) = \left(\hat{f}(\omega)\right)_{\omega \in \Omega} \in \mathbb{C}^{m_1 + \cdots + m_r}.
$$

\pbk \textbf{Decoder:} Given measurements $y = \cE_m(f)$, let $\tilde{f}_m = \sum^{M}_{i=1} \tilde{d}_i \phi_i$, where $\{\phi_i\}^{\infty}_{i=1}$ is the wavelet basis and $\tilde{d} = (d_i)^{M}_{i=1}$ is any minimizer of
\bes{
\min_{z \in \bbC^M} \lambda \nm{z}_{\ell^1_w} +\nm{A z - y}_{\ell^2}.
}

\pbk \textbf{The case of Theorem \ref{t:FourErr2}:} As in the previous section, we now assume $N$ is an input rather than a parameter, and change the definition of $r$ to $r = \lfloor \log_2(N) \rfloor - j_0$.  We now set $M= 2^{j_0+r}$ and note that
\be{
\label{NMtwosided}
N/2 \leq M \leq N,
}
so that the encoder-decoder pair includes all the wavelets up to the scale $j_0+r-1$.  The encoder and decoder are then defined in exactly the same way.

As before, we will prove Theorem \ref{t:FourErr2} first, and then obtain Theorem \ref{t:FourErr} as a corollary.

\subsection{Lower bounds on $m$}
We first need several lower bounds on $m$.  Note that we assume the condition $m \geq c_{p,\alpha,\delta} \cN(f)^2$.  In particular, this implies that
\bes{
m \geq c_{p,\alpha,\delta},
}
and also that
\bes{
m \geq c_{p,\alpha,\delta} \cN(f) (\log(m))^{7+\delta}. 
}
The first inequality is immediate.  The second second follows after noting that $(\log(m))^{7+\delta} \leq (\log(m))^8 \leq \sqrt{m}$ for $m \geq c$.  We will use these inequalities repeatedly in what follows.

\subsection{Technical lemmas}


We first require the following three lemmas.  Proofs are given in Appendix \ref{a:techproofs}.

\lem{
\label{l:balancing}
If $N = M = N_r$ then $U$ has the balancing property with constant $\theta$ satisfying
\bes{
\theta \geq \inf_{|\omega| \leq \pi} | \hat{\varphi}(\omega) |^2 = c_p > 0.
}
}

In particular, the balancing property holds with $\theta = c_p$ depending only on the number of vanishing moments $p$.

\lem{
\label{l:localcoh}
The $(k,l)^{\rth}$ local coherence of $U$ satisfies
\bes{
\mu\left ( U^{(k,l)} \right ) \leq c_p \left \{ \begin{array}{ll} 2^{-(2q+1)(k-l)} & k \geq l \\ 2^{-(2p+1)(l-k)} & k < l \end{array} \right . .
}
}

\lem{
\label{l:Atailbound}
Let $d \in \ell^2(\bbN)$, $e = P_{\Omega} D U P^{\perp}_M d$, where $D$ is as in \R{Ddef} and $M = N$.  Then
\bes{
\nmu{e}_{\ell^2} \leq C_p \nmu{P^{\perp}_M d}_{\ell^1}.
}
}

The remainder of the proof is based on Theorem \ref{t:infrecovmain1}.  We first verify that conditions (A)--(D) hold for the various choices of parameters.  Then we estimate the recovery error in \R{inferr2}.

\subsection{Step 1.\ Verification of conditions (A)--(D)}

\lem{
\label{l:sparsitydef}
Define local sparsities $\mb{s} = (s_1,\ldots,s_r)$ by
\bes{
s_{k} = M_k - M_{k-1},\quad k = 1,\ldots,\bar{r},
}
and
\bes{
s_k = s_{*} = \left \lfloor \frac{m}{\bar{L} r} \right \rfloor,\quad k = \bar{r}+1,\ldots,r.
}
Then $1 \leq s_k \leq M_k - M_{k-1}$ for all $m  \geq c_{p,\alpha,\delta} (\log(m))^{7+\delta}$, and conditions (A)--(D) of Theorem \ref{t:infrecovmain1} hold for this choice of $\mb{s}$ with constants $c_1 = c_{1,p}$ and $c_2 = c_{2,p}$ depending on $p$ only.
}
\prf{

First, recalling the definition of $\bar{r}$, observe that $s_{k} \leq M_k - M_{k-1}$ provided
\bes{
\frac{m}{\bar{L} r} \leq M_{\bar{r} +1}-M_{\bar{r}}
= 2^{j_0 + \bar{r}}  \leq \frac{m}{\bar{L}^{\frac{1}{2(q+1)}} } 
}
or equivalently $\bar{L}^{\frac{2q+1}{2q+2}} r \geq 1$.   This holds whenever $\sqrt{\bar{L}} r \geq 1$, which, since $r \geq c_{\alpha} \log(m) - c_{p}$, is implied by $m \geq c_{p,\alpha}$.  Conversely, $s_{*} \geq 1$ provided $m \geq \bar{L} r$, which, since $r \leq c_{\alpha} \log(m)$ is implied by the condition $m \geq c_{\alpha} \log(m)^{7+\delta}$.

Lemma \ref{l:balancing} implies that (A) of Theorem \ref{t:infrecovmain1} holds with $\theta \geq c_p$.  
Now consider (B).  First, observe that 
\bes{
s =2^{j_0+\bar{r}} + (r-\bar{r}) s_{*} 
\leq m \bar{L}^{-\frac{1}{2(q+1)}} + m \bar{L}^{-1} 
\leq 2 m \bar{L}^{-\frac{1}{2(q+1)}},
}
and conversely $s \geq \frac12 m \bar{L}^{-\frac{1}{2(q+1)}}$, which gives
\be{
\label{stwosided}
\frac12 m \bar{L}^{-\frac{1}{2(q+1)}} 
\leq s 
\leq 2 m \bar{L}^{-\frac{1}{2(q+1)}}.
}
We also have
\bes{
\frac12 m \bar{L}^{-1} r^{-1} 
\leq s_* 
\leq  m \bar{L}^{-1} r^{-1},
}
for all $m \geq \bar{L} r$ (that holds thanks to the assumption on $m$), and therefore
\bes{
\frac12 \bar{L}^{\frac{2q+1}{2q+2}} r 
\leq \frac{s}{s_k} 
\leq 4 \bar{L}^{\frac{2q+1}{2q+2}} r,\qquad k = \bar{r}+1,\ldots,r.
}
Hence $\frac{1}{\sqrt{2}} w^{(k)} \leq \sqrt{s/s_k} \leq 2 w^{(k)}$, which gives (B) for $k=\bar{r}+1,\ldots,r$.  For $k = 1,\ldots,\bar{r}$, we note that 
\bes{
2^k \leq s_k \leq c_p 2^{k},
}
for some $c_p$, and therefore
\bes{
c_p \frac{m}{2^k \bar{L}^{\frac{1}{2(q+1)}}} \leq \frac{s}{s_k} \leq \frac{m}{2^k \bar{L}^{\frac{1}{2(q+1)}}},\qquad k = 1,\ldots,\bar{r}.
}
Hence $c_{1,p} w^{(k)} \leq \sqrt{s/s_k} \leq c_{2,p} w^{(k)}$ in this case as well.

In a similar manner we also note that (D) follows immediately from the definition of $\lambda$ and the fact that $m \geq c_p s$ for all $m \geq c_p$.

Finally, consider condition (C).  By definition, the first $\tilde{r}$ levels are fully saturated.  Note that $\tilde{r} \geq \bar{r}$ for all $m \geq c_p$.  Now consider the unsaturated levels $\tilde{r} < k \leq r$.  Then, by definition of the $s_k$ and Lemma \ref{l:localcoh}, we have
\eas{
 \sum^{r}_{l=1} s_l \mu \left (U^{(k,l)} \right ) &\leq c_p \left ( \sum^{\bar{r}}_{l=1} 2^l 2^{-(2q+1)(k-l)} + \sum^{k}_{l=\bar{r}+1} 2^{-(2q+1)(k-l)} s_* + \sum^{r}_{l = k+1} 2^{-(2p+1)(l-k)} s_* \right )
 \\
 & \leq c_p \left ( 2^{(2q+2) \bar{r}} 2^{-(2q+1) k} + s_{*} \right )
 \\
 & = c_p \left ( 2^{(2q+1)(\bar{r}-k)} 2^{\bar{r}} + s_{*} \right ).
} 
Now consider the log factor $L$ in Theorem \ref{t:infrecovmain1}. Since $r \leq s \leq m$, $r \leq c_{p} \log(m)$,  and $\log(s) \leq \log(m)$ we have
\eas{
L & =  r^2 \cdot \log(m) \cdot \log^2(c_p r s / \theta) \cdot \log(M) + r \cdot \log(\varepsilon^{-1})\\
 & \leq c_{p} \left ( \log^6(m) + \log(m) \cdot \log(\varepsilon^{-1}) \right ) \\
 & \leq c_p (\log(m))^6,
}
where we have also used that $\log(c_p s^2/\theta) \leq c_p \log(s) \leq c_p \log(m)$, thanks to (A), and the condition $(\log(m))^5 \geq \log(\varepsilon^{-1})$. Hence, writing $c'>0$ for the universal constant understood in condition (C), we deduce that
\eas{
c' \cdot \left(\frac{c_{2,p}}{c_{1,p}}\right)^2 \cdot \theta^{-2} &\cdot \left ( \sum^{r}_{l=1} s_l \mu \left (U^{(k,l)} \right ) \right ) \cdot L 
\\
&\leq c_p \left ( 2^{(2q+1)(\bar{r}-k)} 2^{\bar{r}} + s_{*} \right ) \left( \log^6(m) + \log(m) \log(\varepsilon^{-1}) \right )
\\
& \leq c_p \left ( m^{2q+2} 2^{-(2q+1) k} + m / r \right )  \frac{ \log^6(m) + \log(m) \log(\varepsilon^{-1}) }{\bar{L}}
\\
& \leq m_k,
}
which is ensured for $m \geq c_{p,\delta}$.  Hence (C) holds.  }

\subsection{Step 2.\ Estimation of the approximation error}

Having verified conditions (A)--(D) of Theorem \ref{t:infrecovmain1},  and using Lemma~\ref{l:Atailbound}, we now get
\ea{
\nmu{f - \tilde{f}}_{L^2} &\leq \nmu{P^{\perp}_M d}_{\ell^2} + \nmu{P_M d - \tilde{d}}_{\ell^2}  \nn
\\
& \leq \nmu{P^{\perp}_M d}_{\ell^2} + C_p \frac{\sigma_{\mb{s},\mb{M}}(P_M d)_{\ell^1_w}}{r^{1/4} \sqrt{s}} 
+ C_p\frac{1}{r^{1/4} \sqrt{s} \lambda} \nmu{P^{\perp}_M d}_{\ell^1} \nn
\\
& = E_1 + E_2 + E_3. \label{gettingthere}
}
Note that $\nmu{P^{\perp}_M d}_{\ell^2} = e_{M}(f)_{L^2}$, where $e_{M}(f)_{L^2}$ is as in \R{linnonlin}.  Hence Theorem \ref{t:waveletsterm} gives
\be{
\label{taill2wave}
E_1 
\leq C_{p,\alpha} \sqrt{\cN(f)} \nm{f}_{PC^{\alpha}} / \sqrt{M}.
}
Arguing similarly as in the proof of Theorem \ref{t:waveletsterm}, we also see that
\be{
\label{taill1wave}
\nmu{P^{\perp}_M d}_{\ell^1} 
\leq C_{p,\alpha} \nm{f}_{PC^{\alpha}} \left ( M^{-(\alpha-1/2)} + \cN(f) / \sqrt{M} \right ),
}
where we use the fact that $\alpha > 1/2$.  Hence, using the values of $\lambda$, $s$ and $r$, and the bound \R{stwosided}, we deduce that the final term of \R{gettingthere} satisfies 
\ea{
\label{E3bd}
E_3 
\leq C_p \frac{\sqrt{r m}}{r^{1/4} \sqrt{m}} \bar{L}^{\frac{1}{4(q+1)}} \nmu{P^{\perp}_M d}_{\ell^1} 
\leq  C_{p,\alpha} (\log(m))^{\frac14 + \frac{6+\delta}{4(q+1)}} \nm{f}_{PC^{\alpha}} \left ( \frac{1}{M^{\alpha-1/2}} + \frac{\cN(f)}{\sqrt{M}} \right ).
}
Next, consider $\sigma_{\mb{s},\mb{M}}(P_M x)_{\ell^1_w}$.  We first recall that $s_k = M_k - M_{k-1}$ for $k = 1,\ldots,\bar{r}$.  Moreover, $s_k \geq (2p)^2 \cN(f)$ for all $k >\bar{r}$ (recall that $(2p)^2\cN(f)$ is the maximum number of the wavelets at any fixed scale $k$ whose support contains a discontinuity of $f^{\mathrm{ext}}$; see the proof of Theorem~\ref{t:waveletsterm})  by construction and thanks to the assumption $m /(\log(m))^{7+\delta} \geq c_{p} \cN(f)$, which implies that $m \geq c_p \cN(f) \bar{L} r$. Therefore, the coefficients corresponding to the discontinuities of $f^{\mathrm{ext}}$ are excluded from the approximation error.  Applying Lemma \ref{l:PLipalpha_coeff} and the definition of the weights $w^{(k)}$, we deduce that 
\be{
\label{omghard}
\frac{\sigma_{\mb{s},\mb{M}}(P_M d)_{\ell^1_w}}{r^{1/4}\sqrt{s}} 
\leq C_{p,\alpha} \nm{f}_{PC^{\alpha}} \sum^{r}_{j = \bar{r}+1} \frac{1}{r^{1/4}\sqrt{s_*}} 2^{-(\alpha-1/2) j} .
}
Recalling that $\alpha > 1/2$, we have
\eas{
 \sum^{r}_{j 
 = \bar{r}+1} \frac{1}{r^{1/4}\sqrt{s_*}} 2^{-(\alpha-1/2) j} 
\leq C_\alpha \frac{2^{-(\alpha-1/2)(\bar{r}+1)}}{r^{1/4}\sqrt{s_*}}.
}
Using the definitions of $\bar{r}$, $r$ and $s_{*}$ we have
\eas{
\frac{2^{-(\alpha-1/2)(\bar{r}+1)}}{r^{1/4}\sqrt{s_*}} & \leq C_{p,\alpha} \frac{m^{-(\alpha-1/2)} \bar{L}^{\frac{\alpha-1/2}{2(q+1)}}}{(\log(m))^{-1/4} \sqrt{m} / \sqrt{\bar{L}}} 
\\
& = C_{p,\alpha} m^{-\alpha} (\log(m))^{1/4} \bar{L}^{\frac{q+\alpha+1/2}{2(q+1)}}
 \leq  C_{p,\alpha} m^{-\alpha} (\log(m))^{\frac{1}{4} + \frac{(6+\delta)(q+\alpha+1/2)}{2(q+1)}}.
}
Therefore, combining this with the previous estimate, we deduce that
\be{
\label{nearlydone}
E_2 
\leq C_{p,\alpha} \nm{f}_{PC^{\alpha}}  m^{-\alpha} (\log(m))^{\frac{1}{4} + \frac{(6+\delta)(q+\alpha+1/2)}{2(q+1)}} .
}

\subsection{Step 3.\ Concluding the proof}
Substituting \R{nearlydone}, \R{taill2wave} and \R{E3bd} into \R{gettingthere}, we finally deduce that
\eas{
\nmu{f - \tilde{f}}_{L^2} \leq C_{p,\alpha} \nm{f}_{PC^{\alpha}} \Bigg [ & \left ( m^{-\alpha}   + M^{-(\alpha-1/2)} + \cN(f) / \sqrt{M} \right ) (\log(m))^{\frac{1}{4} + \frac{(6+\delta)(q+\alpha+1/2)}{2(q+1)}} 
\\
& + \sqrt{\cN(f)} / \sqrt{M} \Bigg ].
}
To obtain \R{FourErrBd1alt}, we merely use \R{NMtwosided} and the fact that $\cN(f) \geq 1$.  For \R{FourErrBd2alt}, we note that if
\bes{
q \geq 6 \frac{\alpha-1/2}{\delta} + \alpha - \frac{3}{2}
}
Then 
\bes{
\frac14 +\frac{(6+\delta)(q+\alpha+1/2)}{2(q+1)} \leq \frac{13}{4}+\delta.
}

\subsection{Proof of Theorem \ref{t:FourErr}}
We let $M = 2^{j_0+r}$, where $r = \lfloor \max\{ 2 \alpha + 1 , \frac{\alpha}{\alpha-1/2} \} \log_2(m) \rfloor -j_0$.  This gives 
\bes{
M  \geq \frac12 \max \left \{ m^{2 \alpha+1} , m^{\frac{\alpha}{\alpha-1/2}} \right \}.  
}
This, and the condition $m \geq \cN(f)^2$, give 
\bes{
\nmu{f - \tilde{f}}_{L^2} \leq C_{p,\alpha}  (\log(m))^{\frac{1}{4} + \frac{(6+\delta)(q+\alpha+1/2)}{2(q+1)}}  m^{-\alpha},
}
as required.  For \R{FourErrBd2} we argue as above.

\section{Conclusions and challenges}
\label{s:conclusions}

We conclude this paper by listing a number of open problems.

\pbk
\textit{1.\ Reducing the log factor.}\ The log factor in Theorem \ref{t:FourErr} has the potential to be decreased.  There are several ways to do this.  First, reducing the log factor $L$ in Theorem \ref{t:infrecovmain1} (or, more specifically, in Theorem \ref{t:uniform_sparsity_guarantee}).  This is related to, although more general than, the question in compressed sensing of when a subsampled Fourier matrix has the RIP of order $s$.  Theorem \ref{t:uniform_sparsity_guarantee} (specialized to $r = 1$ level) implies this whenever $m \gtrsim s \cdot \log(m) \cdot \log^2(s) \cdot  \log(N)$.  However, it is known  \cite{ChkifaDownwardsCS} that this can be achieved under the weaker condition
\be{
\label{websterbd}
m \gtrsim s \cdot \log^2(s) \cdot \log(N).
}
If such arguments could be generalized to sparsity in levels, we could save one $\log(m)$ factor in Theorem \ref{t:uniform_sparsity_guarantee}.  This would decrease the $\log(m)$ exponent in Theorem \ref{t:FourErr} by $1/2$.  Another way to reduce this exponent would be to remove the dependence on $r$ in Theorem \ref{t:infrecovmain1} (recall that $r \approx \log(m)$ in the setting of Theorem \ref{t:FourErr}).  Further improvements, however, would seemingly necessitate improving \R{websterbd}, which is a challenging open problem in compressed sensing theory.

\pbk
\textit{2.\ Standard decoders.}\ Since they are most commonly used in practice, it is desirable to have guarantees for the standard (unweighted) LASSO and QCBP decoders.  Unfortunately, it is not clear how to avoid using a weighted $\ell^1$-norm without the recovery guarantee being ruined by the interferences (see \S\ref{ss:weightedSRLASSO}).  For some initial work in this direction, see \cite{ABBLocalGap}.

\pbk
\textit{3.\ Higher dimensions and other function classes.}  We have chosen to study the class $PC^{\alpha}$ to avoid additional technical challenges.  An interesting problem is to extend this work to Besov spaces.
Another open problem, motivated by the application to compressive imaging (recall Figure~\ref{f:FourierGauss}), is the extension to higher dimensions.  While a direct extension using wavelets may be not be too challenging for suitable analogues of the class $PC^{\alpha}$, it is well known that wavelet are generally not optimal in higher dimensions.  More interesting challenges involve extending this work to, for instance, shearlets.  We note that sparse recovery guarantees of a similar flavour to \R{KWFHRIP} for Fourier sampling with shearlets have been shown in \cite{KutyniokLimShearletFourier}.  However, to extend our results, we require local recovery guarantees.  See \cite{PoonFrames} for some work in this direction. 

A further direction is to adapt this work to Total Variation (TV) minimization, or its various higher-order generalizations.  For existing recovery guarantees for Fourier sampling with TV, see \cite{PoonTV}.

\pbk
\textit{4.\ Binary measurements.}\ While many imaging modalities employing Fourier sampling, others (in particular, optical imaging) are constrained to acquire binary measurements.  In practice, one can design structure-exploiting binary measurements by replacing the Fourier transform with the Walsh (also known as Hadamard) transform \cite{OptimalSamplingQuest,AAHWalshWavelet}.  As with Fourier sampling, this performs significantly better than random Bernoulli sampling (the binary analogue of Gaussian sampling) \cite{AsymptoticCS}.  Unfortunately, for Walsh sampling with wavelets, the corresponding local coherences $\mu\left ( U^{(k,l)} \right ) \leq c_p 2^{-|k-l|}$ decay at a rate independent of the wavelet order \cite{AAHWalshWavelet}, unlike in Fourier sampling (Lemma \ref{l:localcoh}).  In our proof of Theorem \ref{t:FourErr} fast decay of the coherences is critical.  Hence it remains an open problem to extend our analysis to binary sampling.

\pbk
\textit{5.\ Optimal algorithms.}\ Finally, we remark that our decoders $\cD_m$ are not algorithms per se, since they involve the exact solution of certain convex optimization problems.  An open problem is to design an algorithm that takes inputs $m$, $\alpha$ and the measurements $\cE_m(f)$, and then computes the approximation $\tilde{f}_m$ in polynomial time in $m$.

\section*{Acknowledgements}

The authors extend their thanks to Vegard Antun (University of Oslo), who performed the experiment in  Fig.~\ref{f:FourierGauss}.  They also would like to thank Anders C.\ Hansen, Bradley J.\ Lucier and Clarice Poon. S.B.\ acknowledges the support of the PIMS Postdoctoral Training Centre in Stochastics, the Department of Mathematics of Simon Fraser University, NSERC, the Faculty of Arts and Science of Concordia University, and the CRM Applied Math Lab. This work was supported by the PIMS CRG in ``High-dimensional Data Analysis'' and by NSERC through grant R611675.  

\appendix

\section{Fourier transform and series}\label{s:Fourier}

Given $f \in L^1(\bbR) \cap L^2(\bbR)$ we define the Fourier transform as
\bes{
\hat{f}(\omega) = \int^{\infty}_{-\infty}f(t) \E^{-\I \omega t} \D t.
}
If $f \in L^2([0,1])$ then we can write $f$ as its Fourier series
\bes{
f = \sum_{n \in \bbZ} \ip{f}{\gamma_n}_{L^2} \gamma_n,
}
where
\be{
\label{FourBasis}
\gamma_n(t) = \E^{2 \pi \I n t},\quad n \in \bbZ,
}
is the Fourier basis for $L^2([0,1])$.  If we consider $f$ as a function in $L^2(\bbR)$ that is zero outside $[0,1]$, then $\ip{f}{\gamma_n}_{L^2} = \hat{f}(2 \pi n)$.
For convenience, we also re-index this basis over $\bbN$ as follows:
\be{
\label{FourBasisreindex}
\gamma_{2n-1} = \E^{-2 \pi \I (n-1) t},\qquad \gamma_{2n} = \E^{2 \pi \I n t},\quad n \in \bbN.
}

\section{Orthogonal wavelet bases of $L^2([0,1])$}\label{s:wavelets}

Let $\varphi$ and $\psi$ be the scaling function and mother wavelet, respectively, of the Daubechies' wavelet with $p \geq 1$ vanishing moments.  Write 
\bes{
\varphi_{j,k}(x) = 2^{j/2} \varphi(2^j x - k),\ \psi_{j,k}(x) = 2^{j/2} \psi(2^j x - k),\quad j , k \in \bbZ.
}
Since we work with functions on the interval $[0,1]$, we need an orthonormal wavelet basis of $L^2([0,1])$.  We construct this via periodization (see \eqref{def_periodization} and \cite[Sec.\ 7.5.1]{mallat09wavelet} for more details).  Define the coarsest scale
\be{
\label{j0def}
j_0 = \left \{ \begin{array}{ll} 0 & p =1 \\ \lceil \log_2(2p) \rceil & p \geq 2 \end{array} \right . ,
}
(in general, one could allow any fixed $j_0$ greater than or equal to the right-hand side.  However, this does not affect any of the results in the paper, hence we simply specify $j_0$ exactly). We recall that Daubechies' wavelets with $p$ vanishing moments have the smallest possible support, of length $2p-1$. We assume the scaling function $\varphi$ and the mother wavelet $\psi$ to be supported on $[0, 2p-1]$ and $[-p+1,p]$, respectively.  
Then the set of functions
\be{
\label{wavebasis}
\{ \varphi^{\mathrm{per}}_{j_0,k} : k = 0,\ldots,2^{j_0}-1 \} \cup \{ \psi^{\mathrm{per}}_{j,k} : k = 0,\ldots,2^{j-1}, \ j \geq j_0 \},
}
is an orthonormal basis of $L^2([0,1])$, referred to as the \textit{periodized Daubechies wavelet basis}.  We note in passing that
\bes{
\psi^{\mathrm{per}}_{j,k} = \psi_{j,k},\quad \varphi^{\mathrm{per}}_{j,k} = \varphi_{j,k},\quad k = p-1,\ldots,2^j-p,
}
that is, wavelets that are fully supported in $[0,1]$ are unchanged, and
\eas{
\varphi^{\mathrm{per}}_{j,k} &= \varphi_{j,k} + \varphi_{j,2^j+k},\quad \psi^{\mathrm{per}}_{j,k} = \psi_{j,k} + \psi_{j,2^j+k},\qquad k = 0,\ldots,p-2,
\\
\varphi^{\mathrm{per}}_{j,k} &=  \varphi_{j,k} + \varphi_{j,2^j-p-k},\quad \psi^{\mathrm{per}}_{j,k} =  \psi_{j,k} + \psi_{j,2^j-p-k},\qquad k = 2^j-p+1,\ldots,2^j-1,
}
where the functions in the right-hand sides are implicitly restricted to $[0,1]$. As needed, we order the basis \R{wavebasis} in the usual way, rewriting it as $\{ \phi_n \}_{n \in \bbN}$, where
\be{
\label{DBbasis}
\begin{split}
\phi_{n+1} &= \varphi^{\mathrm{per}}_{j_0,n},
\quad n = 0,\ldots,2^{j_0}-1
\\
\phi_{2^{j} + n + 1} &= \psi^{\mathrm{per}}_{j,n},
\quad n = 0,\ldots,2^j-1,\ j \geq j_0.
\end{split}
}

\section{Proof of Theorem \ref{t:infrecovmain1}}
\label{a:recovproof}

The technical tools we need to prove this theorem were introduced in \cite{AAHWalshWavelet}, where a similar result was proven for the weighted quadratically-constrained basis pursuit decoder.  

We require several concepts from \cite{AAHWalshWavelet}.  First, we introduce several additional pieces of notation.  Given sparsity levels $\mb{M} = (M_1,\ldots,M_r)$ and local sparsities $\mb{s} = (s_1,\ldots,s_r)$, let
\bes{
D_{\mb{s},\mb{M}} = \left \{ \Delta \subseteq \{1,\ldots,N\} : | \Delta \cap \{ M_{k-1}+1,\ldots,M_k \} | \leq s_k \right \},
}
be the set of all possible supports of an $(\mb{s},\mb{M})$-sparse vector.  Given positive weights $w = (w_i)^{M}_{i=1} \in \bbC^{M}$ and a set $\Delta \subseteq \{1,\ldots,M\}$, we define its weighted cardinality as follows:
\bes{
|\Delta|_w = \sum_{i \in \Delta} (w_i)^2.
}
The conventional tool in compressed sensing for establishing recovery guarantees is the so-called Restricted Isometry Property (RIP).  In our case, we require an generalized version of the RIP.  This takes into account the sparsity in levels structure, and the fact that the measurement matrix $A$ satisfies \R{AstarAexp}, rather than the more standard condition $\bbE(A^*A) = I$.

\defn{[G-adjusted RIP in Levels]
Let $\mb{M} = (M_1,\ldots,M_r)$ be sparsity levels, $\mb{s} = (s_1,\ldots,s_r)$ be local sparsities and $G \in \bbC^{M \times M}$ be invertible, where $M = M_r$ is the sparsity bandwidth.  The $(\mb{s},\mb{M})^{\rth}$ \textit{$G$-adjusted Restricted Isometry Constant in Levels (G-RICL)} $\delta_{\mb{s},\mb{M},G}$ of a matrix $A \in \bbC^{m \times M}$ is the smallest $\delta \geq 0$ such that
\bes{
(1-\delta) \nm{G x}^2_{\ell^2} \leq \nm{A x}^2_{\ell^2} \leq (1+\delta) \nm{G x}^2_{\ell^2},\quad \forall x \in \Sigma_{\mb{s},\mb{M}}.
}
If $0 < \delta_{\mb{s},\mb{M},G} < 1$ then the matrix is said to have the \textit{$G$-adjusted Restricted Isometry Property in levels (G-RIPL)} of order $(\mb{s},\mb{M})$.
}

In our setting, if $N$, $M$ are such that $P_N UP_M$ is full rank (in particular, if the balancing property holds), then $G$ will be taken as the unique positive definite square-root of the positive definite matrix $P_M U^* P_N U P_M$.  We write $G = \sqrt{P_M U^* P_N U P_M}$ in this case.

The following result \cite[Thm.\ 3.6]{AAHWalshWavelet} gives conditions under which the matrix $A$ satisfies the G-RIPL:

\thm{
\label{t:uniform_sparsity_guarantee}
Let $0 < \delta,\varepsilon <1$, $M \geq 2$, $1 \leq \tilde{r} \leq r \leq N$ and $\mb{M} = (M_1,\ldots,M_r)$ and $\mb{s} = (s_1,\ldots,s_r)$ be sparsity levels and local sparsities respectively, where $s = s_1+\ldots+s_r \geq 2$ and $M_r = M$.  Let $\Omega$ be an $(\mb{N},\mb{m})$-multilevel random subsampling pattern with $r$ levels and saturation $\tilde{r}$, and $N = N_r$.  Suppose that $N$, $M$ are such that $P_N U P_M$ is full rank, where $U$ is as in \R{Udef} and consider the matrix $A$ given by \R{Adef}.  If
\bes{
m_k \gtrsim \delta^{-2} \cdot \nm{G^{-1}}^2_{\ell^2} \cdot \left ( \sum^{r}_{k=1} s_k \mu \left( U^{(k,l)} \right ) \right ) \cdot L,\qquad k = \tilde{r}+1,\ldots,r,
}
where
\bes{
L =  r \cdot \log(m)\cdot \log^2(s) \cdot \log(N) + \log(\varepsilon^{-1}),
} 
then, with probability at least $1-\varepsilon$, $A$ satisfies the G-RIPL of order $(\mb{s},\mb{M})$ with constant $\delta_{\mb{s},\mb{M},G} \leq \delta$ and $G$ given by $G = \sqrt{P_M U^* P_N U P_M}$.
}

In order to establish Theorem \ref{t:infrecovmain1}, we next show that the G-RIPL implies stable and robust recovery.  To do so, we first introduce the following generalization of the so-called robust Null Space Property (rNSP):

\defn{
Let $\mb{M} = (M_1,\ldots,M_r)$ be sparsity levels, $\mb{s} = (s_1,\ldots,s_r)$ be local sparsities and $w \in \bbC^{M}$ be positive weights, where $M = M_r$.  A matrix $A \in \bbC^{m \times M}$ has the \textit{weighted robust null space property in levels (weighted rNSPL)} of order $(\mb{s},\mb{M})$ with constants $0 < \rho < 1$ and $\gamma > 0$ if
\bes{
\nm{P_{\Delta} x}_{\ell^2} \leq \frac{\rho \nm{P^{\perp}_{\Delta} x}_{\ell^1_w}}{\sqrt{|\Delta|_w}} + \gamma \nm{A x}_{\ell^2},
}
for all $x \in \bbC^M$ and $\Delta \in D_{\mb{s},\mb{M}}$.
}

Suppose the weights $w = (w_i)^{M}_{i=1}$ are of the form \eqref{weightsinf1}, i.e.\ constant on the sparsity levels, and define
\begin{equation}
\label{def_xi_zeta}
\xi = \xi(\mb{s},\mb{w}) = \sum^{r}_{k=1} (w^{(k)})^2 s_k,\qquad
\zeta = \zeta(\mb{s},\mb{w}) = \min_{k=1,\ldots,r} \left \{ (w^{(k)})^2 s_k \right \}.
\end{equation}
The following combines Lemmas 5.2 and 5.3 of \cite{AAHWalshWavelet}:

\lem{
\label{l:rNSPLerrbd}
Suppose that $A$ has the weighted rNSPL of order $(\mb{s},\mb{M})$ with constants $0 < \rho < 1$ and $\gamma > 0$.  Let $x,z \in \bbC^M$.  Then
\bes{
\nm{z - x}_{\ell^1_w} \leq \frac{1+\rho}{1-\rho} \left ( 2 \sigma_{\mb{s},\mb{M}}(x)_{\ell^1_w} + \nm{z}_{\ell^1_w} - \nm{x}_{\ell^1_w} \right ) + \frac{2 \gamma}{1-\rho} \sqrt{\xi} \nm{A (z-x)}_{\ell^2}, 
}
and
\bes{
\nm{z - x}_{\ell^2} \leq \left ( \rho + (1+\rho) (\xi / \zeta )^{1/4} / 2 \right ) \frac{\nm{z-x}_{\ell^1_w}}{\sqrt{\xi}} + \left ( 1 + (\xi / \zeta )^{1/4} / 2 \right ) \gamma \nm{A(z-x)}_{\ell^2}.
}
}


The G-RIPL implies the weighted rNSPL (see \cite[Thm.\ 5.5]{AAHWalshWavelet}):  
\thm{
\label{t:GRIPLimpliesrNSPL}
Let $A \in \bbC^{m \times M}$ and $G \in \bbC^{M \times M}$ be invertible.  Let $\mb{M} = (M_1,\ldots,M_r)$ and $\mb{s} = (s_1,\ldots,s_r)$ be sparsity levels and local sparsities respectively, and $\mb{w}$ be positive weights of the form \R{weightsinf1}. Let $0 < \rho < 1$, and suppose that $A$ has the G-RIPL of order $(\mb{t},\mb{M})$ and constant $1/2$, where $\mb{t} = (t_1,\ldots,t_r)$ satisfies
\be{
\label{tldef}
t_l = \min \left \{ 2 \left \lceil 3 \frac{\kappa(G)^2}{\rho^2} \frac{\xi(\mb{s},\mb{w})}{(w^{(l)})^2} \right \rceil , M_l - M_{l-1} \right \},\quad l =1,\ldots,r,
}
and $\kappa(G)=\|G\|_{\ell^2}\|G^{-1}\|_{\ell^2}$ is the condition number of $G$ with respect to the $\ell^2$-norm. Then, there exists $0 < \gamma \leq \sqrt{2} \nm{G^{-1}}_{\ell^2}$ such that $A$ has the weighted rNSPL of order $(\mb{s},\mb{M})$ with constants $\rho$ and $\gamma$. 
}


Finally, we are now ready to prove Theorem \ref{t:infrecovmain1}:

\prf{[Proof of Theorem \ref{t:infrecovmain1}]
Recall that $G^2 = P_M U^* P_N U P_M$.  Hence $G$ is invertible since $U$ has the balancing property \eqref{balprop}, and moreover, we have
\be{
\label{Ginvbd}
\nm{G^{-1}}_{\ell^2} \leq 1/ \sqrt{\theta}.
}
We also have $\nm{G}_{\ell^2} \leq 1$ since $U$ is unitary, and therefore $\kappa(G) \leq 1/\sqrt{\theta}$.

Let $t_l$ be given by \R{tldef} with $\rho = 1/2$. Recalling \eqref{weightsinf2} and \eqref{def_xi_zeta}, observe that
\bes{
t_l \leq 48 \frac{c_2^2 r s_l}{c_1^2 \theta}.
}
Therefore
\bes{
t = t_1+\ldots+t_r \leq 48 \frac{c_2^2 r }{c_1^2 \theta} s,
}
and
\eas{
\nmu{G^{-1}}^2_{\ell^2} &\cdot \left ( \sum^{r}_{k = 1} t_l \mu \left(  U^{(k,l)} \right ) \right ) \cdot \left ( r \cdot \log(m) \cdot \log^2(t) \cdot \log(M) + \log(\varepsilon^{-1}) \right ) 
\\
& \lesssim \theta^{-2} \frac{c_2^2 r}{c_1^2} \cdot \left ( \sum^{r}_{k = 1} s_l \mu \left(  U^{(k,l)} \right ) \right ) \cdot \left ( r \cdot \log(m) \cdot \log^2(c_2^2 r s / (c_1^2 \theta )) \cdot \log(M) + \log(\varepsilon^{-1}) \right ) .
}
Hence, condition (C) and Theorem \ref{t:uniform_sparsity_guarantee} imply that the matrix $A$ has the G-RIPL of order $(\mb{t},\mb{M})$ with constant $\delta_{\mb{t},\mb{M},G} \leq 1/2$.  It now follows from Theorem \ref{t:GRIPLimpliesrNSPL} that $A$ has the weighted rNSPL of order $(\mb{s},\mb{M})$ with constants $\rho = 1/2$ and $\gamma \leq \sqrt{2} \nmu{G^{-1}}_{\ell^2} \leq \sqrt{2/\theta}$.

To complete the proof we use Lemma \ref{l:rNSPLerrbd} with $z = \hat{x}$.
Using this, \R{Ginvbd} and the bounds
\be{
\label{xizetabds}
c_1^2 r s \leq \xi \leq c_2^2 r s,\qquad c_1^2 s \leq \zeta \leq c_2^2 s.
}
we see that
\eas{
\nm{\hat{x} - x}_{\ell^2} \leq& \left ( 1/2 + 3/4 (c_2^2 r / c_1^2)^{1/4} \right ) \frac{\nm{\hat{x} - x}_{\ell^1_w}}{c_1\sqrt{r s}} + (1 + (c_2^2 r/c_1^2)^{1/4} /2) \sqrt{2/\theta} \nm{A (\hat{x} - x)}_{\ell^2}
\\
\leq & \left ( 1+ (c_2^2 r / c_1^2)^{1/4} \right ) \left [  \frac{\nm{\hat{x} - x}_{\ell^1_w}}{c_1\sqrt{r s}}  + \sqrt{2/\theta} \nm{A (\hat{x} - x)}_{\ell^2} \right ]
\\
\leq & \left ( 1+ (c_2^2 r / c_1^2)^{1/4} \right ) \left [ \frac{3}{c_1\sqrt{r s}} \left ( 2 \sigma_{\mb{s},\mb{M}}(x)_{\ell^1_w} + \nm{\hat{x}}_{\ell^1_w} - \nm{x}_{\ell^1_w} \right ) \right . 
\\
& \left . + 5 \sqrt{2/\theta} (c_2/c_1) \nm{A (\hat{x} - x) }_{\ell^2} \right ].
}
We now use the fact that $\hat{x}$ is a minimizer, and therefore
\bes{
\nm{\hat{x}}_{\ell^1_w} - \nm{x}_{\ell^1_w} \leq \frac{1}{\lambda} \left ( \nm{A x - y}_{\ell^2} - \nm{A  \hat{x} - y}_{\ell^2} \right ),
}
Writing $\nm{A (\hat{x} - x) }_{\ell^2} \leq \nm{A  \hat{x} - y}_{\ell^2} + \nm{A x - y}_{\ell^2}$ and combining with the previous inequality now yields
\eas{
\nm{\hat{x} - x}_{\ell^2} \leq& \left ( 1+ (c_2^2 r / c_1^2)^{1/4} \right ) \left [ \frac{6 \sigma_{\mb{s},\mb{M}}(x)_{\ell^1_w} }{c_1 \sqrt{ r s}} + \left  ( 5 \sqrt{2/\theta} (c_2/c_1)  +\frac{3}{c_1\sqrt{r s} \lambda} \right ) \nm{A x - y}_{\ell^2} \right .
\\
& \left .  +  \left  (5 \sqrt{2/\theta} (c_2/c_1) -  \frac{3}{c_1\sqrt{r s} \lambda} \right ) \nm{A  \hat{x} -y }_{\ell^2} \right ]
}
The result now follows from the bound (D) on $\lambda$ and the fact that $e = y - A x$.
}

\section{Proofs of Lemmas \ref{l:balancing}, \ref{l:localcoh} and \ref{l:Atailbound}}
\label{a:techproofs}

\prf{[Proof of Lemma \ref{l:balancing}]

We first observe that $\theta = \inf_{|\omega| \leq \pi} | \hat{\varphi}(\omega) |^2 > 0$ for the Daubechies wavelet basis \cite[Remark 7.1]{AHPWavelet}.  Now let $x = (x_n)^{N}_{n=1} \in \bbC^N$ with $\nm{x}_{\ell^2} = 1$ and write $g = \sum^{N}_{n=1} x_n \phi_n$ for the corresponding finite wavelet expansion.  
Observe that $ \nm{g}^2_{L^2([0,1])} = \nm{x}^2_{\ell^2} = 1$.
Let $V^{\mathrm{per}}_{j} = \spn \{ \varphi_{j,n} : n = 0,\ldots,2^j-1 \}$ and $W^{\mathrm{per}}_{j} = \spn \{ \psi_{j,n} : n = 0,\ldots,2^j-1 \}$.  Then
\bes{
g \in V^{\mathrm{per}}_{j_0} \oplus W^{\mathrm{per}}_{j_0} \oplus \cdots \oplus W^{\mathrm{per}}_{j_0+r-1} = V^{\mathrm{per}}_{j_0+r},
}
and conversely every $g \in V^{\mathrm{per}}_{j_0+r}$ with $ \nm{g}^2_{L^2([0,1])} = 1$ is equivalent to a vector of coefficients $x \in \bbC^M$ with $\nm{x}_{\ell^2} = 1$.  
Note also that
\bes{
\nm{P_N U P_N x}^2_{2} = \sum^{N}_{n=1} | \ip{g}{\gamma_n} |^2.
}
Hence
\be{
\label{driveway}
\begin{split}
\inf_{\substack{x \in \bbC^N \\ \nm{x}_{\ell^2} = 1}} & \nm{P_N U P_N x}^2_{\ell^2} 
 = \inf \left \{ 
\sum^{N}_{n=1} | \ip{g}{\gamma_n} |^2 : g \in V^{\mathrm{per}}_{j_0+r},\  \nm{g}_{L^2([0,1])} = 1 \right \}.
\end{split}
}
Fix a $g \in V^{\mathrm{per}}_{j_0+r}$ with $\nm{g}_{L^2([0,1])} = 1$ and write
\bes{
g = \sum^{N-1}_{k = 0} z_k \varphi^{\mathrm{per}}_{r+j_0,k},
}
where $ \nm{z}_{\ell^2} = \nm{g}_{L^2(0,1)} = 1$ and $z = (z_k)^{N-1}_{k=0}$.  Then, for any integer $n$,
\eas{
 \hat{g}(2 \pi n) &= N^{-1/2} \hat{\varphi} (2 \pi n / N) \sum^{N-1}_{k=0} z_k \E^{-2 \pi \I n k /N}  
= N^{-1/2} \hat{\varphi} (2 \pi n / N) G( n/N),
}
where $G(x) = \sum^{N-1}_{k=0} z_k \E^{-2 \pi \I k x}$ is a 1-periodic function. In the first equality, we have used that 
\begin{equation}
\label{Fourier_transf_phi}
\widehat{\varphi^{\mathrm{per}}_{j,k}}(\omega) 
= \widehat{\varphi_{j,k}}(\omega)
= 2^{-j/2}\hat{\varphi}(\omega/2^j) \E^{-\I\omega k/2^j}, \quad \forall j, k \in \mathbb{Z}, \; \forall \omega \in 2\pi \mathbb{Z},
\end{equation}
and that $N = 2^{j_0+r}$. 
Hence,
\be{
\label{shovel}
\sum^{N}_{n=1} | \ip{g}{\gamma_n} |^2
=\sum^{N/2}_{n=-N/2+1} | \hat{g}(2 \pi n) |^2 = N^{-1} \sum^{N/2}_{n=-N/2+1} \left | \hat{\varphi}(2 \pi n/N) \right |^2 \left | G(n / N) \right |^2.
}
Using the fact that $G$ is $1$-periodic we deduce that
\bes{
\sum^{N}_{n=1} | \ip{g}{\gamma_n} |^2  \geq \inf_{|\omega| \leq \pi} \left | \hat{\varphi}(\omega ) \right |^2 N^{-1} \sum^{N-1}_{n=0} |G(n/N)|^2.
}
Now, since $G$ is a trigonometric polynomial, it follows that
\bes{
N^{-1} \sum^{N-1}_{n=0} |G(n/N)|^2 = \nm{G}^2_{L^2([0,1])} = \nm{z}^2_{\ell^2} = \nm{g}^2_{L^2([0,1])} = 1.
}
Therefore
\bes{
\sum^{N}_{n=1} | \ip{g}{\gamma_n} |^2  \geq \inf_{|\omega| \leq \pi} \left | \hat{\varphi}(\omega ) \right |^2 = \theta > 0.
}
Since $g$ was arbitrary, we deduce that
\bes{
\inf_{\substack{x \in \bbC^N \\ \nm{x}_{\ell^2} = 1}} \nm{P_N U P_N x}^2_{\ell^2} \geq \theta.
}
To complete the proof, we first recall that $P_N - P_N U^* P_N U P_N$ is positive semidefinite (since $U$ is unitary), and therefore
\eas{
\nm{P_N - P_N U^* P_N U P_N}_{\ell^2} &= \sup_{\substack{x \in \bbC^N \\ \nm{x}_{\ell^2} = 1}} \ip{(P_N - P_N U^* P_N U P_N )x}{x}
\\
& = 1 - \inf_{\substack{x \in \bbC^N \\ \nm{x}_{\ell^2} = 1}} \nm{P_N U P_N x}^2_{\ell^2}
\\
& \leq 1 - \theta,
}
as required.
}

For Lemma \ref{l:localcoh}, we first require the following:

\lem{
\label{l:localcoh1Dgeneralstep}
The $(k,l)^{\rth}$ local coherence satisfies
\bes{
\mu\left ( U^{(k,l)} \right ) \leq 2^{1+ k-l} \max_{\omega \in B_k } \left | \widehat{\psi}(2\pi \omega/2^{l+j_0-1}) \right |^2,\quad l > 1,
}
and 
\bes{
\mu\left ( U^{(k,1)} \right ) \leq 2^{k}  \max \left \{ \max_{\omega \in B_k} \left | \widehat{\psi}(2\pi \omega/2^{j_0}) \right |^2 , \max_{\omega \in B_k} \left | \widehat{\varphi}(2\pi \omega/2^{j_0}) \right |^2 \right \}.
}
}
\prf{
By definition,
\bes{
\mu\left ( U^{(k,l)} \right ) =|B_k|  \max_{\omega \in B_k} \max_{0 \leq n < 2^{j_0+l-1}} \left | \widehat{\psi^{\mathrm{per}}_{j_0+l-1,n}}(2\pi \omega) \right |^2,\quad l > 1,
}
and
\bes{
\mu\left ( U^{(k,1)} \right ) =|B_k| \max \left \{  \max_{\omega \in B_k}  \max_{0 \leq n < 2^l} \left | \widehat{\psi^{\mathrm{per}}_{j_0,n}}(2\pi \omega) \right |^2 ,  \max_{\omega \in B_k}  \max_{0 \leq n < 2^l} \left | \widehat{\varphi^{\mathrm{per}}_{j_0,n}}(2\pi \omega) \right |^2  \right \}.
}
Recall that $|B_k| \leq 2^{j_0+k}$.
Moreover, recall relation \eqref{Fourier_transf_phi} and note that an analogous formula holds for $\widehat{\psi_{j,k}^{\mathrm{per}}}$.
Since $B_k$ is a set of integers, the result now follows immediately. 

}

\prf{[Proof of Lemma \ref{l:localcoh}]
By the previous lemma, it suffices to estimate the Fourier transform of the wavelet and scaling function in different regions of frequency space.  First, suppose that $k \geq l \geq 1$.  Then $| \omega | \geq 2^{j_0+k-1}$ for $\omega \in B_k$, and the smoothness conditions \R{smoothness} give
\bes{
| \hat{\psi}(2\pi\omega/2^{l+j_0-1}) | \lesssim 2^{-(q+1)(k-l)},\qquad | \hat{\varphi}(2\pi \omega/2^{l+j_0-1}) | \lesssim 2^{-(q+1)(k-l)}. 
}
The first estimate now follows from Lemma \ref{l:localcoh1Dgeneralstep}.

For the second estimate, we need to bound $| \hat{\psi}(2 \pi \omega) |$ for $| \omega | \ll 1$.  For this, we recall that $\hat{\psi}(z) = (-\I z)^{p} \chi_p(z)$ for some bounded function $\chi_p(z)$ \cite[Thm.\ 7.4]{mallat09wavelet}.  Hence
\bes{
| \hat{\psi}(2 \pi \omega) |^2 \leq c_p |\omega|^{2p}.
}
If $l > k \geq 1$ then this and the previous lemma give
\bes{
\mu\left ( U^{(k,l)} \right ) \leq 2^{1+k-l} \max_{|\omega | \leq 2^{j_0+k}} | \hat{\psi}(2\pi\omega/2^{l+j_0-1}) |^2 \lesssim c_p 2^{k-l} 2^{2p(k-l)}. 
}
The result now follows immediately. 
}

\prf{[Proof of Lemma \ref{l:Atailbound}]
By direct calculation
\bes{
\nm{P_{\Omega} D U P^{\perp}_M d}^2_{\ell^2} \leq \sum^{r}_{k=1} \frac{N_k - N_{k-1}}{m_k} m_k \max_{N_{k-1} < i \leq N_k} | \ip{u_i}{P^{\perp}_M d} |^2,
}
where $u_i = U^* e_i$ is the $i^{\rth}$ row of $U$.  Observe that
\bes{
| \ip{u_i}{P^{\perp}_M d} |^2 = \left | \sum_{j > M} u_{ij} d_j \right |^2 \leq \max_{j > M} |u_{ij} |^2 \nm{P^{\perp}_M d}^2_{\ell^1}.
}
Hence
\eas{
\nm{P_{\Omega} D U P^{\perp}_M d}^2_{\ell^2} & \leq \sum^{r}_{k=1} (N_k - N_{k-1}) \max_{\substack{N_{k-1} < i \leq N_k \\ j > M}} |u_{ij} |^2  \nm{P^{\perp}_M d}^2_{\ell^1} = \sum^{r}_{k=1} \mu \left (P^{N_{k-1}}_{N_k} U P^{\perp}_M \right ) \nm{P^{\perp}_M d}^2_{\ell^1},
}
which gives
\bes{
\nmu{P_{\Omega} D U P^{\perp}_M d}_{\ell^2} \leq \left ( \sum^{r}_{k=1} \mu \left ( P^{N_{k-1}}_{N_k} U P^{\perp}_M \right ) \right )^{1/2} \nmu{P^{\perp}_M d}_{\ell^1}.
}

Since $M = M_r$, we now apply Lemma \ref{l:localcoh} to get
\bes{
\mu \left ( P^{N_{k-1}}_{N_k} U P^{\perp}_M \right ) = \sup_{l > r} \mu \left ( U^{(k,l)} \right ) \leq c_p  2^{-(2p+1)(r-k)}.
}
Hence 
\bes{
\sum^{r}_{k=1} \mu \left ( P^{N_{k-1}}_{N_k} U P^{\perp}_K \right ) \leq c_p \sum^{r}_{k=1} 2^{-(2p+1)(r-k)} \leq c_p  .
}
The result now follows.
}

\section{Numerical experiments}\label{s:NumExp}

\begin{figure}[t]
\centering
\includegraphics[height = 5.3cm]{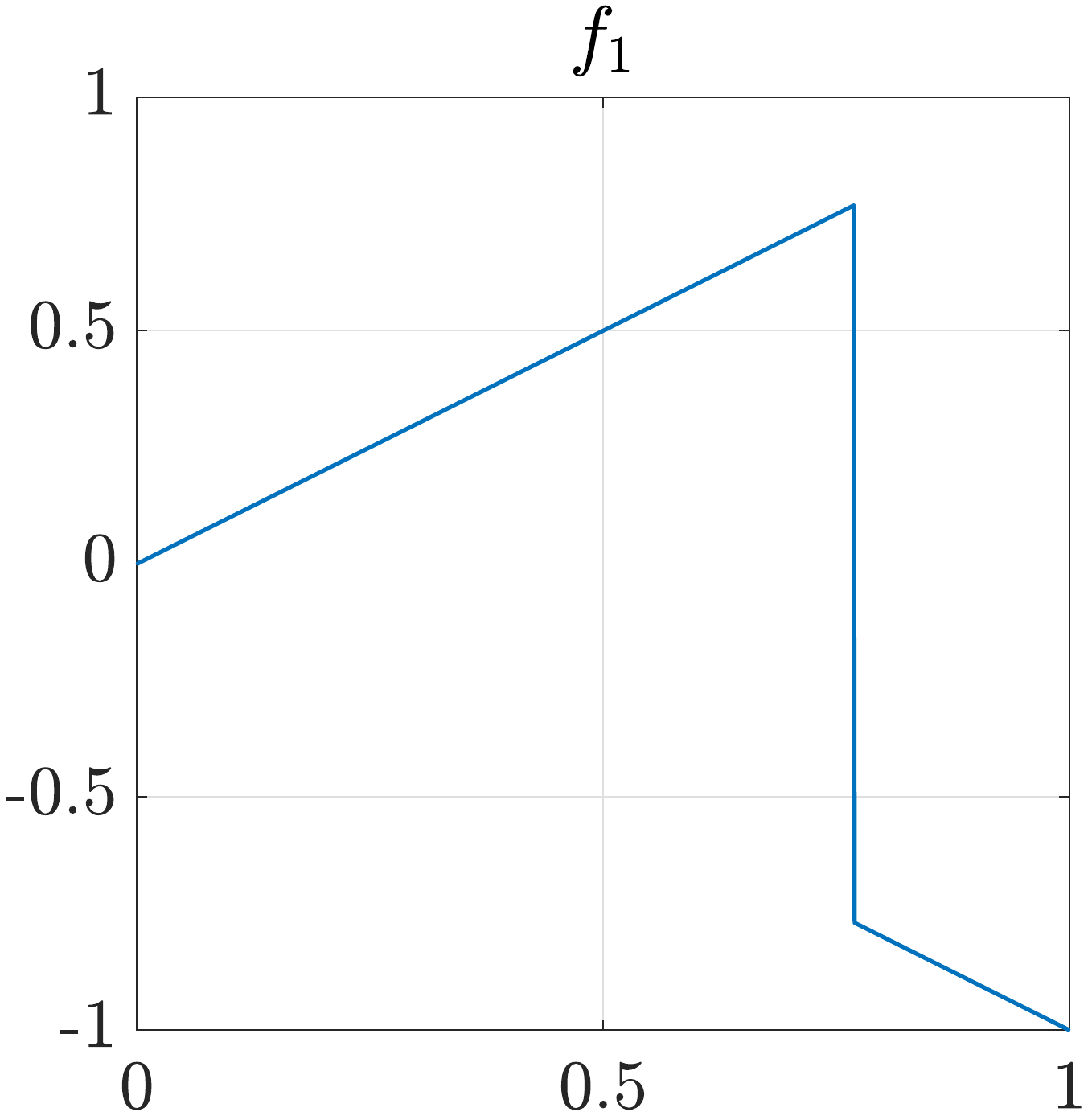}
\includegraphics[height = 5.3cm]{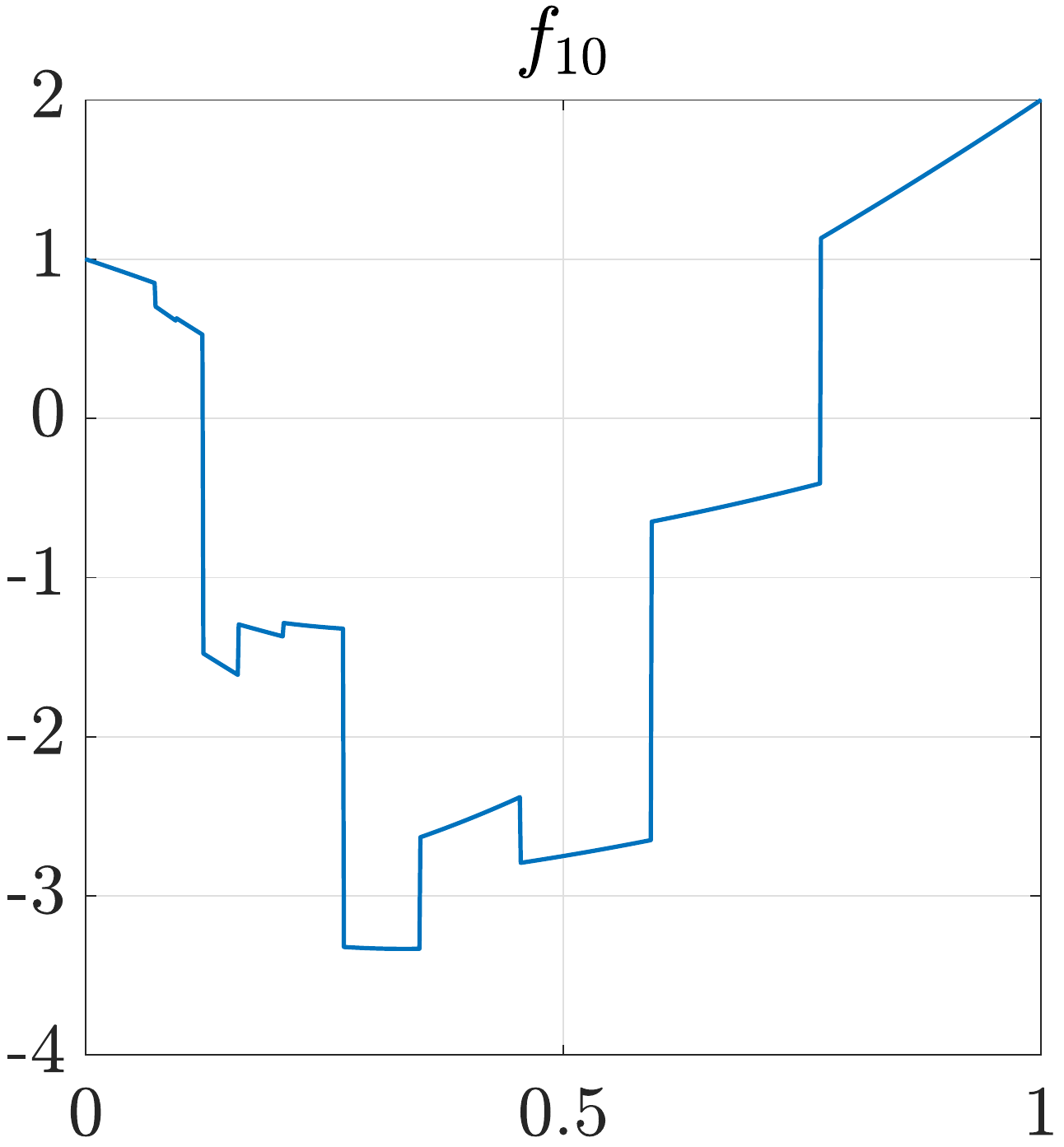}
\includegraphics[height = 5.3cm]{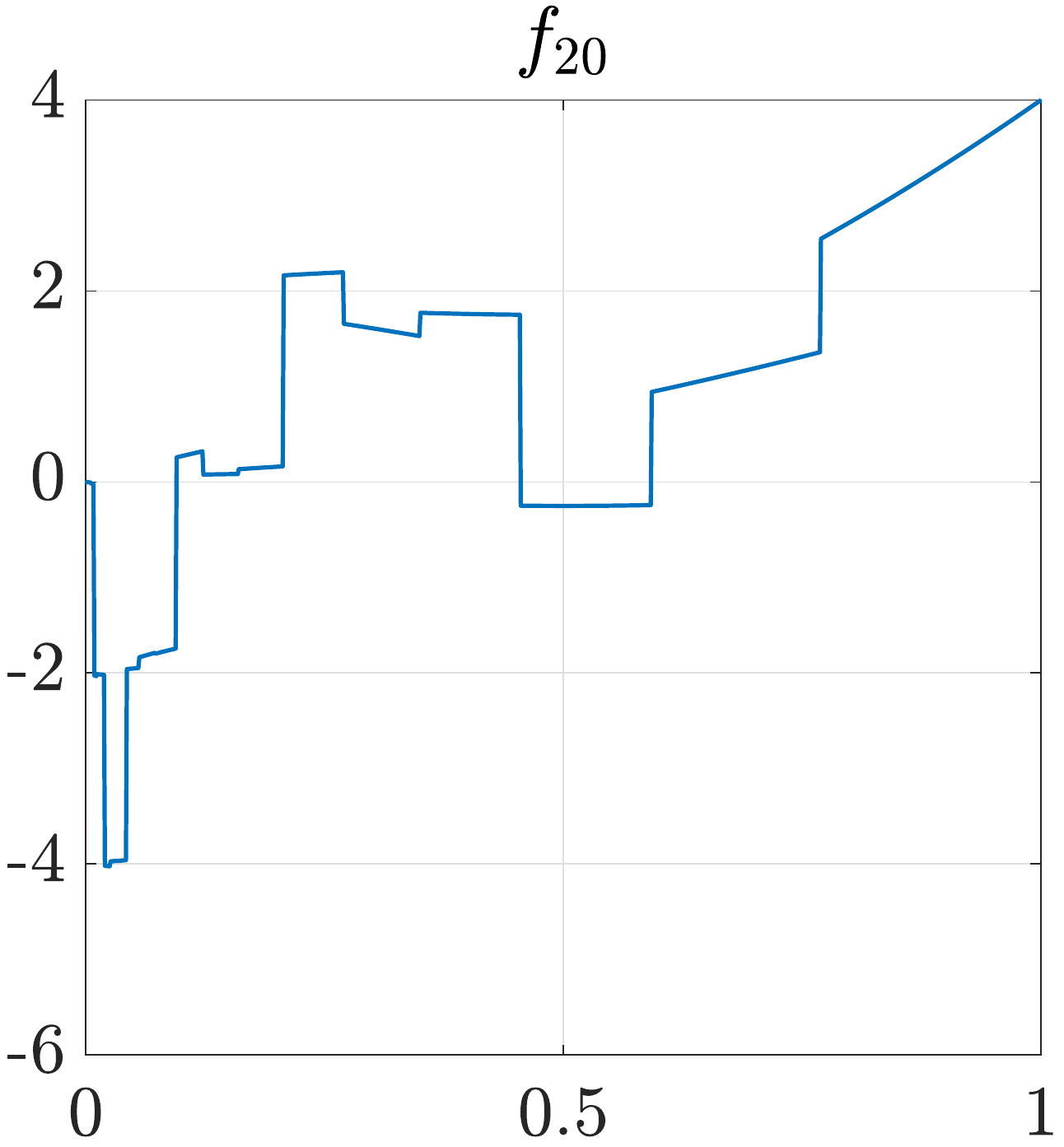}
\caption{\label{fig:test_functions}The function $f_K$ defined as in \eqref{eq:def_fk} for $K=1,10,20$.}
\end{figure}

In this section, we discuss some technical details behind Fig.~\ref{fig:intro_comp}. Moreover, we provide further numerical evidence to support the comparison shown therein.
We consider the function
\begin{equation}
\label{eq:def_fk}
f_K(x) = 
\sum_{i = 1}^{K} (-1)^{\text{mod}(i, 5)} \; x^{\text{mod}(i,3)} \; \text{sign}(x-(1.3)^{i-9}), \quad 0 \leq x \leq 1 .
\end{equation}
This funciton has $K$ discontinuities in $(0,1)$ and its plot is shown in Fig.~\ref{fig:test_functions}. 
We approximate $f_K$ for $K = 1,10,20$ using the four different encoder-decoder pairs described below.
\begin{figure}[t!]
\centering
\newcommand{\figsize}{0.4\textwidth}
\includegraphics[width = \figsize]{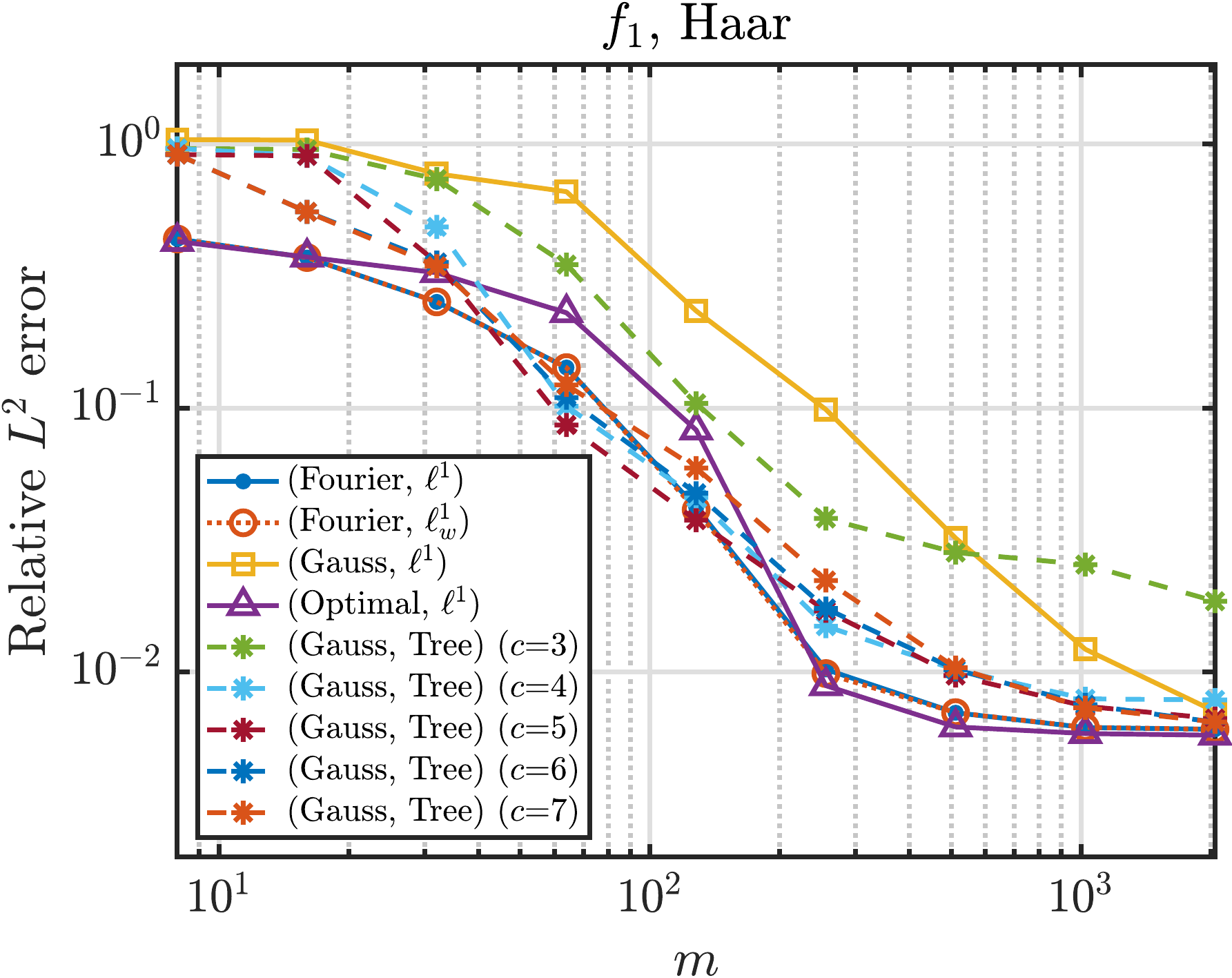}
\includegraphics[width = \figsize]{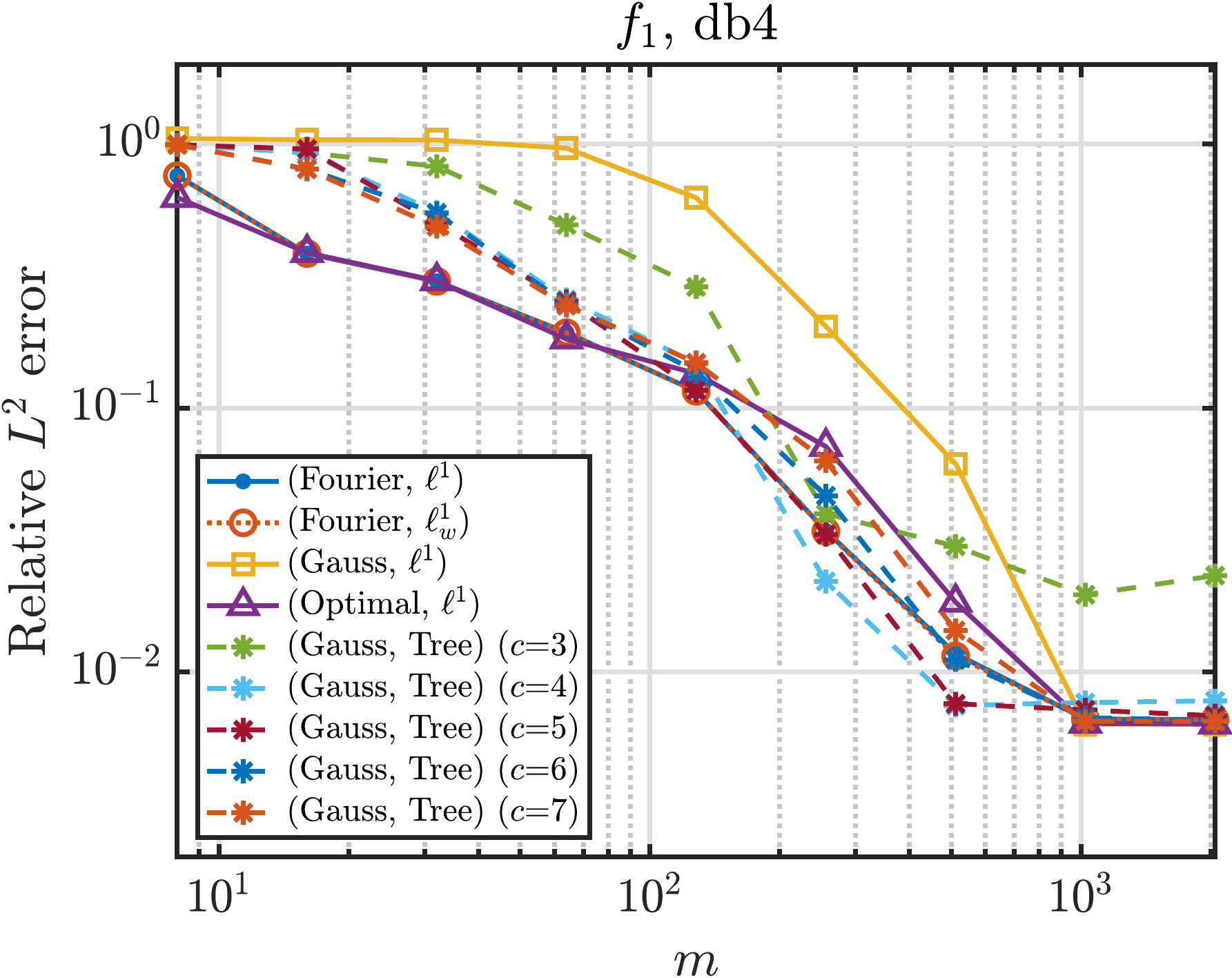}\\
\includegraphics[width = \figsize]{fig/PLOT_f10_haar_rmax_15_rmax_m_11_it_100-eps-converted-to.pdf}
\includegraphics[width = \figsize]{fig/PLOT_f10_db4_rmax_15_rmax_m_11_it_100-eps-converted-to.pdf}\\
\includegraphics[width = \figsize]{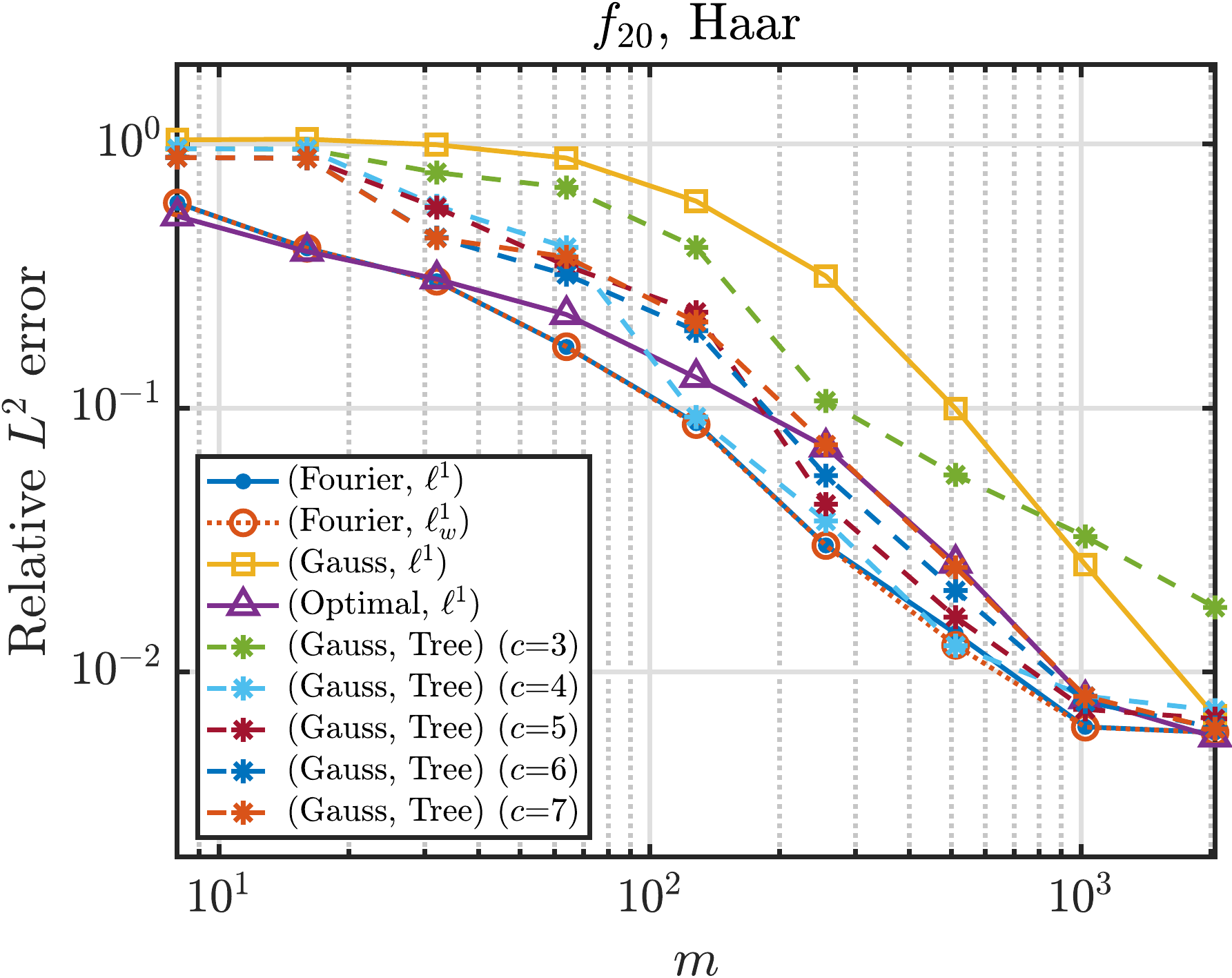}
\includegraphics[width = \figsize]{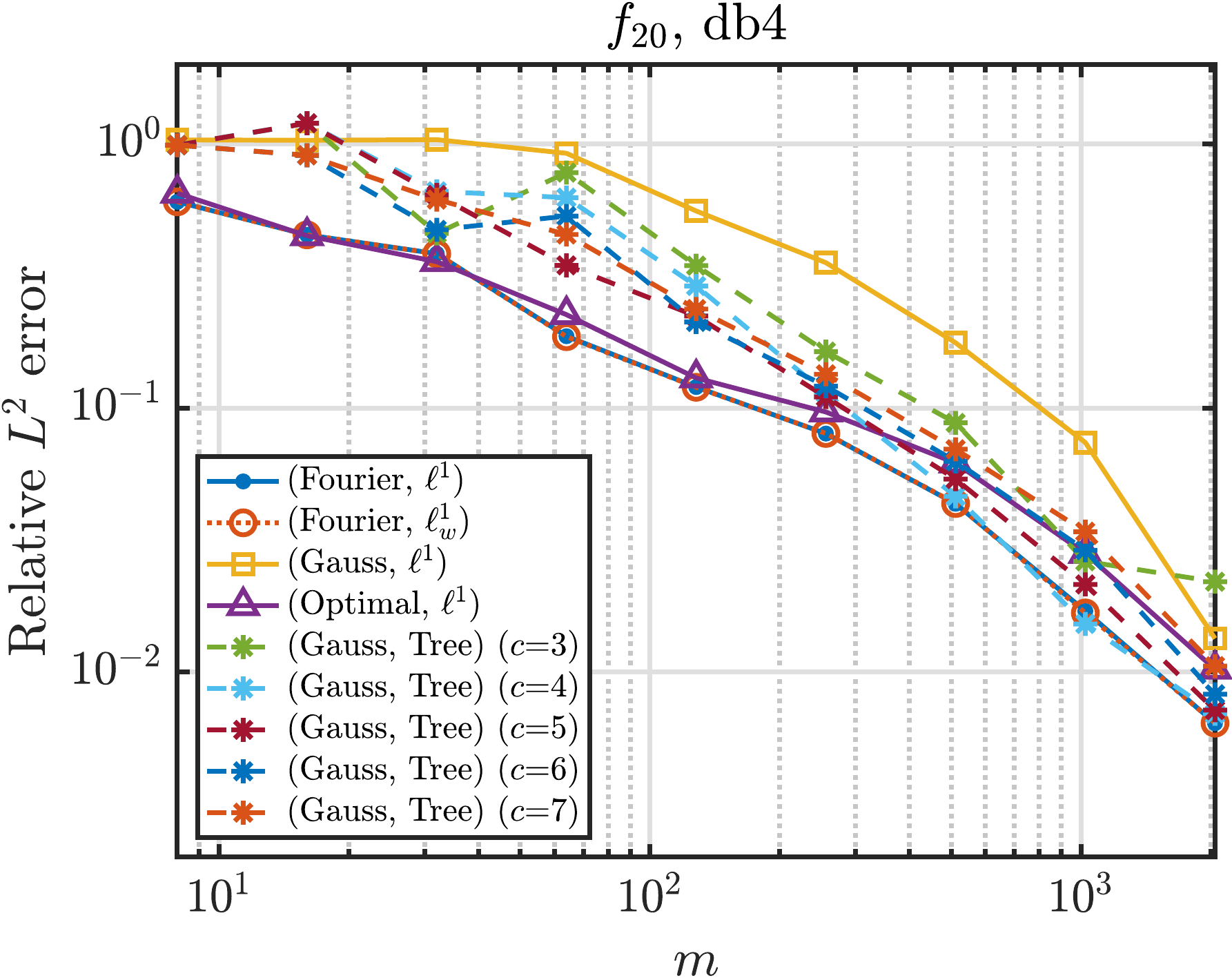}
\caption{\label{fig:comparison_N_32768}Comparison of different encoder-decoder pairs for the approximation of the function $f_K$ defined in \eqref{eq:def_fk}, using Haar (left) and db4 wavelets (right) and for $K = 1$ (top), $K=10$ (center), and $K = 20$ (bottom).}
\end{figure}

\pbk
\textbf{(Fourier,~$\ell^1$):}
This strategy corresponds to the setting of Theorems~\ref{t:FourErr} and \ref{t:FourErr2} and to the error bound \eqref{FourErr}, up to a few minor technical modifications. The Fourier sampling strategy is as follows. We divide the frequency space into dyadic bands and consider a sampling scheme analogous to the $(\mb{N},\mb{m})$-multilevel random subsampling strategy with saturation $\tilde{r}$ described in Definition~\ref{def:sampling_pattern}, where symmetry of the samples is enforced in every frequency band. In particular, $\mb{N}$ is defined as in \eqref{Nkvalues}, the saturation level is $\tilde{r} = \text{round}(\log_2(m/2))$, and the local numbers of measurements are
$$
m_k = 2\left\lfloor \frac{m}{4(r-r_0)}\right\rfloor, \quad k = \tilde{r} +1, \ldots, r-1,
$$ 
where, in the last frequency band, we let $m_r = m-(m_1 +\cdots + m_{r-1})$ in order to reach a total budget of $m$ measurements exactly. The samples are then computed as follows. The first $\tilde{r}$ dyadic bands are saturated. For every $k > \tilde{r}$, we pick $m_k/2$ samples  uniformly at random from the $k$-th frequency semiband (corresponding to positive frequencies) and we choose frequencies in the opposite semiband (corresponding to negative frequencies) in a symmetric way.   The wavelet coefficients of $f$ are recovered via  basis pursuit \eqref{BP}.
 Numerically, \eqref{BP} is solved using the Matlab package SPGL1 (see \cite{spgl1:2007,BergFriedlander:2008}) with parameters $\texttt{bpTol}$ = 1e-6, $\texttt{optTol}$= 1e-6, and a maximum of $10000$ iterations.\footnote{The entries of the  cross-Gramian matrix $U$ \eqref{Udef} used in this sampling strategy are computed by applying the inverse discrete wavelet and Fourier transforms to the first $N$ elements of the canonical basis of the augmented space $\mathbb{R}^{16N}$. Then, only the $N$ entries corresponding to the frequencies of interest are kept. This augmentation makes the computation of $U$ more accurate.}

\pbk
\textbf{(Fourier,~$\ell^1_w$):} This strategy is almost identical to (Fourier,~$\ell^1$). The only difference is that wavelet coefficients are recovered via weighted (as opposed to unweighted) basis pursuit, i.e., by solving \eqref{BP} where the $\ell^1$-norm is replaced with the weighted $\ell^1_w$-norm.
The weights $w$ are set according to the recipe described in \S\ref{sec:recipe_Fourier} with $\delta = 10^{-5}$. Weighted basis pursuit is numerically solved using the Matlab package SPGL1 as in the previous case.

\pbk
\textbf{(Gauss,~$\ell^1$):}
This is the standard encoder-decoder pair of compressed sensing with random Gaussian measurements, corresponding to the setting of Theorems~\ref{t:GaussErr} and \ref{t:GaussErr2} and to the error bound \eqref{GaussErr}. The vector  $d \in \mathbb{R}^N$ of wavelet coefficients of $f$ is explicitly computed and then encoded as $y = A d$, where $A \in \mathbb{R}^{m \times N}$ has i.i.d.\ entries drawn from the normal distribution with mean zero and variance $1/m$. The function is recovered by means of the basis pursuit decoder \eqref{BP}, numerically solved via SPGL1 as in the previous cases.\footnote{In order to avoid discretization effects related to the wavelet crime, the vector $d$ of wavelet coefficients is computed by sampling the function $f$ on a uniform grid of $16 N$ points, applying the discrete wavelet transform, and then keeping the first $N$ of entries of the resulting vector.}


\pbk
\textbf{(Optimal,~$\ell^1$):}
This strategy corresponds to the setting of Theorems~\ref{t:GaussErr} and \ref{t:GaussErr2} and to the optimal error bound \eqref{GaussErr}. As in the previous case, we compute the vector $d\in \mathbb{R}^N$ of wavelet coefficients of $f$. Then, the first $m_1 = \text{round}(m/2)$ entries of $d$ are directly encoded into $y^{(1)} \in \mathbb{R}^{m_1}$. The remaining $m_2 = m - m_1$ measurements are computed as $y^{(2)} = A (d_n)_{n = m_1+1}^{N}$, where $A \in \mathbb{R}^{m_2 \times (N-m_1)}$ has i.i.d.\ entries drawn from the normal distribution with mean zero and variance $1/m_2$. We consider the basis pursuit decoder \eqref{BP}, numerically solved using SPGL1 as in the previous cases.


\pbk
\textbf{(Gauss,~Tree):}
This encoder-decoder pair corresponds to the model-based  compressive sensing strategy proposed in \cite{BaranuikModelCS}. The encoder identical to (Gauss,~$\ell^1$), and the decoder explicitly promotes tree-structured sparsity in the recovered function using the model-based CoSaMP algorithm \cite{BaranuikModelCS}. This strategy requires tuning a parameter $c$, which links $m$ to the desired tree-sparsity level $s$ as $m = c s$. In the numerical tests, we consider $c = 3,4,5,6,7$. We employ the Model-based Compressive Sensing Toolbox v1.1 provided by the authors of \cite{BaranuikModelCS}. The maximum number of iterations for the outer loop of CoSaMP is set to 100. 

\pbk
These four encoder-decoder pairs are compared with $N = 2^{15} = 32768$ and values of $m$ ranging from $2^3 = 8$ to $2^{11} = 2048$. We employ Haar and db4 wavelets, having $p= 1$ and $p = 2$ vanishing moments, respectively. In this setting, the weights used in (Fourier,~$\ell^1_w$) are constant for all $m \leq 256$.
The relative $L^2$ error is computed using the wavelet coefficients of $f$, approximated as in the strategies (Gauss,~$\ell^1$), (Optimal,~$\ell^1$), and (Gauss,~Tree). 

In Fig.~\ref{fig:comparison_N_32768} the encoder-decoder pairs (Fourier,~$\ell^1$) and (Fourier,~$\ell^1_w$) have almost identical performances and they consistently outperform all the other strategies, with only a few exceptions. Moreover, this behaviour is independent of the number of discontinuities $K$. It is remarkable that (Fourier,~$\ell^1$) and (Fourier,~$\ell^1_w$) are able to numerically outperform even the theoretically-optimal pair (Optimal,~$\ell^1$). Although our theory prescribes the use of weighted square-root LASSO decoder in the Fourier case, the numerics show that employing (weighted or unweighted) basis pursuit \eqref{BP} is enough to numerically outperform the other strategies. 

\bibliographystyle{abbrv}
\small
\bibliography{biblio}

\end{document}